%% file: main.tex
\documentclass[aps,prx,reprint,showpacs,amsmath,amssymb,floatfix,noeprint,nofootinbib]{revtex4-2}

\usepackage{dsfont}
\usepackage{placeins}
\usepackage{xcolor}

\definecolor{blue}{RGB}{0, 0, 128}
\definecolor{teal}{RGB}{0, 100, 100}
\definecolor{darkred}{RGB}{139, 0, 0}
\usepackage{hyperref}
\hypersetup{linktocpage,colorlinks,citecolor={blue},pdfdisplaydoctitle=true,pdfpagemode=UseOutlines,bookmarksnumbered=true,urlcolor={teal},linkcolor={darkred}}

\usepackage{graphicx} 
\usepackage[percent]{overpic}
\usepackage{lipsum}
\usepackage{mathrsfs}
\usepackage{physics}
\usepackage{siunitx}
\DeclareSIUnit{\belmilliwatt}{Bm}
\DeclareSIUnit{\dBm}{\deci\belmilliwatt}
\usepackage{tensor}

\definecolor{hhcolor}{RGB}{30,144,255}
\definecolor{atcolor}{RGB}{0,100,0}

\usepackage{stackengine}
\stackMath

\newcommand{\e}{\mathrm{e}}
\newcommand{\brho}{\boldsymbol{{\rho}}}

\newcommand{\Ahat}{\hat{A}}
\newcommand{\Adag}{\hat{A}^\dagger}
\newcommand{\Bhat}{\hat{B}}
\newcommand{\Bdag}{\hat{B}^\dagger}

\newcommand{\ahat}{\hat{a}}
\newcommand{\adag}{\hat{a}^\dagger}
\newcommand{\bhat}{\hat{b}}
\newcommand{\bdag}{\hat{b}^\dagger}

\renewcommand{\ketbra}[2]{|#1\rangle \langle#2|}
\renewcommand{\braket}[2]{\langle#1|#2\rangle}

\newcommand{\drm}{\mathrm{d}}

\newcommand{\Ccal}{\mathcal{C}}

\newcommand{\Dcal}{\mathcal{D}}

\newcommand{\Lhat}{L}
\newcommand{\Hhat}{\hat{H}}
\newcommand{\Uhat}{\hat{U}}
\newcommand{\Ghat}{G}

\newcommand{\intinf}{\int_{-\infty}^{+\infty}}

\newcommand{\pref}[1]{(\ref{#1})}

\newcommand{\vac}{\ket{\mathrm{vac}}}
\newcommand{\vacd}{\bra{\mathrm{vac}}}

\newcommand{\customcaption}[2]{\mathbf{\hat{\sigma}_+}}

\newcommand{\lbb}{\{\hspace{-0.38em}\{}
\newcommand{\rbb}{\}\hspace{-0.38em}\}}

\makeatletter
\newsavebox\myboxA
\newsavebox\myboxB
\newlength\mylenA

\newcommand*\xoverline[2][0.9]{%
    \sbox{\myboxA}{$\m@th#2$}%
    \setbox\myboxB\null
    \ht\myboxB=\ht\myboxA%
    \dp\myboxB=\dp\myboxA%
    \wd\myboxB=#1\wd\myboxA
    \sbox\myboxB{$\m@th\overline{\copy\myboxB}$}
    \setlength\mylenA{\the\wd\myboxA}
    \addtolength\mylenA{-\the\wd\myboxB}%
    \ifdim\wd\myboxB<\wd\myboxA%
       \rlap{\hskip 0.5\mylenA\usebox\myboxB}{\usebox\myboxA}%
    \else
        \hskip -0.5\mylenA\rlap{\usebox\myboxA}{\hskip 0.5\mylenA\usebox\myboxB}%
    \fi}
\makeatother

\begin{document}
\renewcommand\stackalignment{l}
\title{Calculus of Robinet: completely positive reconstruction of time-averaged diffusive quantum trajectories}

\author{Hector Hutin}
\author{Antoine Tilloy}
\affiliation{Laboratoire de Physique de l’École Normale Supérieure - PSL, Centre Automatique et Systèmes Mines Paris - PSL, CNRS, Inria, PSL Research University, Paris, France}

\date{\today}
\begin{abstract}
Truly continuous quantum trajectories, obtained from homodyne or heterodyne readouts, can only ever be reconstructed approximately. The continuous measurement signal, needed for exact reconstruction, is averaged over bins of finite time $\Delta t$ during any analog to digital conversion step. The best reconstruction possible, knowing only this discrete record, was introduced recently and dubbed the Robinet state. In this article, we show how the Robinet state can be computed with a numerical discretization scheme that is completely positive, accurate to arbitrarily high order in $\Delta t$, and that does not rely on any other external solver. Our derivation relies on a dilation of the stochastic master equation into a system + transmission line setup, constructed in such a way that measuring what we call the ``zero mode'' of the line yields the Robinet state. We test the method on a challenging example with random Hamiltonian and jump operator, and verify its accuracy up to order $10$. Apart from its numerical interest, our approach provides a wealth of physical insights, extending in particular recent results on purity obtained by Wonglakhon, Chantasri, and Wiseman, that would be difficult to obtain in any other way.
\end{abstract}
\maketitle

\section{Introduction}
\subsection{Context and motivation}
From quantum optics to superconducting circuits, one can now measure quantum systems continuously in the lab~\cite{weber2014nature,negretti2013,six2015cdc,philippe2015,hutin2024}. In such a situation, a measurement apparatus provides a continuous signal, from which one can in principle reconstruct the state of the system in real time. The standard formalism to describe this situation, and our starting point in this article, is the Stochastic Master Equation (SME) \cite{davies1976,gardiner1985book,carmichael1993,belavkin1999,breuer2002,jacobs2006,wiseman2009,jacobs2010}.

To be concrete, we consider a finite-dimensional quantum system with density matrix $\rho_t$, with Hamiltonian $H$ (self-adjoint), that is continuously measured with an operator $L$ (not necessarily self-adjoint) with efficiency $\eta \leq 1$. We assume that there is no additional (unmonitored) dissipation acting on the quantum system. The SME is a pair of It\^o stochastic differential equations (SDE), relating the state $\rho_t$ and the signal. The first SDE gives the evolution of the system state $\brho_t$ given the stochastic signal $\frac{\drm \boldsymbol{y}_t}{\drm t}$:
\begin{align}\label{eq:SME-homodyne}
	\drm \brho_t & = -i[H, \brho_t]\,\drm t + \Dcal[L](\brho_t) \drm t \nonumber                                                                   \\
	            & \quad + \sqrt{\eta}\,\mathcal{M}[L](\brho_t) \left(\drm \boldsymbol{y}_t - \sqrt{\eta}\, \langle L + L^\dagger\rangle_{\brho_t}\drm t\right)
\end{align}
where we have introduced, as is customary,
\begin{equation}
	\Dcal[L](\rho) := L \rho L^\dagger - \frac{1}{2}\left(L^\dagger L \rho + \rho L^\dagger L\right) \, ,
\end{equation}
which encodes the decoherence and dissipation introduced by measurement,
and
\begin{equation}
	\mathcal{M}[L](\rho) := L \rho + \rho L^\dagger - \tr\left[(L+L^\dagger)\rho\right] \rho\,
\end{equation}
sometimes called the \emph{stochastic innovation}, which encodes the back-action of the measurement on the system state. We also used the compact notation $\langle A\rangle_{\rho} = \tr(\rho A)$. Here, we distinguish a stochastic process (denoted in bold font $\boldsymbol{W}_t$) from a realization of this process $W_t$ (denoted in normal font).
If $\eta=1$, the SDE \eqref{eq:SME-homodyne} preserves the purity of the quantum state. The second SDE relates the statistics of the measurement record $\frac{\drm y_t}{\drm t}$ to the system state
\begin{align}\label{eq:SME-signal}
	\drm \boldsymbol{y}_t = \sqrt{\eta}\, \langle L + L^\dagger\rangle_{\brho_t}\drm t + \drm \boldsymbol{W}_t
\end{align}
where $\boldsymbol{W}_t$ is a Wiener process (equivalently a $1$-dimensional Brownian motion).

There are many ways to derive the pair of SDE \eqref{eq:SME-homodyne} and \eqref{eq:SME-signal}. The technically easiest (and perhaps most pedagogical) derivation is to take a continuum limit of repeated weak interactions with an ancilla, that is then projectively measured \cite{attal2006,pellegrini2008,pellegrini2009,pellegrini2010,jacobs2006,ciccarello2022}. Alternatively, the SME can also be derived straight from the continuum, by considering a system coupled to a transmission line -- modeled by a continuum of harmonic oscillators -- which is then measured at its end~\cite{gardiner1985, clerk2010,abdo2013,flurin2014,hutin2024d}. This latter derivation is slightly more demanding technically, but closer to the physics of experiments where continuous measurements are realized. Further, as we will see, this approach yields results beyond the SME that we would not know how to obtain in any other way.

\subsection{Robinet}
A crucial point is that in practice, the state $\rho_t$ cannot be reconstructed exactly. Indeed, one only ever has access to a sequence of \emph{time-averaged} (or digitized) signals
\begin{equation}\label{eq:digitized}
	I_m = \int_{(m-1)\Delta t}^{m\Delta t} {\drm y_t}\, ,
\end{equation}
where $\Delta t$ is the time resolution of some ultimate analog-to-digital conversion. This averaging induces an inevitable loss of information that one needs to take into account. Note that to solve the SDE, one would anyway need to discretize time in some way. But the step $\drm t$ needed to solve the SDE in a precise and stable way need not be related to the time-bin $\Delta t$ that is relevant experimentally (usually $\Delta t \gg \drm t$).

This led to the introduction of the so-called \emph{Robinet state} $\bar{\rho}_n$ in \cite{guilmin2025}, the object we focus on in this article. It is the quantum state obtained by averaging over all the possible continuous signals compatible with a given digitized record $\{I_k\}$:
\begin{equation}\label{eq:robinet_def}
	\bar{\rho}_n := \mathbb{E}\big[\brho_{n\Delta t} \,|\, I_1, I_2, \dots, I_n\big] \, .
\end{equation}
It was shown in \cite{guilmin2025} that $\bar{\rho}_n$ can be obtained recursively from $\bar{\rho}_{n-1}$:
\begin{align}\label{eq:robinet-recursion}
	\bar{\rho}_n = \frac{\mathcal{K}_{I_n}(\bar{\rho}_{n-1})}{\tr\left[\mathcal{K}_{I_n}(\bar{\rho}_{n-1})\right]},
\end{align}
where $\mathcal{K}_{I_n}$ is a completely positive (CP) map, \textit{i.e.} an element of a quantum instrument. Since the average map must be CPTP, we also have $\int \drm I\, \mathcal{K}_I = \mathds{1}$. Perhaps surprisingly, the map $\mathcal{K}_I$ can be computed explicitly using only stochastic calculus techniques, starting from \eqref{eq:SME-homodyne} and \eqref{eq:SME-signal}, and \cite{guilmin2025} gave:
\begin{equation}\label{eq:Kcal}
	\mathcal{K}_I = \frac{1}{2\pi} \int_\mathbb{R} \drm p\, \e^{i p I - \Delta t\frac{p^2}{2}} \e^{\Delta t(\mathcal{L} -ip\,\mathcal{C})},
\end{equation}
with the Lindblad (super)operator $\mathcal{L}(\rho):= -i[H,\rho] + \mathcal{D}[L](\rho)$ and $\mathcal{C}(\rho) := \sqrt{\eta}\,(L\rho + \rho L^\dagger)$. To compute this map numerically, one option used in \cite{guilmin2025} is to evaluate the integral in $p$ with quadratures, and compute the exponential $\e^{\Delta t(\mathcal{L} -ip_k\,\mathcal{C})}\cdot \rho_0$ at each quadrature point $p_k$ with an external Krylov or Runge-Kutta solver.

Alternatively, one can expand the map in powers of $\Delta t$ (taking into account the fact that $I$ is of order $\sqrt{\Delta t}$ for $\Delta t$ smaller than other dynamical timescales). This lets us evaluate it without relying on an external solver, and provides a natural high-order discretization scheme. One can stop at any desired order to get a numerical scheme with an error of order $\Delta t^N$. However, stopping at any finite order yields a map that is not exactly CP, and with an average over the signal that is not exactly CPTP. We thus lose the structural properties of $\mathcal{K}_I$. This is a major drawback of this expansion: at every time-bin, we may get a state $\bar{\rho}_n$ that is slightly unphysical (for example with tiny negative eigenvalues). This could be projected away, but it is known that schemes that preserve CP typically behave better both in terms of precision and stability \cite{rouchon2015,robin2025,cao2025a,appelo2025}.

At sufficiently low order, and for efficiency $\eta = 1$, one can reshuffle the expansion into an explicitly CP form. In \cite{guilmin2025}, it was found that up to order $\sqrt{\Delta t}^{5}$, the map can be written with a single Kraus operator $B_I$:
\begin{equation}
	\mathcal{K}_I(\bar{\rho}) := B_I \bar{\rho} B_I^\dagger + O\left(\sqrt{\Delta t}^6\right)
\end{equation}
Surprisingly, since it is made of a single Kraus operator, it preserves the purity of an initially pure state at an order higher than naively expected. In \cite{wonglakhon2026}, Wonglakhon, Chantasri, and Wiseman (WCW) showed that adding an integral $J_n$ of the measurement record against a first-order polynomial, on top of the information contained in $I_n$, reaches order $\Delta t^{3}$ while preserving purity.

\subsection{Goals and results}
In this article, our main objective is to construct an explicitly CP scheme to obtain $\mathcal{K}_I$ to arbitrarily high order, with an exactly CPTP average. In doing so, we obtain a powerful and stable numerical scheme as well as a wealth of unexpected physical results.

While \cite{guilmin2025} used purely stochastic calculus techniques, starting directly from the SME, we found it necessary to take a step back in its derivation, and start from a unitary dilation where the system interacts with a continuous harmonic bath. This comes with a small technical cost, but provides a physical intuition for some results that might have otherwise seemed puzzling.

The first result that naturally comes off this approach is the derivation of arbitrarily high-order expansions of the map $\mathcal{K}_I$ in Kraus form, thus preserving its CP property exactly at each order. Using standard techniques, we also get a map that is exactly CPTP on average. We use this to build integration schemes to reconstruct $\bar{\rho}_n$ from a given sequence of binned signals $I_1,\dots, I_n$. In practice, for a single measurement channel and efficiency $\eta=1$, we can compute terms up to and including order $10$, for homodyne detection. We verify the order of our scheme by testing it against the numerically exact method of cascaded quantum systems (analogous to the quadrature approach of \cite{guilmin2025}). We can readily extend the method to heterodyne detection and finite efficiency $\eta$. It is also possible to include multiple measurement channels, time-dependent Hamiltonian and measurement operators, although in these cases we can reach only lower orders numerically.

Further, the tools we develop also let us sample binned signals with a probability distribution accurate to the same order. This, in turn, makes it possible to sample quantum trajectories $n\mapsto \{\bar{\rho}_n,I_n\}$ that are physical and accurate in law to arbitrarily high order in $\Delta t$.

As a byproduct of these numerical developments, our method rederives the result of WCW \cite{wonglakhon2026} in a compact way. We also see that, in general, there is no way to extend their result to order $\Delta t^4$ by adding a \emph{finite} number of integrals of the signal against appropriate functions. For example, adding a record $\{J_n^{(k)}\}_{k\leq k_\text{max}}$ corresponding to integrals of the signal against all polynomials of degree $k\leq k_\text{max}$ still yields a loss of purity at order $\Delta t^4$. This does not prevent the reconstruction of the Robinet state (or WCW state) to arbitrarily high order in $\Delta t$, but simply means that the ``sharp'' state $\rho_t$ cannot be approached faster asymptotically by adding such information.

\subsection{Strategy}
Let us first explain the intuition of our derivation. The map $\mathcal{K}_I$ of \eqref{eq:Kcal} is CP but not obviously so. Ideally, we would like to write an expansion of $\mathcal{K}_I$ up to a fixed order $\Delta t^{2N}$ exactly in Kraus form, which would enforce CP:
\begin{equation}
	\mathcal{K}_I(\rho) = \sum_{\mu} B_\mu(I)\rho B_{\mu}^\dagger(I) + o(\Delta t^{2N})\, ,
	\label{eq:K_krauss_form}
\end{equation}
where the sum over $\mu$ would be finite.

Of course, one could try to reshuffle the expansion of \cite{guilmin2025} into this form. We instead follow a more direct and systematic route: we construct the operators $B_{\mu}(I)$ from a \emph{dilation}, that is, from a unitary evolution on a larger Hilbert space in which $\mu$ appears as an additional discrete measurement outcome on top of the continuous outcome $I$. Since dilations of the SME \eqref{eq:SME-homodyne} in terms of quantum noise, or equivalently a continuum of harmonic oscillators, are standard, they provide a natural way to obtain the explicit CP form \eqref{eq:K_krauss_form}.

For simplicity, we work on a single time bin and initialize the system at $t=0$. We then enlarge the Hilbert space by adding a bosonic Hilbert space
\begin{equation}
\mathscr{H}_\text{line}= \mathcal{F}(L^2([0,\Delta t])),
\end{equation}
the symmetric Fock space on the interval $[0,\Delta t]$. Because the dynamics is Markovian, this is equivalent, up to a time shift, to describing the evolution from $t$ to $t+\Delta t$: the state on this interval depends only on the system state at time $t$, not on the earlier history.

This constitutes a standard \emph{dilation} that naturally recovers the Lindblad dynamics together with a CP representation of the integrated evolution on $[0,\Delta t]$. In particular, if the initial system-line state is $\ket{\Psi_0} = \ket{\psi_0} \otimes \vac$, a generic unitary evolution brings it to:
\begin{equation}
	\ket{\Psi_{\Delta t}} = \sum_{\mu} M_\mu \ket{\psi_0} \otimes \ket{g_\mu}\label{eq:fullstate_line}
\end{equation}
where we assume only that the states of the line are orthogonal $\langle g_\mu |g_\nu\rangle = \delta_{\mu\nu}$. This series \eqref{eq:fullstate_line} is infinite, but assuming the order of $M_\mu$ in $\Delta t$ grows with $\mu$, we can truncate $\mu$ to $\mu_\text{max}$ to get the state up to a finite order $\Delta t^{N}$. Finally, if we trace over the line we get
\begin{align}
	\varrho_{\Delta t} : & = \tr_{\text{line}} \left[\ket{\Psi_{\Delta t}}\bra{\Psi_{\Delta t}}\right]       \\
	                       & = \sum_{\mu=0}^{\mu_\text{max}} M_\mu\ket{\psi_0}\bra{\psi_0} M_\mu^\dagger +o (\Delta t^{2N})\, ,
\end{align}
which generalizes to an initially mixed state of the system $\ket{\psi_0}\bra{\psi_0}\rightarrow \varrho_0$. This is an explicitly CP form as required.

This construction can be generalized from the Lindblad case to the Robinet state. To this end, one observes that the Hilbert space of the line has a natural tensor product structure in terms of a zero mode and all the rest of $L^2$: $\mathscr{H}_\text{line} = \mathscr{H}_0 \otimes \mathscr{H}_\text{rest}$. We choose a standard interaction (which we detail below) such that measuring a quadrature on the zero mode Hilbert space gives us the Robinet state.
Assuming as before that the initial system + line state is $\ket{\Psi_0} = \ket{\psi_0} \otimes \ket{0} \otimes \vac$, a generic unitary evolution brings it to:
\begin{equation}
	\ket{\Psi_{\Delta t}} = \sum_{\mu,k} P_{\mu,k} \ket{\psi_0} \otimes \ket{k} \otimes \ket{h_\mu}
	\label{eq:fullstate}
\end{equation}
where $\ket{k}$ is just the Fock state with $k$ excitations, and $\ket{h_\mu}$ are orthogonal states of $\mathscr{H}_\text{rest}$: $\langle h_\mu |h_\nu\rangle = \delta_{\mu\nu}$. As before, the orthogonalization of $\mathscr{H}_\text{rest}$ is the crucial step needed to get $\mathcal{K}_I$ in Kraus form.

Measuring the projector $\ket{I}\bra{I}$ on $\mathscr{H}_0$ ---which, as we detail later, corresponds to a measurement of a quadrature averaged over a constant bin--- gives the (unnormalized) state:
\begin{equation}
	\ket{\tilde{\Psi}_{\Delta t}(I)} = \sum_{\mu,k} P_{\mu,k} \ket{\psi_0} \otimes \langle I \ket{k} \ket{I} \otimes \ket{h_\mu}
\end{equation}
with probability density $\langle \tilde{\Psi}_{\Delta t}(I) |\tilde{\Psi}_{\Delta t}(I)\rangle$. The (unnormalized) reduced density matrix of the system $\tilde{\rho}_{\Delta t}(I)$ post measurement is thus:
\begin{equation}
	\tilde{\rho}_{\Delta t}(I)=\tr_{\mathscr{H}_\text{line}}\left[\ket{\tilde{\Psi}_{\Delta t}(I)}\bra{\tilde{\Psi}_{\Delta t}(I)}\right]
\end{equation}
which gives:
\begin{align}
	\tilde{\rho}_{\Delta t}(I): & = \mathcal{K}_I(\rho_0)                                                                                                           \\
	                              & = \sum_{\mu} \left(\sum_k\langle I |k\rangle P_{\mu,k}\right) \rho_0 \left(\sum_{k'}\langle I |k'\rangle P_{\mu,k'}\right)^\dagger\, ,
	\label{eq:eqfullstate_I}
\end{align}
with $\rho_0 := \ket{\psi_0}\bra{\psi_0}$ (the generalization to an initially mixed system state is immediate as well).

Now, $\mathcal{K}_I$ is explicitly in Kraus form, with $B_{\mu}(I) = \sum_{k}\langle I|k\rangle P_{\mu,k}$. If $P_{\mu,k}$ can be expanded in $\Delta t$, and if terms with larger $\mu$ or $k$ are of increasing order in $\Delta t$, then we can keep only a finite number of terms in the sum to reach a given order in $\Delta t$ for $\mathcal{K}_I$, which gives the desired result \eqref{eq:K_krauss_form}. Finally, to get a map that is also exactly CPTP (and not just CP), we need to use a standard normalization trick, which we explain in Sec.~\ref{sec:renormalization_procedure}.

\section{Unitary description of continuous measurement}
In what follows, we first explain more precisely how the Hilbert space of the line is rigorously constructed, and how it relates to the stochastic representation of the SME \eqref{eq:SME-homodyne}. We then recall the standard unitary evolution between system and line that recovers the dynamics we desire, and then explain how this yields the joint system-line wave-function \eqref{eq:fullstate_line} exactly. Finally, we expand this representation in $\Delta t$ to get the explicit CP form of $\mathcal{K}_I$.
\subsection{Fock space of the line}
The Hilbert space $\mathscr{H}_\text{line}$ is a Fock space that we can construct from singular distribution valued creation and annihilation operators $\adag(t),\ahat(t)$ such that $[\ahat(t), \adag(u)] = \delta(t-u)$ acting on a vacuum state $\ket{\text{vac}}$ such that $\forall t,~ \ahat(t) \ket{\text{vac}}=0$. We get regular (unnormalized) creation and annihilation operators by smearing the sharp $\ahat(t)$ with a function in $L^2([0,\Delta t],\mathbb{C})$
\begin{align}
	A(f) = \int_0^{\Delta t}\, f^*(t)\,  \ahat(t)\,  \drm t,
\end{align}
which gives $\big[A(f),A^\dagger(g)\big] = \int_0^{\Delta t} f^*\, g$. This latter commutation relation can be used to \emph{define} $A$ in a rigorous way, without reference to $\ahat$. We will however stick to the standard physics level of rigor and manipulate distribution valued operators for convenience.

Given $g^{(1)}$ a normalized function in $L^2([0,\Delta t], \mathds{C})$, we can construct a one-photon state in the line by applying the creation operator $A[g^{(1)}]$ to the vacuum state $\vac$:
\begin{align}
	\ket{g^{(1)}} = A^\dagger[g^{(1)}] \vac = \int_0^{\Delta t}\drm t \, g^{(1)}(t)\, \adag(t) \vac \, .
\end{align}
More generally, an $n$-photon state is built from a symmetric normalized $n$-dimensional function $g^{(n)}\in L^2([0,\Delta t]^n, \mathds{C})$ as
\begin{align}
	\begin{split}
		  & \ket{g^{(n)}}:=                                                                                                                                   \\& \int_{[0,\Delta t]^n}\!\!\! \frac{\drm t_1\cdots \drm t_n}{n!}\, g^{(n)}(t_1, \dots, t_n) \adag(t_1)\cdots \adag(t_n)\vac\\
		= & \int_{\Delta t\geq t_1 \geq \cdots \geq t_n\geq 0}\hskip-1.4cm \drm t_1\cdots \drm t_n\, g^{(n)}(t_1, \dots, t_n) \adag(t_1)\cdots \adag(t_n)\vac
	\end{split}
\end{align}
The scalar product of two $n$-photon states is simply:
\begin{align}
	\begin{split}
		\langle f^{(n)}|g^{(n)}\rangle = \int_{\mathscr{S}_n^{\Delta t}} f^{(n)*} g^{(n)},
	\end{split}
\end{align}
where we have written $\mathscr{S}_n^{\Delta t}$ the simplex $\{t_1,\cdots,t_n\,  | \, \Delta t\geq t_1 \geq \cdots \geq t_n\geq 0\}$. A generic state in the transmission line is a superposition of states with different photon numbers, and two states with different numbers of photons are orthogonal.

\subsection{Joint system-line evolution}
We now describe the interaction between the system and such a transmission line, that ultimately corresponds to the SME \eqref{eq:SME-homodyne}. To this end, one introduces the Hamiltonian
\begin{align}
	H_\mathrm{tot} = H + H_{\mathrm{int}}(t),
\end{align}
where $H_{\mathrm{int}}(t)$ is the interaction Hamiltonian between the system and the line, given by
\begin{align}\label{eq:continuous-Hint}
	H_\mathrm{int}(t) = i L\adag(t) -iL^\dagger\ahat(t).
\end{align}
The corresponding unitary evolution $U$ over a time $\Delta t$ is
\begin{equation}\label{eq:unitary_def}
	U := \mathcal{T}\exp\left\{\int_0^{\Delta t}\!\!\drm t -i H + L \adag(t) - L^\dagger \ahat(t)\right\}
\end{equation}
Working with such unitary operators obtained from a time-ordering is notoriously subtle. Indeed, since $[\ahat(t),\adag(u)] = \delta(t-u)$, $\ahat(t)\, \drm t$ scales like $\sqrt{\drm t}$, and the order of $\adag$ and $\ahat$ at equal time thus matters. This is the quantum equivalent of the distinction between Stratonovich and It\^o conventions in stochastic calculus \cite{gardiner1985, zoller1997a}.

In practice, we will only act with $U$ on states of the form $\ket{\psi}\otimes \vac$, that is states where the line is in its vacuum state. Using quantum stochastic calculus, in either of its flavors, one can show \cite{vonwaldenfels2017}
\begin{equation}\label{eq:unitary_on_vac}
	U \ket{\psi} \otimes \vac = \mathcal{T}\exp\left\{\int_0^{\Delta t}\!\! \drm t\; G+ L a^\dagger(t) \right\} \ket{\psi} \otimes \vac\\
\end{equation}
with $G = -i H - \frac{1}{2}L^\dagger L$. This form \eqref{eq:unitary_on_vac} is now unambiguous, because $\adag(t)$ commutes with itself.

One way to understand this result intuitively, without going into the formalism of quantum stochastic calculus, is to use the Baker-Campbell-Hausdorff (BCH) formula to bring each term in the time-ordered product \eqref{eq:unitary_def} in normal-ordered form:
\begin{align}
	\e^{\drm t [L \adag - L^\dagger\ahat ]} & \simeq \e^{\drm t L\adag} \e^{-\drm t L^\dagger \ahat} \e^{-\frac{\drm t^2}{2} [L\adag, L^\dagger\ahat]} \\
	                                        & \simeq  \e^{\drm t L\adag- \frac{L^\dagger L}{2}} \e^{- \drm t L^\dagger \ahat}
\end{align}
where we have used $[\ahat,\adag] = 1/\drm t$ and discarded all terms of order higher than $\drm t$. Acting on the vacuum removes the rightmost exponential and yields formula \eqref{eq:unitary_on_vac}.

\subsection{Recovering the Lindblad equation, continuous measurement, and the Robinet state} \label{sec:recovering}
We now rederive the standard results on continuous measurement with this unitary formalism, starting with the Lindblad equation, obtained by tracing over the line\footnote{The reader familiar with the continuous tensor network literature will note that this problem is formally identical to the computation of the norm of a CMPS, as done \textit{e.g.} in~\cite{haegeman2013}.}.

Starting from $\varrho_0 = \rho_0\otimes \vac \vacd$ and applying \eqref{eq:unitary_on_vac} (the anti-time-ordering of $U^\dagger$ becomes time-ordering in the right-multiplication convention), we get $\varrho_{\Delta t} = U \varrho_0 U^\dagger$, and $\rho_{\Delta t}$ as
\begin{align}\label{eq:lindblad_trace}
	 \rho_{\Delta t} =& \tr_\text{line} \left[\varrho_{\Delta t}\right]                                          \nonumber\\
	 =& \tr_\text{line} \left[U\left(\rho_0 \otimes \vac \vacd\right)U^\dagger\right]                                          \nonumber\\
	                                           = &\vacd \mathcal{T}\exp\left\{\int_0^{\Delta t} \!\!\drm t\, G_r^\dagger + L_r^\dagger a (t) \right\}\nonumber \\
	                                           & \times \mathcal{T}\exp\left\{\int_0^{\Delta t}\!\! \drm t\, G_\ell + L_\ell a^\dagger (t) \right\} \vac\cdot\rho_0 \, ,
\end{align}
where we have introduced the superoperator notations $A_\ell \cdot \rho = A\rho$, $A_r \cdot \rho = \rho A$. 

To evaluate the right-hand side of eq. \eqref{eq:lindblad_trace}, we apply the same BCH technique as for \eqref{eq:unitary_on_vac} to each infinitesimal factor in the time-ordered products. Since $A_\ell$ and $B_r$ commute, the only non-trivial commutator at the $\drm t$ scale is $[L_r^\dagger \ahat, L_\ell \adag] = L_r^\dagger L_\ell\, [\ahat,\adag] = L_r^\dagger L_\ell / \drm t$. Fusing the two exponentials by BCH and normal-ordering (each step contributing $\frac{1}{2}L_r^\dagger L_\ell\, \drm t$, for a total of $L_r^\dagger L_\ell\, \drm t$) yields:
\begin{align}\label{eq:bch_lindblad}
	\e^{\drm t\, (G_r^\dagger + L_r^\dagger a)}\, \e^{\drm t\, (G_\ell + L_\ell a^\dagger)} &\simeq \e^{\drm t\, L_\ell a^\dagger}\, \e^{\drm t\,\mathcal{L}}\, \e^{\drm t\, L_r^\dagger a}.
\end{align}
Using $\e^{\drm t\, L_\ell a^\dagger} \vac = \vac$, $\vacd \e^{\drm t\, L_r^\dagger a}  = \vacd$,  leaving only $\e^{\drm t\, \mathcal{L}}$ at each step, and integrating finally gives $\mathcal{T}\e^{\int_0^{\Delta t} \drm t\, \mathcal{L}} = \e^{\Delta t\, \mathcal{L}}$ (since $\mathcal{L}$ is time-independent):
\begin{align}
	\tr_\text{line} \left[U\left(\rho_0 \otimes \vac \vacd\right)U^\dagger\right] &= \exp(\Delta t\, \mathcal{L})\cdot \rho_0 = \mathbb{E}\left[\brho_{\Delta t}\right]
\end{align}
with the Lindblad superoperator
\begin{equation}\label{eq:lindblad_superop}
	\mathcal{L}(\rho) := G\rho + \rho G^\dagger + L\rho L^\dagger= -i[H,\rho] + \Dcal[L](\rho).
\end{equation}

We now show that homodyne measurement can be recovered as well. A homodyne readout of the line corresponds to a measurement of the quadrature operator $\ahat(t) + \adag(t)$ at every time $t\in [0,\Delta t]$, and yields a measurement record $t\mapsto I_t = \frac{\drm y_t}{\drm t}$ (which is technically a distribution, $y_t$ has the regularity of the Brownian motion). The state conditioned on this readout is exactly the one given by the SME \eqref{eq:SME-homodyne}.

This can be proven using standard techniques of quantum stochastic calculus \cite{barchielli1988,zoller1997,wiseman1993}. We propose a quick alternative proof using the characteristic function of the measurement record. Namely, we show that the statistics of the measurement record is given by the same \emph{tilted Liouvillian} as the one obtained from the SME \cite{guilmin2025}. This approach has the advantage of highlighting the connection between the characteristic function in the sense of probability theory, and the quantum characteristic function of the field, which is a more standard object in quantum optics.

To this end, we first define:
\begin{align}
D^{\ahat}_{\Delta t}(\beta) =& \mathcal{T}\exp{\int_0^{\Delta t} \beta(t)\adag(t) - \beta^*(t) \ahat(t)\drm t}
\end{align}  
the generalized displacement operator. In the following, we use its anti-normal-ordered form, which can be obtained by applying the BCH formula to each infinitesimal factor in the time-ordered product:
\begin{align}
	\begin{split}
	D^{\ahat}_{\Delta t}(\beta) = \e^{\frac{1}{2}\int_0^{\Delta t} |\beta(t)|^2 } \;  \e^{-\int_0^{\Delta t} \beta^*\ahat}
	\; \e^{\int_0^{\Delta t}  \beta \adag}\, .
	\end{split}
\end{align}
For $\beta = i J$ with $J$ a real-valued function, $D^{\ahat}_{\Delta t}(iJ) = \mathcal{T}\exp\{\int iJ(a^\dagger + a)\}$ is the Weyl operator for the quadrature $\ahat + \adag$. Inspired by the standard derivation of signal correlation functions~\cite{barchielli1993,zoller1997,tilloy2018exact}, it is natural to introduce a tilted density matrix:
\begin{align}\label{eq:tilted_rho_from_trace}
	\rho_{\Delta t}^{\beta} := & \tr_\text{line} \left[D^{\ahat}_{\Delta t}(\beta) \varrho_{\Delta t} \right] \\
	= & \tr_\text{line} \left[D^{\ahat}_{\Delta t}(\beta) U\left(\rho_0 \otimes \vac \vacd\right)U^\dagger\right].
\end{align}
Intuitively, this object contains all the information about the statistics of $\ahat(t)+ \adag(t)$ jointly with the system state.

Inserting the anti-normal-ordered form of $D^{\ahat}_{\Delta t}(\beta)$ between the vacuum and the two time-ordered exponentials from \eqref{eq:lindblad_trace}, the annihilation part $\mathcal{T}\exp\{-\int \beta^*\, a\}$ can be absorbed into the $r$-exponential (shifting $L_r^\dagger \to L_r^\dagger - \beta^*$), while the creation part $\mathcal{T}\exp\{\int \beta\, a^\dagger\}$ is absorbed into the $\ell$-exponential (shifting $L_\ell \to L_\ell + \beta$). This gives
\begin{align}
	 \rho_{\Delta t}^{\beta}                                          & = \vacd \mathcal{T}\exp\left\{\int_0^{\Delta t} \!\!\drm t\, G_r^\dagger +  \left(L_r^\dagger - \beta^*(t)\right) a (t) \right\} \nonumber \\
	                                           & \times \mathcal{T}\exp\left\{\int_0^{\Delta t}\!\! \drm t\, G_\ell + \left(L_\ell +  \beta(t)\right)a^\dagger (t) \right\} \nonumber\\&\times  \exp{\frac{1}{2}\int_0^{\Delta t} \drm t\, |\beta|^2(t)}\vac \cdot \rho_0.     
\end{align}
Applying the same BCH formula and vacuum-contraction procedure as for \eqref{eq:lindblad_trace}, with the substitution $L_\ell \to L_\ell + \beta$ and $L_r^\dagger \to L_r^\dagger - \beta^*$, the vacuum contraction now produces $(L_r^\dagger -\beta^*)(L_\ell + \beta) = L_r^\dagger L_\ell - \beta^* L_\ell + \beta L_r^\dagger - |\beta|^2$ at each time step, yielding:
\begin{align}
	 \rho_{\Delta t}^{\beta}                                          & =\mathcal{T}\exp{\int_{0}^{\Delta t} \drm t\mathcal{L}_t^{\beta}}\cdot \rho_0
\end{align}
where $\mathcal{L}^{\beta}$ denotes the \emph{tilted Liouvillian}~\cite{tilloy2018exact,guilmin2023,landi2024current}
\begin{align}
  \mathcal{L}_t^{\beta} := \mathcal{L} + \left[\beta(t)L_r^\dagger - \beta^*(t)L_\ell \right] - \frac{1}{2}|\beta(t)|^2 \, .
\end{align}
The joint statistics of the observable $-i\int_{0}^{\Delta t}\drm t\beta(t)\adag(t) - \beta^*(t)\ahat(t)$ and the system state are thus encoded in the tilted Liouvillian $\mathcal{L}^{\beta}$ applied to the initial state. 

For $\beta =  i J$ with $J$ a real function, we have $\mathcal{L}_t^{iJ} := \mathcal{L} + iJ \Ccal - \frac{1}{2}J^2$
and $D^{\ahat}_{\Delta t}(iJ) = \mathcal{T}\exp{\int_0^{\Delta t} iJ(\ahat(t) + \adag(t))\drm t}$ which encodes the statistics of the measurement of the $\ahat + \adag$ quadrature. 
Thus, in that case, the expression for the tilted density matrix defined from the trace over the line  in \eqref{eq:tilted_rho_from_trace} should match the one obtained from stochastic calculus
\begin{align}\label{eq:tilte_rho_from_sto}
	\rho^{iJ}_{\Delta t} = \mathbb{E}\left[\exp{\int_0^{\Delta t} iJ(t)\drm \boldsymbol{y}_t} \boldsymbol{\rho}_{\Delta t} \right],
\end{align}
where $\brho$ and $\boldsymbol{y}_t$ follow the SME defined in Eq.~\pref{eq:SME-homodyne} and Eq.~\pref{eq:SME-signal}, and generate the whole joint statistics of $\boldsymbol{\rho}_t$ and $\drm \boldsymbol{y}_t$. This is indeed the case, as shown \textit{e.g.} in~\cite{guilmin2025}. This establishes that both approaches yield the same measurement-record statistics, as well as the same correlations between the measurement record and the system state.

From here, it is possible to carry on the derivation of the Robinet state done in \cite{guilmin2025}, but using the unitary dilation formalism. Let the integrated measurement record be $I_f = \int_0^{\Delta t} f(t)\frac{\drm y_t}{\drm t}\drm t$ for some real-valued function $f$. This corresponds to a measurement of the quadrature $\Ahat[f] + \Adag[f]$ of the line. The corresponding back-action on the system is given by the projection $\Pi_{I_f}$ of the joint system-line state, which we can write as 
\begin{align}\label{eq:homodyne_projection}
	\Pi_{I_f} = \delta(I_f-(\Ahat[f] + \Adag[f])) &= \int_\mathbb{R} \frac{\drm p}{2\pi} \e^{i p (I_f-(\Ahat[f] + \Adag[f]))}\\
	& = \int_\mathbb{R} \frac{\drm p}{2\pi} \e^{i p I_f} D(-i p f)
\end{align}
Taking the trace on the environment of the projected state thus gives the unnormalized post-measurement state of the system:
\begin{align}
	\tilde{\rho}_{\Delta t}(I_f) & = \tr_\text{line} \left[\Pi_{I_f} U\left(\rho_0 \otimes \vac \vacd\right)U^\dagger\right]                                          \\
	& = \int_\mathbb{R} \frac{\drm p}{2\pi} \e^{i p I_f} \rho_{\Delta t}^{-i p f}\\
	& = \int_\mathbb{R} \frac{\drm p}{2\pi} \e^{i p I_f} \mathcal{T}\exp{\int_{0}^{\Delta t} \drm t\mathcal{L}_t^{-ipf}}\cdot \rho_0
\end{align}
where we used the cyclic property of the trace and $\Pi_{I_f}^2 = \mathds{1}$. For $f = 1$, we get the Robinet state, obtained by applying $\mathcal{K}_I$ defined in Eq.~\pref{eq:Kcal} for $\eta = 1$.

A similar procedure yields the Robinet state for heterodyne readout by measuring in a coherent overcomplete basis instead of a quadrature basis. We present it in Appendix~\ref{app:heterodyne_robinet}.

\section{Explicit CP form for the Lindblad equation}
\noindent In this section, we first explain how an explicitly CP time discretization of the Lindblad equation can be obtained using our dilation technique. This recovers known results in a compact and very general way, and serves as a warm-up before the extension to the Robinet state. The code used for the automated derivation of the results presented in this section and in Sec.~\ref{sec:DSME} is available on GitHub and has been permanently archived on Zenodo~\cite{hutin2026robinetcalculus}.

\subsection{Expanding the state of the line}
Our objective is now to expand $U$ applied to the initial state of the system $\ket{\Psi(0)} = \ket{\psi_0}\otimes \vac$, up to a given order in $\Delta t$, to write the final state $\ket{\Psi(\Delta t)}$ in the form
\begin{align}
	\ket{\Psi(\Delta t)} = \sum_{\mu=0}^{\mu_\text{max}} M_\mu \ket{\psi_0}\ket{g_\mu} + O(\Delta t^{N+\frac{1}{2}})
\end{align}
where we recall that $\{M_\mu\}$ act only on the system, and $\{\ket{g_\mu}\}$ are some (multi-photon) states of the line.

We call here $N$ the ``half-order'', as $2N$ is the order of the associated CP map on the system density matrix (see Sec.~\ref{sec:Lindblad}). The number $N$ can be an integer or a half-integer.

Expanding the time-ordered exponential in \eqref{eq:unitary_on_vac} gives
\begin{align}
	\begin{split}
		U \ket{\Psi(0)}     & = \sum_{n=0}^{+\infty} \int_{\Delta t \geq t_1 \geq  t_2 \geq  \cdots \geq t_n} \drm t_1  \drm t_2 \cdots \drm t_n \\
		e^{G(\Delta t-t_1)} & L\ahat(t_1)^\dagger e^{G(t_1-t_2)}\ldots L \ahat(t_n)^\dagger e^{G t_n}\ket{\Psi(0)}.
	\end{split}
\end{align}
We get ``propagators'' $\e^{(t_j-t_{j+1} ) G}$ separated by operator insertions of $L$. In the context of quantum measurement, this formula already appears in Barchielli's work \cite{barchielli1988}. It is also a well-known formula for tensor network states, where it represents the wave-function expansion of a continuous matrix product state (CMPS) \cite{verstraete2010,haegeman2013} (in this context, the notation is usually $G\rightarrow Q$ and $L\rightarrow R$).

We may further expand the remaining exponentials in a Taylor series:
\begin{align}
	\begin{split}
		 & U \ket{\Psi(0)} =
		\\ &\sum_{n=0}^{+\infty}\sum_{k_0, k_1, \ldots, k_n = 0}^{+\infty} \int_0^{\Delta t} \drm t_1 \int_0^{t_1} \drm t_2 \ldots \int_0^{t_{n-1}}\drm t_n  \\
		 & \times F_{k_0, k_1, \ldots, k_n}(t_1, t_2, \ldots, t_n)\ahat(t_1)^\dagger\ldots\ahat(t_n)^\dagger\ket{\Psi(0)},
	\end{split}
\end{align}
where we have defined the operator $F$ acting on the system
\begin{align}
	\begin{split}
		 & F^{(n)}_{k_0, k_1, \ldots, k_n}(t_1, t_2, \ldots, t_n)                                                                                           \\
		 & ~~~~:= \frac{(\Delta t-t_1)^{k_0}}{k_0!}\Ghat^{k_0}\Lhat\frac{(t_1-t_2)^{k_1}}{k_1!}\Ghat^{k_1}\Lhat\ldots\Lhat\frac{t_n^{k_n}}{k_n!}\Ghat^{k_n} \\
	\end{split}
\end{align}
The time-dependent part and system operator part factorize and we can write $F^{(n)}_{k_0,\dots, k_n}= O_{k_0, \ldots, k_n}f_{k_0, \ldots, k_n}(t_1, t_2, \ldots, t_n)$ with
\begin{align}
	{O}_{k_0, \ldots, k_n} = \Ghat^{k_0}\Lhat\Ghat^{k_1}\ldots\Lhat \Ghat^{k_n}
\end{align}
and
\begin{align}
	\begin{split}
		f^{(n)}_{k_0, \ldots, k_n}(t_1, t_2, & \ldots, t_n)= \\ &\frac{(\Delta t-t_1)^{k_0}}{k_0!}\Big(\prod_{i = 1}^{n-1}\frac{(t_i-t_{i+1})^{k_i}}{k_i!}\Big)\frac{t_n^{k_n}}{k_n!}.
	\end{split}
\end{align}
This expansion makes the order of $f^{(n)}_{k_0,\dots,k_n}$ in $\Delta t$ explicit. Recall that $f$ can be identified with the wave-function of an unnormalized $n$-photon state in the line, with scalar product $\langle f^{(n)} | g^{(m)}\rangle = \delta_{mn}\int_{\mathscr{S}_n^{\Delta t}} f^*\, g$. Since $O$ is of order $1$, we thus have that the state $F^{(n)}_{k_0, \ldots, k_n}\ket{\Psi(0)}$ is of order $n/2+\sum_i k_i$ in $\Delta t$. Since this is growing both with $k_i$ and $n$, we have a well-defined and finite hierarchy of states to keep to reach a given order in $\Delta t$ for $\ket{\Psi(\Delta t)}$.

\subsection{Orthogonalizing the states of the line}

Once we have the previous expansion up to a certain order, the second step is to orthogonalize the line states. This is achieved simply by Gram-Schmidt orthonormalization, which can be done automatically with a computer algebra software~\cite{hutin2026robinetcalculus}. We ultimately obtain the representation we advertised, for any half-order $N$
\begin{align}
	\ket{\Psi(\Delta t)} = \sum_{\mu=0}^{\mu_\text{max}} M_\mu \ket{\psi_0}\ket{g_\mu^{(n_\mu)}} + O(\Delta t^{N + \frac{1}{2}})
\end{align}
where the sum is finite, and the states $\{g_\mu^{(n_\mu)}\}$ form an orthonormal family. The superscript $n_\mu$ is simply aimed at making explicit the number of photons of the corresponding line states.

In practice, to obtain the full expansion up to half-order $N$, we thus need to follow these three steps:
\begin{enumerate}
	\item Choose a half-order $N$ (can be half-integer).
	\item Generate all the functions $f^{(n)}_{k_0, \ldots, k_n}$ such that $n/2+\sum_i k_i \leq N$.
	\item Perform a Gram-Schmidt orthonormalization procedure on the corresponding states $\ket{f^{(n)}_{k_0, k_1, \ldots, k_n}}$, which gives a finite orthonormal family $\{g_\mu^{(n_\mu)}\}$.
\end{enumerate}
The result of this procedure up to half-order $N = 2$ is an orthonormal family of 8 functions~\cite{hutin2026robinetcalculus}:
\begin{align}
	\begin{split}
		 & N=0~:~ g_0^{(0)} = 1~,                                                                                        \\
		 & N=\frac{1}{2}~:~  g_1^{(1)}(t_1) = \frac{1}{\sqrt{\Delta t}}~,                                                \\
		 & N=1~:~ g_2^{(2)}(t_1, t_2) = \frac{\sqrt{2}}{\Delta t}~,                                                      \\
		 & N=\frac{3}{2}~:~ g_3^{(1)}(t_1) = \frac{\sqrt{3} \left(- \Delta t + 2 t_{1}\right)}{\Delta t^{\frac{3}{2}}}~, \\
		 & N=\frac{3}{2}~:~ g_4^{(3)}(t_1, t_2, t_3) = \frac{\sqrt{3!}}{\Delta t^{\frac{3}{2}}}~,                        \\
		 & N=2~:~ g_5^{(2)}(t_1, t_2) = \frac{2 \left(- \Delta t + 3 t_{2}\right)}{\Delta t^{2}}~,                       \\
		 & N=2 ~:~ g_6^{(2)}(t_1, t_2) = \frac{2 \sqrt{3} \left(- \Delta t + 2 t_{1} - t_{2}\right)}{\Delta t^{2}}~,     \\
		 & N=2~:~g_7^{(4)}(t_1, t_2, t_3, t_4) = \frac{\sqrt{4!}}{\Delta t^{2}}~,
	\end{split}
\end{align}
where the superscript $(n)$ again indicates the number of variables (and thus of photons).

Analysing terms up to half-order $N=2$ is already instructive. Five of these states are particularly simple: the functions $g_0^{(0)}, g_1^{(1)}, g_2^{(2)}, g_4^{(3)}, g_7^{(4)}$ correspond to the zero-mode of the line populated with 0, 1, 2, 3 and 4 photons respectively:
\begin{align}
	\ket{g_0^{(0)}} & = \frac{1}{\sqrt{0!}} A^{\dagger 0}[g_1^{(1)}]\vac = \ket{0}_0 \otimes \vac_\text{rest},    \\
	\ket{g_1^{(1)}} & = \frac{1}{\sqrt{1!}}A^{\dagger 1}[g_1^{(1)}]\vac = \ket{1}_0\otimes \vac_\text{rest},      \\
	\ket{g_2^{(2)}} & =\frac{1}{\sqrt{2!}} A^{\dagger 2}[g_1^{(1)}]\vac = \ket{2}_0\otimes \vac_\text{rest},      \\
	\ket{g_4^{(3)}} & =\frac{1}{\sqrt{3!}} A^{\dagger 3}[g_1^{(1)}]\vac =\ket{3}_0 \otimes \vac_\text{rest},      \\
	\ket{g_7^{(4)}} & = \frac{1}{\sqrt{4!}}A^{\dagger 4}[g_1^{(1)}]\vac = \ket{4}_0 \otimes \vac_\text{rest} \, ,
\end{align}
where we are using the decomposition of the Fock space $\mathscr{H}_\text{line}$ into $\mathscr{H}_0 \otimes \mathscr{H}_\text{rest}$ that we advertised before.
Up to half-order $N = 1$, all the terms are of this form, and thus the environment can be truncated to a single mode, represented by the function $g_1^{(1)}$. This lets us quickly foreshadow our results on the Robinet state, discussed in the next section. The Robinet state is obtained by measuring a quadrature of this zero mode, and thus keeps the state pure up to this order. At the density matrix level, this implies that the purity is preserved up to order $2N = 2$ (instead of naively $1$, which, as we will see in Sec.~\ref{sec:multiple_jump_operators_robinet}, is the order one recovers for joint non-commuting measurements). This provides a particularly transparent physical intuition for these mathematical results derived in~\cite{guilmin2025}.

Starting with half-order $N = 3/2$, we need to add the state $\ket{g_3^{(1)}}$, which is a 1-photon state occupying the mode with wave-function given by the first shifted Legendre polynomial over the interval $[0, \Delta t]$ (which we call simply the first mode). Hence, at half-order $N = 3/2$, we need two modes of the line to represent the state exactly. Measuring these two modes makes the system state pure. This explains why, at order $2N = 3$, one can still preserve the purity of the system by joint homodyne measurement of two distinct modes, as found by WCW~\cite{wonglakhon2026}. The extra measurement proposed by WCW is the integral of the signal against a ramp, which, upon keeping the orthogonal to the zero mode only, corresponds to measuring a quadrature of the mode with wave-function $g_3^{(1)}$.

Finally, at half-order $N=2$, we obtain two 2-photon states $g_5^{(2)}$ and $g_6^{(2)}$ which are not symmetric. It is convenient to rotate them by $\pi/3$, which gives
\begin{align}
	g_{5}'^{(2)} & = \frac{1}{2}\left(\sqrt{3}g_5^{(2)} + g_6^{(2)}\right) = \frac{2\sqrt{3}\left(- \Delta t + t_{1} + t_{2}\right)}{\Delta t^{2}}~, \\
	g_{6}'^{(2)} & = \frac{1}{2}\left(g_5^{(2)} - \sqrt{3}g_6^{(2)}\right) = \frac{2\left(-\Delta t + 3 |t_1-t_2|\right)}{\Delta t^{2}}.
\end{align}
The states $g_{5}'^{(2)}$ and $g_{6}'^{(2)}$ are now symmetric functions of $t_1, t_2$. The state $g_{5}'^{(2)}$ is a particularly simple product of $1$ photon in the zero mode, and $1$ photon in the first mode:
\begin{align}
	\ket{g_{5}'^{(2)}} & =A^{\dagger}[g_1^{(1)}]A^{\dagger}[g_3^{(1)}]\vac.
\end{align}
This state does not introduce new modes, and can be captured as before: measuring the zero mode and first mode as WCW suggested would still give a pure state even if $g_{5}'^{(2)}$ is included.

However, $\ket{g_{6}'^{(2)}}$ is fundamentally different. It cannot be written as a product of single modes acting on the vacuum, \textit{i.e.} it is \emph{entangled}. Worse, it has a kink at $t_1 = t_2$, which implies that it cannot even be written as a \emph{finite} sum of monomials of creation operators acting on different modes:
\begin{equation}
	\ket{g_6'^{(2)}}  \neq \sum_{\alpha,\beta} A^{\dagger}[f_\alpha]A^{\dagger}[f_\beta]\vac \, ,
\end{equation}
for any finite sum over $\alpha$ and $\beta$ 
\footnote{Entanglement that survives any change of mode basis is called \emph{mode-independent} or \emph{mode-intrinsic} entanglement \cite{sperling2019, lopetegui2025} and is a potential resource in quantum computing. Note that here the important property is the multimode nature of the state, not its entanglement.}
.

In other words, we suddenly need an infinite number of modes to express this state at half-order $N = 2$. At this order, the environment cannot be truncated to a finite number of modes if we want to keep all the information on the evolution of the system. This is somewhat counter-intuitive: we could recover the half-order $N = 3/2$ by adding one mode, but the half-order $N = 2$ is beyond any finite truncation of the environment into modes. The consequence is that there is no way to preserve the purity of the system at half-order $N=2$ using only a finite number of integrals of the signal, and the result of WCW~\cite{wonglakhon2026} cannot be generalized beyond the order they found (except as we will see, if the operator in front of this problematic state happens to be zero).

A rigorous proof based on the one-body correlation function, its spectral decomposition, and the trace identity is provided in Appendix~\ref{sec:appendix_g6_rigorous}.

\subsection{Kraus operators for the Lindblad equation}\label{sec:Lindblad}
To obtain the Kraus operators $M_\mu$ appearing in the explicit CP form of the integrated Lindblad dynamics, we simply need to project the states ${F}_{k_0,k_1,\ldots,k_n}\ket{\Psi(0)}$ onto the orthonormal eigenvectors we previously computed. Importantly, carrying the orthonormalization and obtaining the eigenvectors $\ket{g_\mu^{(n_\mu)}}$ up to order $N$ is sufficient to get the correct CP map up to order $2N$, because $M_\mu$ appears squared:
\begin{align}
	\varrho(\Delta t) = \sum_{\mu=0}^{\mu_\text{max}} M_\mu \varrho(t)M_\mu^\dagger + O(\Delta t^{2N + 1}).
\end{align}
However, we do need to know $M_\mu$ itself to a higher order: if $M_\mu$ is of leading order $K$, then we do need to expand it to order $2N-K$ to get all the terms up to order $2N$ in the $M_\mu\varrho(t) M_\mu^\dagger$ term above.

Consequently, after obtaining the eigenvectors up to order $N$, we follow the next two steps:
\begin{enumerate}
	\setcounter{enumi}{3}
	\item Keep on generating the functions $f_{k_0, k_1, \ldots, k_n}$, up to order $n/2+\sum_i k_i \leq 2N$ (instead of $N$ during the previous orthonormalization procedure),
	\item Obtain the Kraus operators $M_\mu$ by projecting  $F_{k_0, k_1, \ldots, k_n}\ket{\Psi(0)}$ onto the orthonormal family $\{\ket{g_\mu}\}$ computed before:    \begin{equation}
		      M_\mu = \sum_{k_0,k_1,\ldots,k_{n_\mu}} \big\langle g_\mu^{(n_\mu)} \big|F_{k_0,\ldots,k_{n_\mu}}\big|\Psi(0)\big\rangle + O(\Delta t^{2N+1})\, ,
	      \end{equation}
	      and if $M_\mu$ is of leading order $K$, we keep only the terms of order at most $2N-K$ in the sum.
\end{enumerate}

At order $2N = 4$, the eight Kraus operators are given by~\cite{hutin2026robinetcalculus}:
\begin{align*}
	M_0 & = \sum_0^4 \frac{\Delta t^k}{k!}G^k~                                                                     \\
	M_1 & = \sum_0^3 \frac{\Delta t^{k+ 1/2}}{(k+1)!}\lbb G^{k} L\rbb ~                                            \\
	M_2 & = \sqrt{2}\sum_0^2 \frac{\Delta t^{k+ 1}}{(k+2)!}\lbb G^{k} L^2\rbb ~                                    \\
	M_3 & = \frac{\Delta t^{3/2}}{2\sqrt{3}} [L, G]+ \frac{\Delta t^{5 /2}}{4\sqrt{3}}[L, G^2]                     \\
	M_4 & = \sqrt{3!}\sum_0^1 \frac{\Delta t^{k+ 3/2}}{(k+3)!}\lbb G^{k} L^3\rbb ~                                 \\
	M_5 & = \frac{\Delta t^2}{12}(2L[L, G]+ [L, G]L)                                                               \\
	M_6 & = \frac{\Delta t^2}{4 \sqrt{3}}[L, G]L                                                                   \\
	M_7 & = \frac{\Delta t^2}{2\sqrt{6}}L^4 = \sqrt{4!}\sum_0^0 \frac{\Delta t^{k+ 2}}{(k+4)!}\lbb G^{k} L^4\rbb ~ \\
\end{align*}
where $\lbb A^k, B^l\rbb $ denotes all the possible permutations of $k$ operators $A$ and $l$ operators $B$. For example, $\lbb A^2, B\rbb  = A^2 B + ABA + BA^2$.

We can replace $M_5$ and $M_6$ by $M'_5$ and $M'_6$, by the same operations as for $g_5^{(2)}$ and $g_6^{(2)}$, which gives
\begin{align*}
	M'_5 & = \frac{\sqrt{3} \Delta t^2}{12}[L^2, G]~, \\
	M'_6 & = \frac{\Delta t^2}{12}[L, [L, G]]~,
\end{align*}
Hence, it tells us that the problematic contribution preventing one from going to half-order $N=2$ with the technique shown by WCW~\cite{wonglakhon2026} vanishes if $M'_6$ is zero, \textit{i.e.} if the commutator $[L, [L, G]]$ is zero. This is the case for example for a linear cavity decaying in a transmission line.

Interestingly, the number of Kraus operators at order $2N$ is given by $\mathcal{F}_{2N+2}$, the $(2N+2)$-th Fibonacci number (starting with $\mathcal{F}_0 = 0$ and $\mathcal{F}_1 = 1$). Indeed, to build all the states of order up to $N$, we first need all the states of order up to $N-1/2$, plus all the states (multivariate functions of $t_2, \dots$) that we can multiply by a certain monomial $(t-t_1)^k$ to get a state of order exactly $N$ (if it gives a lower order, it means that it is already contained in the orders up to $N-1/2$), which are exactly the states of half-order $N-1$. This is exactly the Fibonacci recurrence.

Symbolic expressions for orders up to 8 can be obtained in a few minutes on a personal computer with \texttt{SymPy}~\cite{meurer2017}. We provide an example explicitly in Appendix~\ref{app:explicit_expansions}, and the code used to generate these expressions is available on GitHub and has been permanently archived on Zenodo~\cite{hutin2026robinetcalculus}.

\section{Completely positive integration of the Robinet state}
\label{sec:DSME}

\subsection{Splitting the line into zero mode and rest}
In the previous section, we have written the system + line state:
\begin{equation}
	\ket{\Psi(\Delta t)} = \sum_{\mu = 0}^{\mu_\text{max}} M_\mu \ket{\psi_0} \ket{g_\mu^{(n_\mu)}} + O(\Delta t^{N + \frac{1}{2}})
\end{equation}
where $\ket{g_\mu^{(n_\mu)}}$ are orthogonal states of the line, and we recall that $n_\mu$ is just a label to explicitly keep track of the particle number of the state. As we argued, to obtain a completely positive map for the Robinet state, we will measure a quadrature of the zero mode.  A first step in this direction is to explicitly split the line states into zero mode and rest
\begin{equation}\label{eq:line_split_zero_rest}
	\ket{g^{(n_\mu)}_\mu} = \sum_{k=0}^{n_\mu} \ket{k}_0 \ket{\tilde{h}_{\mu,k}^{(n_\mu-k)}}_\text{rest}
\end{equation}
where $\ket{k}_0$ is the zero mode populated with $k$ excitations
\begin{equation}
	\ket{k}_0 \ket{0}_\text{rest} = \frac{A[g_1^{(1)}]^{\dagger k}}{\sqrt{k!}}\vac\, .
\end{equation}
The states $\tilde{h}$ are obtained by projecting the states $\ket{g^{(n_\mu)}_\mu}$ we computed before onto $\ket{k}_0$, which can be done recursively. Let us do it for a generic line state with $n$ particles $\ket{f^{(n)}} = \sum_k \ket{k}_0 \ket{\varphi^{(n-k)}}$, where $\varphi^{(i)}$ are $i$-particle states of the ``rest'' that we aim to find.

The first step is to determine $\ket{\varphi^{(0)}} = \varphi^{(0)} \vac_\text{rest}$ which is just the vacuum up to a normalization. We obtain this normalization simply by annihilating $n$ particles in the zero mode (which removes all the states in the sum \eqref{eq:line_split_zero_rest} but the last) and projecting onto the vacuum:
\begin{equation}
	\varphi^{(0)} = \bra{\text{vac}} \frac{A[g_1^{(1)}]^{n}}{\sqrt{n!}} |f^{(n)}\rangle
\end{equation}
To get $\ket{\varphi^{(1)}}$, we annihilate $n-1$ particles:
\begin{equation}
	\begin{split}
		A[g_1^{(1)}]^{n-1} \ket{f^{(n)}} = & \sqrt{\frac{(n-1)!}{0!}} \ket{0}_0 \ket{\varphi^{(1)}} \\
		                                   & + \sqrt{\frac{n!}{1!}} \ket{1}_0 \ket{\varphi^{(0)}}
	\end{split}
\end{equation}
which implies:
\begin{equation}
	\ket{0}_0 \ket{\varphi^{(1)}} =\frac{A[g_1^{(1)}]^{n-1} }{\sqrt{(n-1)!}} \ket{f^{(n)}} -\sqrt{n} \ket{1}_0 \ket{\varphi^{(0)}}\, .
\end{equation}
The right-hand side is explicit because we already know $\varphi^{(0)}$.
To obtain the corresponding wave-function, we simply insert an annihilation operator at $t$ and project onto the vacuum:
\begin{equation}
	\varphi^{(1)}(t) = \bra{\text{vac}} \ahat(t) \ket{0}_0 |\varphi^{(1)}\rangle\, .
\end{equation}
To get the other states, we proceed in the same fashion, recursively. Assuming we know all the $\ket{\varphi^{(j)}}$ for $0\leq j \leq k$, we apply $A[g_1^{(1)}]^{n-k-1}$
\begin{align}
	\begin{split}
	A[g_1^{(1)}]^{n-k-1} \ket{f^{(n)}} &= \\\sum_{j=0}^{k+1}&\sqrt{\frac{(n-k-1+j)!}{j!}} \ket{j}_0 \ket{\varphi^{(k+1-j)}}\, .
	\end{split}
\end{align}
This yields
\begin{align}
	\begin{split}
		\ket{0}_0 \ket{\varphi^{(k+1)}} &=  \frac{A[g_1^{(1)}]^{n-k-1} \ket{f^{(n)}} }{\sqrt{(n-k-1)!}} \\ &-  \sum_{j=1}^{k+1}\sqrt{\frac{(n-k-1+j)!}{j!(n-k-1)!}} \ket{j}_0 \ket{\varphi^{(k+1-j)}}\, ,
	\end{split}
\end{align}
which, again, is explicit, as the right-hand side is known. Finally, we get the wave-function representation by inserting annihilation operators at $t$ and then projecting onto the vacuum:
\begin{equation}
	\varphi^{(k)}(t_1, \cdots, t_k) = \bra{\text{vac}} \; \ahat(t_1) \ldots \ahat(t_k) \;  \ket{0}_0|\varphi^{(k)}\rangle \, .
\end{equation}
This is our procedure to obtain the full decomposition of the line states into zero mode + rest, and applying it to the orthogonal states of the line $\ket{g^{(n_\mu)}_\mu}$ gives the explicit decomposition advertised in~Eq.~\eqref{eq:line_split_zero_rest}. 

The final step is to orthogonalize the states of the rest of the line. Like before, this can be done using Gram-Schmidt orthonormalization, which one can carry out on states with different particle numbers independently. Ultimately, this provides us with states $\ket{h_j^{(n_j)}}$, for the rest of the line, which are orthogonal $\langle h_j^{(n_j)} | h_{j'}^{(n_{j'})}\rangle = \delta_{j,j'}$.

\subsection{Expanding the total state into zero mode and rest}

As we did in the Lindblad case, we now expand the total state $\ket{\Psi(\Delta t)}$ into this orthogonal basis $\ket{k}_0 \ket{h_j^{(n_j)}}_\text{rest}$. Concretely, one expands independently each term $F_{k_0,\ldots, k_n}\ket{\Psi(0)}$ into the orthonormal basis, and sums the contributions to each basis state, to obtain an expansion of the form:
\begin{equation} \label{eq:expansion_zero_orthogonal}
  \ket{\Psi(\Delta t)}= \sum_{k,j} P_{k,j} \ket{\psi_0}\ket{k}_0 \ket{h_j^{(n_j)}} \, .
\end{equation}

We must be careful about which order we keep at each step of this procedure. To understand exactly what to keep, it helps to start from the end, and see what we require for the final CP map $\mathcal{K}_I$ corresponding to the Robinet state to be of order $2N$. Recall that this map is of the form 
\begin{equation}
  \mathcal{K}_I (\rho) = \sum_j B_j(I) \rho B_j^\dagger(I) + O(\Delta t^{2N+1})
\end{equation}
with $B_j(I) = \sum_k \langle I \ket{k}_0 P_{k,j}$. First, as for the Lindblad case, we need only construct the orthonormal basis of the line up to order $N$. Any term that is of higher \emph{leading} order will not contribute. Second, we need each $B_j(I)$ of leading order $K$ to be accurate up to order $2N-K$, so that the ``square'' $B_j(I)\rho B_j^\dagger(I)$ remains accurate to order $2N$. This implies that when projecting $\ket{\Psi(\Delta t)}$ onto the orthonormal basis, we need to consider terms $F_{k_0,\ldots,k_n}\ket{\Psi(0)}$ up to order $2N$, \textit{i.e.} with $n/2 + \sum_i k_i \leq 2N$. Finally, when computing $P_{k,j}$, we need to keep terms of order $2N-K$ where $K$ is the leading order of $B_j(I)$, \textit{i.e.} it is not the leading order of $P_{k,j}$ that matters, but the leading order of the sum indexed by $j$ where it appears. As we will see, this brings new terms that do not appear in the Lindblad case at the same order.

We may first instantiate this procedure at a low order. At order $2N = 3$, essentially the lowest non-trivial order, the orthonormal family $\{\ket{h_\nu}\}$ is given by two states, $\ket{h_0^{(0)}} = \ket{0}_\text{rest}$, and $\ket{h_1^{(1)}}$ such that $\ket{0}_0\ket{h_1^{(1)}} = \hat{A}^\dagger[g_3^{(1)}]\vac$, \textit{i.e.} one photon in the mode $g_3^{(1)}$ defined previously. 
This shows that at half-order $N = 3/2$, the environment can be truncated to two modes only: the integration mode, and the first mode. Measuring these two modes jointly thus preserves the purity of the system up to order $3$, as shown in \cite{wonglakhon2026}. The complete state reads
\begin{align}
	\begin{split}
    \ket{\Psi(\Delta t)} = & \sum_{k}  P_{k, 0}\ket{\psi_0} \ket{k}_0\ket{0}_\text{rest}                   \\
    +                      &  P_{0, 1}\ket{\psi_0} \ket{0}_0\ket{h_1^{(1)}}_\text{rest} + O(\Delta t^{2})
	\end{split}
\end{align}
where we have (see~\cite{hutin2026robinetcalculus} for the derivation):
\begin{align}
	 P_{k, 0} & = \sqrt{k!}\sum_{j+k/2\leq 3} \frac{\Delta t^{j+k/2}}{(k+j)!} \lbb G^jL^k\rbb \\
	 P_{0, 1} & = \Big(\frac{\Delta t^{3/2}}{2\sqrt{3}}[L, G]\Big) .
\end{align}

At order $2N = 4$, we get the following non-zero $ P_{k, \nu}$ (see~\cite{hutin2026robinetcalculus}):
\begin{align}
	 P_{k, 0} & = \sqrt{k!}\sum_{j+k/2\leq 4} \frac{\Delta t^{j+k/2}}{(k+j)!} \lbb \Ghat^j\Lhat^k\rbb                        \\
	 P_{0, 1} & = \Big(\frac{\Delta t^{3/2}}{2\sqrt{3}}[\Lhat, \Ghat]+ \frac{\Delta t^{5/2}}{4\sqrt{3}}[\Lhat, \Ghat^2]\Big) \\
	 P_{1, 1} & = \frac{\Delta t^{2}}{4\sqrt{3}}[\Lhat^2, \Ghat]                                                             \\
	 P_{2, 1} & = \frac{\Delta t^{5/2}}{20\sqrt{6}}\Big(3[\Lhat^3, \Ghat]+ [\Lhat, \Lhat\Ghat \Lhat]\Big)                    \\
	 P_{1, 2} & = \frac{\Delta t^2}{12 \sqrt{5}} [\Lhat, [\Lhat, \Ghat]]                                                     \\
	 P_{0, 3} & =\frac{\Delta t^2}{6 \sqrt{5}} [\Lhat, [\Lhat, \Ghat]].
\end{align}
Note here that for $ P_{k,0}$, $k$ can run up to $8$, which means terms up to order $L^8$ appear, which was not the case for Lindblad.

\subsection{Measuring the zero mode}

As shown in Sec.~\ref{sec:recovering}, the Kraus operators for an integrated measurement record correspond to the projection of the corresponding mode onto the quadrature state $\ket{x}$ (for homodyne) or the coherent state $\ket{\alpha}$ (for heterodyne). Since we have now isolated the zero mode, we can directly write this procedure in the basis we just constructed. For homodyne detection, given the rescaled measurement record $x = I_{g_1^{(1)}} = \int_0^{\Delta t} \frac{\drm y}{\drm t}(t)\drm t/\sqrt{\Delta t}$, the Kraus operator is given by Eq.~\eqref{eq:homodyne_projection} as the projection of the zero mode onto the quadrature state $\ket{x}$:
\begin{align}
	\Pi_{x} = \delta(x-(\Ahat[g_1^{(1)}] + \Adag[g_1^{(1)}])) \\
	= \mathds{1}_S\otimes\ketbra{x}{x}_0\otimes \mathds{1}_\text{rest}
\end{align}
As for heterodyne detection, given a rescaled complex measurement record $\alpha = \int_0^{\Delta t} \frac{\drm y}{\drm t}(t)\drm t/\sqrt{\Delta t}$, the Kraus operator is given by
\begin{align}
	\Pi_{\alpha} = \frac{1}{\sqrt{\pi}}\mathds{1}_S\otimes\ketbra{\alpha}{\alpha}_0\otimes \mathds{1}_\text{rest}.
\end{align}

Now that we have the right form for $\ket{\Psi(\Delta t)}$ given by Eq.~\eqref{eq:expansion_zero_orthogonal}, we can directly apply the measurement operator $\Pi_x$ or $\Pi_\alpha$ to it, and obtain the Kraus operators corresponding to the measurement of the zero mode only. For homodyne detection, we get

\begin{align}
	\ketbra{x}{x}_0\ket{\Psi(\Delta t)} = \sum_{\mu, k} P_{k, \mu}\ket{\psi_0}_S \braket{x}{k}\ket{x}_0\ket{h_\mu^{(n_\mu)}}
\end{align}
The Kraus operators on the system are then given by
\begin{align}
	 P_\mu(x)= \sum_k \langle x|k\rangle  P_{k, \mu}
\end{align}
and $\langle x|k\rangle$ is the wave-function of the Fock state $\ket{k}$.

Similarly, for a heterodyne detection, obtaining a rescaled measurement record $\alpha = \int_0^{\Delta t} \frac{\drm y}{\drm t}(t)\drm t/\sqrt{\Delta t}$, the Kraus operators are given by
\begin{align}
	 P_\mu(\alpha)=\frac{1}{\sqrt{\pi}} \sum_k \langle \alpha|k\rangle  P_{k, \mu}
\end{align}
and $\langle \alpha|k\rangle$ is the overlap between the coherent state $\ket{\alpha}$ and the Fock state $\ket{k}$, which is simply given by $\langle \alpha|k\rangle = \e^{-|\alpha|^2/2}\alpha^{*k}/\sqrt{k!}$.

We now give a first example of the Kraus operators for homodyne detection at order $2N = 3$, which is the lowest order at which we lose purity if we only measure the zero mode. We have two Kraus operators, $P_0(x)$ and $P_1(x)$, which are given by
\begin{align}
	 P_0(x) = & \sum_{k = 0}^6 \langle x|k\rangle  P_{k, 0} \\
	 P_1(x) = & \langle x|0\rangle  P_{0, 1}
\end{align}
This way of writing the Kraus operators makes an intuitive decomposition: $P_0(x)$ corresponds to the back-action of the measurement on the system when the first mode is in the vacuum, while $P_1(x)$ corresponds to the back-action when the first mode is populated with one photon. If we had a measurement device able to detect the number of photons in the first mode, we could distinguish these two cases, and the system state would be pure after the measurement. 

If we now instead measure the quadrature of the first mode, which amounts to integrating the measurement record against $g_3^{(1)}$, we can also project the first mode onto a quadrature state $\ket{x_1}$, which gives the following measurement operator:
\begin{align}
	\Pi_{x, x_1} = \mathds{1}_S\otimes\ketbra{x}{x}_0\otimes \ketbra{x_1}{x_1}_1\otimes \mathds{1}_\text{rest}
\end{align}
where $\ket{x_1}_1$ is the quadrature state of the first mode. For $2N = 3$, applying this measurement operator to the state $\ket{\Psi(\Delta t)}$ directly gives the Kraus operator of the measurement of the zero mode and the first additional mode:
\begin{align}
	 P(x, x_1) = & \sum_{k = 0}^6   P_{k, 0}\langle x|k\rangle \langle x_1|0\rangle \\
	                    & +  P_{0, 1} \langle x|0\rangle \langle x_1|1\rangle
\end{align}
which is exactly the result obtained in \cite{wonglakhon2026} with one monitored channel with efficiency $\eta = 1$.

We now give the Kraus operators for the measurement of the zero mode at order $2N = 4$, which is the lowest order where new terms appear:
\begin{align}
	 P_0(x) = & \sum_{j+k/2\leq 4}\sqrt{k!} \frac{\Delta t^{j+k/2}}{(k+j)!} \lbb \Ghat^j\Lhat^k\rbb \langle x | k\rangle                                                                                                                                             \\
	\begin{split}
		 P_1(x) = & \Big(\frac{\Delta t^{3/2}}{2\sqrt{3}}[\Lhat, \Ghat]+ \frac{\Delta t^{5/2}}{4\sqrt{3}}[\Lhat, \Ghat^2]\Big)\langle x|0\rangle \\
		               & +\frac{\Delta t^{2}}{4\sqrt{3}}[\Lhat^2, \Ghat] \langle x|1\rangle                                                           \\
		               & +\frac{\Delta t^{5/2}}{20\sqrt{6}}\Big(3[\Lhat^3, \Ghat]+ [\Lhat, \Lhat \Ghat \Lhat]\Big)\langle x|2\rangle
	\end{split} \\
	 P_2(x) = & \frac{\Delta t^2}{12 \sqrt{5}} [\Lhat, [\Lhat, \Ghat]]\langle x|1\rangle                                                                                                                                                                             \\
	 P_3(x) = & \frac{\Delta t^2}{6 \sqrt{5}} [\Lhat, [\Lhat, \Ghat]]\langle x|0\rangle \, .
\end{align}
Note that the Kraus operator $ P_0(x)$ involves powers of $\Lhat$ up to order 8, which do not appear in the Lindblad case. This is due to the fact that $\Lhat^8$ implies an $8$-photon state, which in the Lindblad case forms a separate Kraus operator of order 4 (of order 8 once squared). But if the 8 photons are all in the zero mode, this term is grouped with terms of order 0 in $ P_0$, so it needs to be taken into account at order 4.

In practice, this implies that the Kraus operators for the stochastic master equation are more costly to compute than for the Lindblad case. We ran this algorithm up to order $2N = 10$, which has $107$ Kraus operators for unit quantum efficiency. We did not find any known numerical sequence matching the number of Kraus operators. Taking the full CP map on the state, expanding it, and truncating it to order $2N$ (which breaks exact positivity), we recover exactly the expansion of the Robinet state for homodyne measurement proposed in \cite{guilmin2025}.

\section{Applications}

We now present the applications of the previous expansions to the reconstruction of the Robinet state, its sampling, and to novel stochastic unravelings. We do not present nor benchmark the CP schemes we obtained for the Lindblad equation as warmup, since more general ones have appeared in the literature already \cite{cao2025,appelo2025}. The code to reproduce the results of this section is available on GitHub and has been permanently archived on Zenodo~\cite{hutin2026robinetcalculus}.

\subsection{Probability distribution of the measurement record}

We first showcase our ability to compute the probability distribution of the measurement record at arbitrary order. To this end, we compare the probability distribution obtained at order $2N$ to the exact distribution of integrated measurement records. The latter can be computed with quadratures using the method in~\cite{guilmin2025}, but we use instead the method of cascaded quantum systems~\cite{kiilerich2019}, which lets us compute the full density matrix of the system and the zero mode exactly (see Appendix \ref{sec:cascaded}). The probability distribution of the measurement record is then obtained by projecting the zero mode onto the quadrature state $\ket{x}$, and taking the trace.

For our order $2N$ approximation, the probability density function (pdf) of the measurement record $x$ is given by the trace of the unnormalized state obtained by applying the Kraus operators corresponding to the measurement record $x$ to the initial state $\rho(t)$:
\begin{equation}
	p(x) = \tr\Big(\sum_{\mu} P_{\mu}(x) \rho(t) P^\dagger_{\mu}(x)\Big) + O(\Delta t^{2N+1}).
\end{equation}

To test convergence, we considered a 7-level system with random Hamiltonian and a random jump operator. Intuitively, random models are maximally difficult, and prevent accidental cancellations of Kraus operators of the same order. We implemented our numerical reconstruction schemes using the libraries \texttt{Dynamiqs} \cite{guilmin2025dynamiqs} and \texttt{QuTiP} \cite{johansson2012,johansson2013}. We generated the parameters of the system as follows:
\begin{align}
	H &= X^\dagger X - Y^\dagger Y\\
	L &= Z
\end{align}
where $X$, $Y$, $Z$, are independent random matrices whose entries are independent and identically distributed complex Gaussian entries with variance 2 (the real part and imaginary part have variance 1). The initial state is chosen as a random pure state 
\begin{align}
	\ket{\psi_0} = \frac{\sum_{k=1}^7 c_k \ket{k}}{\sqrt{\sum_{k=1}^7 |c_k|^2}},
\end{align}
with $c_k$ independent and identically distributed complex Gaussian random variables with variance 2.
\begin{figure}[h]
	\centering
	\includegraphics[width=\linewidth]{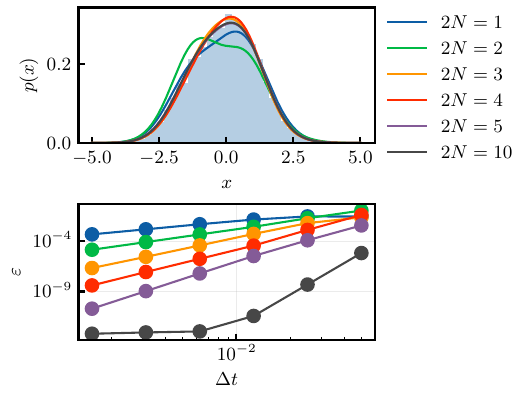}
	\caption{(a) Histogram of the integrated measurement record (blue) and probability density $p$ of the measurement record obtained at various orders between $2N = 1$ and $2N = 10$ and $\Delta t = 0.05$. The first orders are visibly wrong but orders $2N = 5$ and $2N = 10$ are almost superimposed and agree with the histogram. (b) $L^2$ distance $\varepsilon$ between the probability density obtained at order $2N$ and the reference distribution obtained with the cascaded formalism, as a function of $\Delta t$ for various values of $2N$.}
	\label{fig:px}
\end{figure}

We first compute the expected pdf of an integrated measurement record during a single time step of size $\Delta t= 0.05$  at various orders between $2N = 1$ and $2N = 10$. In Fig. \ref{fig:px}a, we compare these pdfs to empirical histograms obtained from sampling $10^4$ measurement records. The latter are computed with a standard discretization of the SME, with much smaller time step $\delta t = 5\times10^{-6}$, and integrated over the same single time bin $\Delta t$. We observe that the distribution converges quickly and matches the histogram obtained with the fine-grained simulation. Further, all the pdfs obtained at order $2N \geq 5$ are superimposed. 

To quantify convergence more precisely, we show in Fig. \ref{fig:px}b the $L^2$ distance  $\varepsilon$ between the pdf obtained at order $2N$ and the \emph{exact} distribution obtained with the cascaded quantum system formalism, as a function of $\Delta t$ and for various values of $2N$. We clearly see the different orders of convergence. For $2N=10$, the saturation at small $\Delta t$ comes from the accumulation of floating-point errors in the exact computation of the pdf with the cascaded formalism.

\subsection{Robinet state reconstruction}

We now verify that we apply the right operators up to the expected order. To this end, we fix some measurement records $I_1,\cdots, I_n$, and compare the approximate map obtained in this article to an exact solution. To generate typical measurement records, we use a standard SME solver running with a small time step. We then time-average these records on bins of finite size $\Delta t$. Note that these integrated measurement records need only be plausible, and do not have to be generated with exactly the right law, since we aim to test the quality of the state reconstruction conditioned on a fixed signal.

At each time step $\Delta t$, we apply the Kraus map 
\begin{equation}
\bar{\rho}_k = \frac{\mathcal{K}_{I_k}(\bar{\rho}_{k-1})}{\tr\left[\mathcal{K}_{I_k}(\bar{\rho}_{k-1})\right]},
\end{equation}
which we approximate at a fixed order $2N$ using the methods presented in this article. We then compare the final state $\bar{\rho}_{n\Delta t}$ for $n\Delta t = T$ to the one obtained with the exact cascaded quantum system solution. Varying the order $2N$, and the bin size $\Delta t$, we expect that the distance between the solution of order $2N$ and the exact solution scales as $O(\Delta t^{2N})$: the error at each step is $O(\Delta t^{2N+1})$, and the number of steps is $O(1/\Delta t)$, so the total error is $O(\Delta t^{2N})$.

\begin{figure}
  \centering
	\begin{minipage}{\linewidth}
		\centering
		\begin{overpic}[width=\linewidth]{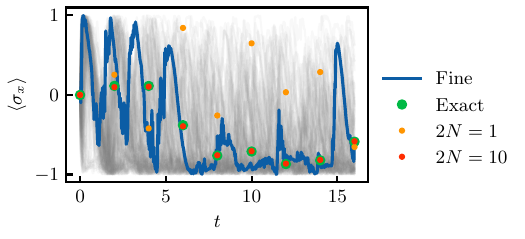}
			\put(0,41){{\small (a)}}
		\end{overpic}
	\end{minipage}
	\hfill
	\begin{minipage}{\linewidth}
		\centering
		\begin{overpic}[width=\linewidth]{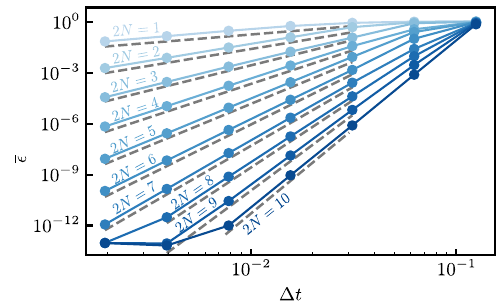}
			\put(0,57){{\small (b)}}
		\end{overpic}
	\end{minipage}
  \caption{(a) Expectation value $\langle \sigma_x \rangle$ as a function of time for fine-grained trajectory (blue and gray), exact coarse trajectory (exact Robinet state, green) and approximate coarse trajectory (approximate Robinet state of order 1, orange, and 10, red). (b) Mean geometric error $\overline{\epsilon}$ as a function of the time step $\Delta t$ for order $2N = 1$ (lighter blue) to $2N = 10$ (darker blue)}
  \label{fig:Fig1_2}
\end{figure}

We first assess convergence visually on a simple qubit example, with Hamiltonian $H$ and jump operator $L$ given by
\begin{align}
	H &= \frac{\omega_x}{2}\sigma_x + \frac{\omega_y}{2}\sigma_y + \frac{\omega_z}{2}\sigma_z\\
	L &= \gamma\sigma_-,
\end{align}
where $\sigma_x$, $\sigma_y$, $\sigma_z$ are the Pauli matrices, and $\sigma_- = (\sigma_x - i\sigma_y)/2$. We choose $\omega_x = 1, \omega_y = 1, \omega_z = 0.5, \gamma = 1$. Results are shown in Fig.\ref{fig:Fig1_2}a. The exact solution is close to the fine-grained trajectory, indicating that the loss of purity is weak in this example. The order $2N = 1$ result deviates substantially, while the order $2N = 10$ result closely follows the exact solution, as expected.

Second, we carry out a more systematic analysis by considering, as before, a 7-level system with random Hamiltonian and jump operators. We simulate $100$ quantum trajectories with a standard solver with $\delta t = 2^{-10}$, and final time $T = 2^{-3}$, giving us $100$ sequences of fine-grained records. We integrate these measurement records over different time bins of size $\Delta t$, from $\Delta t = 2^{-9}$ to $2^{-3}$. For each choice of $\Delta t$ and each set of measurement records, we compute the exact trajectory and the order $2N$ trajectory up to the final time, and extract the error $\epsilon = \|\rho^\text{exact} - \rho^{(2N)}\|_2$. Finally, we take the geometric mean over the $100$ trajectories $\overline{\epsilon}:=\exp\left[1/100 \sum \log (\epsilon)\right]$. With 100 trajectories, we estimate $\overline{\epsilon}$ with a relative precision of a few percent. The results for $\overline{\epsilon}$ as a function of $\Delta t$ for orders $2N$ between 1 and $10$ are shown in Fig.~\ref{fig:Fig1_2}b. The order of convergence is exactly the expected one, as evidenced by the grey dashed lines.

\subsection{Sampling Robinet trajectories}

We finally show that we can sample time-averaged measurement records directly, without needing to resort to a finer-grained simulation. In the same setting as in Fig.~\ref{fig:px}a, we now sample integrated measurement records directly from the probability distribution obtained at order $2N$
\begin{align}
	p(x) = \sum_{\mu} \tr( \rho P_{\mu}(x)^\dagger  P_{\mu}(x)).
\end{align}
Crucially, this pdf has an analytical expression in terms of Hermite functions $\langle x | n \rangle$  and operators (polynomials of $L$ and $G$) average on $\rho(t)$ that we can compute explicitly.
We can thus convert this analytical expression into a function of the form $p(x)=Q(x)e^{-x^2/2}$ where $Q$ is a positive polynomial whose coefficients depend on $\rho(t)$, and use the procedure described in Appendix~\ref{sec:sampling_polynomial_gaussian}. In Fig.~\ref{fig:sampling} we show that a histogram of $10^4$ records sampled this way for $2N = 10$ and $\Delta t = 0.05$ (blue) matches the exact distribution computed with the cascaded formalism (red curve).
\begin{figure}[t]
	\centering
	\includegraphics[width=\linewidth]{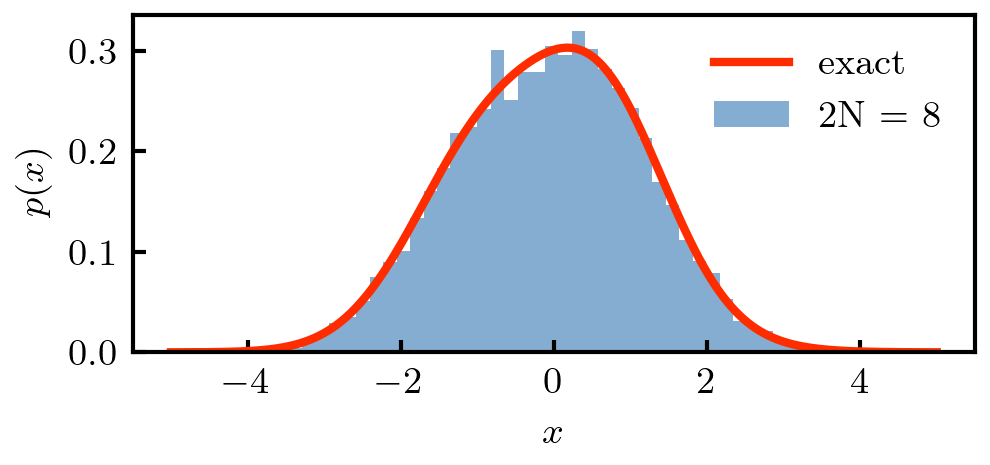}
	\caption{Histogram of integrated measurement record (blue) and probability density $p$ of the measurement record obtained at order $2N = 10$ and $\Delta t = 0.05$ (orange).}
	\label{fig:sampling}
\end{figure}

We now have all we need to sample Robinet trajectories: we can sample a measurement record $x$ at each time, and apply the corresponding back-action on the state (see~\cite{hutin2026robinetcalculus} for the implementation).
\begin{enumerate}
	\item Sample the measurement record $x$ according to the probability distribution $p(x) = \sum_{\mu} \tr( \rho P_{\mu}(x)^\dagger  P_{\mu}(x))$.
	\item Apply the Kraus operators $P_{\mu}(x)$ corresponding to the measurement record $x$.
	\item Renormalize the state (see Sec.~\ref{sec:renormalization_procedure}).
\end{enumerate}
Each step is independently validated.

\subsection{Stochastic Schrödinger equation}
\label{sec:SSE}
Our method can also sample pure quantum trajectories that \emph{unravel} a given Lindblad equation, and sample an integrated measurement record using pure states. Given a state $\ket{\psi(t)}$ at time $t$, the algorithm goes as follows:
\begin{enumerate}
	\item Sample the index $\mu$ according to the probabilities $p_{\mu} = \sum_k \langle \psi(t)|P_{k,  \mu}^\dagger P_{k, \mu}|\psi(t)\rangle$. Note that these probabilities do not sum up to $1$, but to $1 - O(\Delta t^{2N+1})$. See Sec.~\ref{sec:renormalization_procedure} for a discussion on the way to renormalize the Kraus operators to enforce exact normalization.
	\item Sample the measurement record $x$ according to the probability distribution $p_{\mu}(x) = \langle \psi(t)|P^\dagger_{\mu}(x) P_{\mu}(x)|\psi(t)\rangle$ (see appendix~\ref{sec:sampling_polynomial_gaussian}).
	\item Apply the Kraus operator: $\ket{\psi'(t + \Delta t)} = {P}_{\mu}(x)\ket{\psi(t)}$.
	\item Renormalize $\ket{\psi'(t + \Delta t)}$ to get $\ket{\psi(t + \Delta t)} = \frac{\ket{\psi'(t + \Delta t)}}{\sqrt{\langle \psi'(t + \Delta t)|\psi'(t + \Delta t)\rangle}}$.
\end{enumerate}
Formally, the tuple $(\mu, x)$, where $\mu$ is a discrete index and $x$ is a continuous variable, is a measurement record. Of course, the measurement outcome $\mu$, necessary to purify the state, is never accessible to the experimentalist. Numerically, this approach samples an integrated measurement record in a lightweight manner, as we only deal with pure states instead of density matrices, which could be needed for systems with very large Hilbert spaces. An implementation of this algorithm can be found in~\cite{hutin2026robinetcalculus}.

\section{From CP to CPTP maps}\label{sec:renormalization_procedure}
The schemes we presented are \emph{exactly} Completely Positive (CP), but so far trace-preserving (TP) up to order $2N$ only: \textit{e.g.} the sum of the probabilities $p(x)$ to sample all the possible record $x$ verifies $\intinf p(x) \drm x = 1 - O(\Delta t^{2N+1})$. A naive way to enforce trace preservation is to simply renormalize $p$ itself before sampling. This is what we did in this article. When ignoring the measurement outcome (Lindblad equation), it corresponds to renormalizing by $\tr(\rho)$ after applying the Kraus map, as done in~\cite{cao2025,appelo2025}. This method has the disadvantage of rendering the map non-linear, and we lose its contractive property. For completeness, we present the alternative method of~\cite{robin2025}, which restores these two properties, and has the additional advantage of making the underlying schemes more stable.

The trace preservation is lost because the system-line unitary is no longer exactly unitary after expansion and truncation. A natural way to enforce trace preservation is simply to project the resulting operator onto the subspace of unitary operators. In fact, since we only act on a line in its vacuum, we need only enforce unitarity explicitly in this subspace.

The action of a general unitary operator $ U(\Delta t)$ on a system + line \emph{in vacuum} can be written in full generality as 
\begin{align}\label{eq:infinite_sum_U}
  U(\Delta t)_\text{vac} = \sum_\mu M_\mu \otimes \ketbra{\mu}{\mathrm{vac}} \, ,
\end{align}
where the sum is infinite.
Unitarity in the full space implies $U(\Delta t)^\dagger  U(\Delta t) = \mathds{1} \otimes \mathds{1}$, which implies, if we restrict to lines in the vacuum, $ U(\Delta t)^\dagger_\text{vac}  U(\Delta t)_\text{vac} = \mathds{1}\otimes \ketbra{\mathrm{vac}}{\mathrm{vac}}$. This further implies $\sum_{\mu} M_\mu^\dagger M_\mu=\mathds{1}$. If we truncate $U(\Delta t)$ to $U^c(\Delta t)$ by taking a finite sum in \eqref{eq:infinite_sum_U}, we have
\begin{align}
  U^c(\Delta t)^\dagger_\text{vac} U^c(\Delta t)_\text{vac}  =  S \otimes \ketbra{\mathrm{vac}}{\mathrm{vac}}
\end{align}
where $ S = \sum_{\mu} M_{\mu}^\dagger M_{\mu} \neq \mathds{1}$. In general, $S$ is a non-negative Hermitian operator, and it is close to the identity (up to the order in $\Delta t$ at which the sum is truncated). Except in pathological cases, it is thus a positive operator, and we can define $S^{-1/2}$. We finally define a new operator $\tilde{U}^c(\Delta t)$ as
\begin{align}
	\tilde{ U}^c(\Delta t)
  :=  U^c(\Delta t)  S^{-1/2}\otimes \mathds{1}.
\end{align}
And indeed, we have:
\begin{align}
	U^c(\Delta t)  &= \tilde{U}^c(\Delta t)  S^{1/2} \\
	\tilde{ U}^c(\Delta t)^\dagger\tilde{ U}^c(\Delta t)  &= \mathds{1} \otimes \ketbra{\mathrm{vac}}{\mathrm{vac}} \, .
\end{align}
The unitary first line is thus the polar decomposition of $ U^c(\Delta t)$, and the unitary operator $\tilde{U}^c(\Delta t)$ is thus the projection of $ U^c(\Delta t)$ onto the subspace of unitary operators for the Frobenius norm. Ultimately, this condition can be enforced directly on the Kraus operators, by defining new Kraus operators $\tilde{M}_\mu$ as
\begin{align}
	\tilde{M}_\mu :=  M_\mu  S^{-1/2}.
\end{align}

In practice, $S^{-1/2}$ is not determined uniquely. One practical choice (also the one used by \texttt{Dynamiqs}' developers \cite{guilmin2025dynamiqs} for their current ``Rouchon'' integrators) is to:

\begin{itemize}
	\item compute the \emph{Cholesky} decomposition of $ S = T^\dagger T$, which gives a triangular matrix $T$.
	\item renormalize directly $\rho$ by applying the transformation $\rho \rightarrow T^{-1}\rho T^{-1\dagger}$, which can be done by solving two linear triangular systems instead of computing $T^{-1}$ explicitly. It is numerically more stable, and has the same algorithmic complexity as matrix multiplication.
\end{itemize}

We presented this normalization by $S^{-1/2}$ for the Lindblad equation. The same procedure can be applied to the Robinet state, by replacing $M_\mu$ by $P_\mu(x)$ and summing over $x$, to define
\begin{align}
	S = \sum_\mu \intinf P_\mu(x)^\dagger P_\mu(x) \drm x = \sum_{\mu, k}  P_{k, \mu}^\dagger P_{k, \mu} \,.
\end{align}
One then sets
\begin{align}
	\tilde{P}_\mu(x) :=  P_\mu(x)  S^{-1/2}
\end{align}
or equivalently 
\begin{align}
	\tilde{P}_{k, \mu} :=  P_{k, \mu}  S^{-1/2}.
\end{align}
Note that $S$ is in general different between the Lindblad case and the Robinet case.

\section{Extensions}

\noindent So far, for simplicity, we have derived our results for perfect efficiency $\eta = 1$, a single measured operator $L$ (both in the Lindblad and Robinet cases), and time-independent dynamics. We now explain how to use this formalism when we relax some of these hypotheses.

\subsection{Finite quantum efficiency}
Taking into account the finite quantum efficiency $\eta$ of the detector is relatively simple in this formalism. It simply corresponds to applying a beam-splitter transformation between the zero mode and an additional auxiliary environment mode, and tracing over the latter (since we now measure only what remains of the zero mode). This transformation amounts to replacing the Fock states $\{\ket{k}_0\}$ of the zero mode by
\begin{align}
	\{\sum_{j=0}^k \sqrt{\binom{k}{j}}\eta^{(k-j)/2}(1-\eta)^{j/2}\ket{k-j}_0\ket{j}_\mathrm{aux}\}
\end{align}
where $\ket{k}_\mathrm{aux}$ are the Fock states of the additional environment mode. 

If we could measure both modes, the action on the system would be given by the operator $P^\eta_{l, j, \nu}$ \begin{align}
	P^\eta_{l, j, \nu} := \sqrt{\binom{l+j}{j}}\eta^{l/2}(1-\eta)^{j/2} P_{l+j, \nu} \, ,
\end{align}
indexed by the number of photons in the zero mode (index $l$), the number of photons (index $j$) in the auxiliary mode and the state of the rest of the environment (index $\nu)$.

Summing over $l$ gives the Kraus operators $P^\eta_{j, \nu}(x)$ indexed by the number of photons $j$ in the auxiliary mode and the state of the rest of the environment $\nu$:
\begin{align}
	\begin{split}
		P^\eta_{j, \nu}(x) & =  \sum_l P^\eta_{l,j, \nu}(x) \\&= \sum_l \sqrt{\binom{j+l}{j}}\eta^{l/2}(1-\eta)^{j/2} P_{l+j, \nu}\langle x|l\rangle.
	\end{split}
\end{align}

All the previous results hold by replacing the index $\mu$ by a tuple $j, \mu$ where $j$ is the number of photons of the auxiliary mode, now part of the environment that is traced out.

In the case of the Stochastic Schrödinger Equation described in Sec.~\ref{sec:SSE}, accounting for the quantum efficiency only amounts to adding noise to the measurement record $x$ at each time step:
\begin{align}
	x \rightarrow \sqrt{\eta}x + \sqrt{1-\eta}y
\end{align}
which can be done fully as a post-processing step. 

\subsection{Time-dependent Hamiltonian and jump operator}
When the Hamiltonian and jump operators are time-dependent, we need to replace the exponential defining $U(t)$ by a time-ordered exponential. The orthogonal line states, decomposed into zero mode and rest, can be computed like before, but the corresponding operators $P_{k,\nu}$ acting on the system state differ. Writing $h^{'(k+n_{\nu})}_{k, \nu}$ the wave-function of the state $\ket{k}_0\ket{h_\nu^{(n_\nu)}}$, we have:
\begin{align}
  \begin{split}\label{eq:time_dep_P}
		 P_{k, \nu} & := \bra{k}_0\bra{h_\nu^{(n_\nu)}}\Uhat(t)\ket{\psi(0)}                                                                         \\
		                 & = \int_0^t \drm t_1 \ldots \int_0^{t_{n_\nu + k-1}} \drm t_{n_\nu + k} h^{'(k+n_{\nu})*}_{k, \nu}(t_1, \ldots, t_{n_\nu + k}) \\
		                 & \times V(t, t_1)\Lhat(t_1) V(t_1, t_2) \ldots \Lhat(t_{n_\nu + k}) V(t_{n_\nu + k}, 0)
	\end{split}
\end{align}
where $V(t_1, t_2)$ is the no-jump evolution operator between times $t_2$ and $t_1$, verifying, for $t_1 \geq t_2$:
\begin{align}
	\partial_{t_1} V(t_1, t_2) & = -iH(t_1)V(t_1, t_2) - \frac{1}{2}\Lhat^\dagger(t_1)\Lhat(t_1)V(t_1, t_2) \\
	V(t_2, t_2)  & = \mathds{1}.
\end{align}
The multi-dimensional integral \eqref{eq:time_dep_P} can be evaluated at order $2N$ using cubature rules on the standard simplex as in \cite{cao2025}.

\subsection{More jump operators (Lindblad)}
A Lindblad dynamics with multiple jump operators $\{\Lhat_i\}_{i=1\dots \mathfrak{m}}$, can be modeled by coupling the system to $\mathfrak{m}$ independent lines. In that case, our previous scheme is generalized simply. We first define $\Ghat:= -i\Hhat - \frac{1}{2}\sum_i\Lhat^\dagger_i \Lhat_i$. Recall that for a single loss channel, a Kraus operator $M_\mu^{(n)}$ is a sum of terms of the form:
\begin{align}
	M_\mu^{(n)} = \sum_{k_0, k_1, \ldots, k_n} m_{k_0, ..., k_{n}}\Ghat^{k_0}\Lhat\Ghat^{k_1}\ldots \Lhat\Ghat^{k_n}.
\end{align}
When there are $\mathfrak{m}$ lines, this operator is replaced by the $\mathfrak{m}^n$ terms of the form
\begin{align}
  M_{\mu, i_1, \ldots, i_n}^{(n)} = \sum_{k_0, k_1, \ldots , k_n} m^{i_1,\cdots,i_n}_{k_0, ..., k_{n}} \Ghat^{k_0}\Lhat_{i_1}\Ghat^{k_1}\ldots \Lhat_{i_n}\Ghat^{k_n}.
\end{align}
At order $2N = 4$, the number of Kraus operators is thus $F^\mathfrak{m}_{2N} = 1 + 2\mathfrak{m} + 3\mathfrak{m}^2 + \mathfrak{m}^3 + \mathfrak{m}^4$, and in general, it scales like $\mathfrak{m}^{2N}$.

This procedure gives the minimal number of Kraus operators needed to represent the evolution of the system up to order $2N$. In this regard, the numerical schemes proposed in \cite{cao2025} up to $2N = 4$ are optimal, as they use exactly this number of Kraus operators. However, they rely on Gaussian cubatures on the standard simplex, and we believe they will quickly require extra Kraus operators starting from order $2N = 5$.

\subsection{Multiple jump operators (Robinet)}
\label{sec:multiple_jump_operators_robinet}
The present formalism already handles heterodyne detection, which corresponds to two particular homodyne measurements, but the case of multiple generic measurements is more complex.  

We first consider the case of two joint measurements, which is sufficient to understand the generalization. Let the two line operators be $\hat{a}_1(t)$ and $\hat{a}_2(t)$, and the corresponding jump operators be $\Lhat_1$ and $\Lhat_2$. Following the same procedure as in the single jump operator case, we obtain the following Dyson expansion for the unitary operator acting on the initial state
\begin{align}\label{eq:Dyson expansion 2}
	\begin{split}
    \Uhat(t)\ket{\Psi(0)} = {} & \sum_{n\geq 0} \sum_{j_1, j_2, \ldots, j_n \in \{1, 2\}^{n}} \\
	& \int_0^{\Delta t} \drm t_1 \int_0^{t_1} \drm t_2 \ldots \int_0^{t_{n-1}}\drm t_n \\
	& e^{\Ghat(t-t_1)}\Lhat_{j_1}\hat{a}_{j_1}^\dagger(t_1) e^{\Ghat(t_1-t_2)}\ldots \\
	& \Lhat_{j_n}\hat{a}_{j_n}^\dagger(t_n) e^{\Ghat t_n}\ket{\Psi(0)}.
	\end{split}
\end{align}
The states of the environment are given by multivariate functions of the form  $f^{(j_1, j_2, \ldots, j_n)}(t_1, \ldots, t_n)$, where $j_i \in \{1, 2\}$.  Two states are orthogonal unless they have the same number of photons, emitted in the same order in the two lines
\begin{align}
    \begin{split}
	\langle f^{(j_1, j_2, \ldots, j_n)}|g^{(j'_1, j'_2, \ldots, j'_{n'})}\rangle = {} & \delta_{n, n'}\delta_{j_1, j'_1}\ldots \delta_{j_n, j'_n} \\
                                  \int_{\mathscr{S}_{\Delta t}^n} 
	 f^{(j_1, j_2, \ldots, j_n)*}
	& g^{(j'_1, j'_2, \ldots, j'_{n})}.
    \end{split}
\end{align}

At order $2N=1$, we find the following basis:
\begin{align}
    g_0 &= 1\\
    g_1^{1}(t_1) &= \frac{1}{\sqrt{\Delta t}}\\
    g_1^{2}(t_1) &= \frac{1}{\sqrt{\Delta t}}\\
    g_2^{1, 1}(t_1, t_2) &= \frac{\sqrt{2}}{\Delta t}\\
    g_2^{2, 2}(t_1, t_2) &= \frac{\sqrt{2}}{\Delta t}\\
    g_2^{1, 2}(t_1, t_2) &= \frac{\sqrt{2}}{\Delta t}\\
    g_2^{2, 1}(t_1, t_2) &= \frac{\sqrt{2}}{\Delta t}
\end{align}
giving the corresponding states in the two lines:
\begin{align}
    g_0 &= 1\\
    \ket{g_1^{1}(t_1)} &= \int_0^{\Delta t }\frac{\drm t}{\sqrt{\Delta t}}\hat{a}_1^\dagger(t_1)\vac\\
    \ket{g_1^{2}(t_1)} &= \int_0^{\Delta t }\frac{\drm t}{\sqrt{\Delta t}}\hat{a}_2^\dagger(t_1)\vac\\
    \ket{g_2^{1, 1}(t_1, t_2)} &= \int_0^{\Delta t }\int_0^{t_1}\drm t_1 \drm t_2\frac{\sqrt{2}}{\Delta t} \hat{a}_1^\dagger(t_1)\hat{a}_1^\dagger(t_2)\vac\\
    \ket{g_2^{2, 2}(t_1, t_2)} &= \int_0^{\Delta t }\int_0^{t_1}\drm t_1 \drm t_2\frac{\sqrt{2}}{\Delta t} \hat{a}_2^\dagger(t_1)\hat{a}_2^\dagger(t_2)\vac\\
    \ket{g_2^{1, 2}(t_1, t_2)} &=\int_0^{\Delta t }\int_0^{t_1}\drm t_1 \drm t_2\frac{\sqrt{2}}{\Delta t} \hat{a}_1^\dagger(t_1)\hat{a}_2^\dagger(t_2)\vac\\
    \ket{g_2^{2, 1}(t_1, t_2)} &= \int_0^{\Delta t }\int_0^{t_1}\drm t_1 \drm t_2\frac{\sqrt{2}}{\Delta t} \hat{a}_2^\dagger(t_1)\hat{a}_1^\dagger(t_2)\vac
\end{align}

We can perform a change of basis on the last four states, and write
\begin{align}
	\ket{\psi_+} &= \frac{1}{2}\left(\ket{g_2^{1, 1}} + \ket{g_2^{2, 2}} + \ket{g_2^{1, 2}} + \ket{g_2^{2, 1}}\right)\nonumber\\
	&= \frac{1}{\sqrt{2}}\left(\frac{\hat{A}_1^\dagger[g_1^1] + \hat{A}_2^\dagger[g_1^2]}{\sqrt{2}}\right)^2\vac\\
	\ket{\psi_-} &= \frac{1}{2}\left(\ket{g_2^{1, 1}} + \ket{g_2^{2, 2}} - \ket{g_2^{1, 2}} - \ket{g_2^{2, 1}}\right)\nonumber\\
	&= \frac{1}{\sqrt{2}}\left(\frac{\hat{A}_1^\dagger[g_1^1] - \hat{A}_2^\dagger[g_1^2]}{\sqrt{2}}\right)^2\vac\\
	\ket{\psi_\Delta} &= \frac{1}{\sqrt{2}}\left( \ket{g_2^{1, 1}} - \ket{g_2^{2, 2}}\right)\nonumber\\
	& = \frac{1}{\sqrt{2}}\left(\frac{1}{\sqrt{2}}\hat{A}_1^\dagger[g_1^1]^2 - \frac{1}{\sqrt{2}}\hat{A}_2^\dagger[g_1^2]^2\right)\vac\\
	\ket{\psi_\perp} &= \frac{1}{\sqrt{2}}\left( \ket{g_2^{1, 2}} - \ket{g_2^{2, 1}}\right).
\end{align}
This form makes it clear that the states 
$\ket{\psi_+}$, $\ket{\psi_-}$ and $\ket{\psi_\Delta}$ only populate the zero modes $g_1^1$ and $g_1^2$, of the two lines. The last one, $\ket{\psi_\perp}$ needs to be treated more carefully. If we try to decompose it in the zero modes of the two lines by computing $\hat{A}_1[g_1^1]\ket{\psi_\perp}$ and $\hat{A}_2[g_1^2]\ket{\psi_\perp}$, we get
\begin{align}
	\hat{A}_1[g_1^1]\ket{\psi_\perp} & = -\frac{1}{\sqrt{3}}\hat{A}_2^\dagger[g_3^2]\vac
\end{align}
where $\hat{A}_2^\dagger[g_3^2]$ is the creation operator for the first mode $g_3^2$ of line 2 with $g_3^2(t) = \frac{\sqrt{3}}{\Delta t^{3/2}}\left(\Delta t-2t\right)$.
By symmetry, we can thus decompose $\ket{\psi_\perp}$ as
\begin{align}
	\ket{\psi_\perp} & = \frac{1}{\sqrt{3}}\left(\hat{A}_2^\dagger[g_1^2]\hat{A}_1^\dagger[g_3^2]-\hat{A}_1^\dagger[g_1^1]\hat{A}_2^\dagger[g_3^2] \right)\vac\nonumber\\
	& + \frac{1}{\sqrt{3}}\ket{\psi'_\perp}
\end{align}
where $\ket{\psi'_\perp}$ is a state that does not populate either of the two zero modes, and that is orthogonal to the other states. We thus lose information if we only measure the zero modes of the two lines.

Can we restore the purity by considering more modes, as in the single jump case? The answer is no. A rigorous proof that $\ket{\psi_\perp}$ is genuinely multimode, via the eigendecomposition of its one-body correlation function, is given in Appendix~\ref{sec:appendix_psi_perp_multimode}. In short, the correlation function, seen as an operator kernel, has infinitely many non-zero eigenvalues, so infinitely many modes are populated. 

Let us now look at the Kraus operators associated with these states, which will give us more insight on the way the state $\ket{\psi_\perp}$ is populated. Setting $G = -i H - \frac{1}{2} (\Lhat_1^\dagger \Lhat_1 + \Lhat_2^\dagger \Lhat_2)$, the Kraus operators associated with these states are
\begin{align}
    M_0 &= \mathds{1} + G\Delta t + \frac{1}{2}G^2\Delta t^2\\
    M_1^1 &= \sqrt{\Delta t}\Lhat_1 + \frac{{\Delta t}^{3/2}}{2} \left(G\Lhat_1 + \Lhat_1 G\right) \\
    M_1^2 &= \sqrt{\Delta t}\Lhat_2 + \frac{{\Delta t}^{3/2}}{2} \left(G\Lhat_2 + \Lhat_2 G\right) \\
    M_2^{1, 1} &= \frac{\Delta t}{\sqrt{2}}\Lhat_1^2\\
    M_2^{2, 2} &= \frac{\Delta t}{\sqrt{2}}\Lhat_2^2 \\
    M_2^{1, 2} &= \frac{\Delta t}{\sqrt{2}}\Lhat_1\Lhat_2\\
    M_2^{2, 1} &= \frac{\Delta t}{\sqrt{2}}\Lhat_2\Lhat_1
\end{align}

The last four operators can be combined into Kraus operators associated with the states $\ket{\psi_+}$, $\ket{\psi_-}$, $\ket{\psi_\Delta}$ and $\ket{\psi_\perp}$:
\begin{align}
	M_+ &= \frac{M_2^{1, 1} + M_2^{2, 2} + M_2^{1, 2} + M_2^{2, 1}}{2} = \frac{\Delta t}{\sqrt{2}}\left(\frac{\Lhat_1 + \Lhat_2}{\sqrt{2}}\right)^2 \\
	M_- &= \frac{M_2^{1, 1} + M_2^{2, 2} - M_2^{1, 2} - M_2^{2, 1}}{2} = \frac{\Delta t}{\sqrt{2}}\left(\frac{\Lhat_1 - \Lhat_2}{\sqrt{2}}\right)^2 \\
	M_\Delta &= \frac{M_2^{1, 1} - M_2^{2, 2}}{\sqrt{2}} = \Delta t\frac{\Lhat_1^2 - \Lhat_2^2}{2} \\
	M_\perp &= \frac{M_2^{1, 2} - M_2^{2, 1}}{\sqrt{2}} = \frac{\Delta t}{\sqrt{2}}\frac{\Lhat_1\Lhat_2 - \Lhat_2\Lhat_1}{\sqrt{2}}
\end{align}
In this form, we can distinguish three different cases:

If $L_1 = L_2$, then only $M_+$ is non-zero. We can also reshuffle $M_1$ and $M_2$ into $\sqrt{\frac{\Delta t}{2}}\left(\Lhat_1 + \Lhat_2\right)$ and $\sqrt{\frac{\Delta t}{2}}\left(\Lhat_1 - \Lhat_2\right)$, and we recognize the single jump operator case with $L = {\sqrt{2}}L_1$.

If $L_1$ and $L_2$ commute, then $M_\perp$ is zero, and the state $\ket{\psi_\perp}$ is not populated. In this case, the purity of the state can only be lost at order larger than $3$, as in the single jump operator case. This is the case of heterodyne detection, where $L_1 = \frac{1}{\sqrt{2}}L$ and $L_2 = \frac{i}{\sqrt{2}}L$.

Finally, if $L_1$ and $L_2$ do not commute, then $M_\perp$ is not zero, and the state $\ket{\psi_\perp}$ has non-zero weight. In this case, the purity of the state is lost at order 2, and no additional statistics on the measurement record can restore it.

\section{Conclusion}

We have introduced a completely positive discretization scheme to compute the time-averaged quantum trajectories (Robinet states) introduced in~\cite{guilmin2025}. Our derivation relied on dilating the stochastic dynamics of the system into a unitary dynamics on a system + line setup, and isolating the zero mode of the line. Conceptually, the method sheds some light on how purity decreases as a function of averaging time, reproducing and extending previous work by Wonglakhon, Chantasri, and Wiseman~\cite{wonglakhon2026}. Numerically, it provides schemes of arbitrarily high order in time, that can be obtained systematically.

We discussed several applications, but the main one is to reconstruct quantum trajectories from experimental records. Currently, in most experiments, measurement records are averaged over a time-bin, and the state is reconstructed using a naive first order discretization of the continuous SME (which assumes the averaging is infinitesimally small). Using a low order version of our scheme is already a principled drop-in replacement for this naive approach. Increasing the order makes it match the exact quadrature approach of~\cite{guilmin2025} (or the cascaded quantum system formalism of~\cite{kiilerich2019}), without the need to rely on an external solver.

A natural extension of this work would be to develop adaptive algorithms. At each time step, we can compute the joint density matrix of the system and the integration mode after a time $\Delta t$ at all orders. Hence, comparing this density matrix across several orders (\emph{before} sampling the record) gives an estimate of the error made. Intuitively, this density matrix encodes all the information about the measurement record distribution and the back-action on the system, so if it does not change much when increasing the order, it means that the time step is small enough. This can serve two different strategies: an adaptive order (for fixed time step), or an adaptive time step (for fixed order). The first strategy is adapted to the Robinet state when the records are given; computing higher orders, however, quickly becomes costly. The second strategy applies standard time-step techniques from ordinary differential equations, instead of the more subtle ones used for stochastic differential equations. Indeed, we can estimate the error \emph{before} sampling a record and not \emph{after}, and thus without the risk of biasing the stochastic dynamics.

Continuous measurement theory and continuous matrix product states share the same mathematics, especially when the first is written in precisely the dilated form we used. We suspect that the discretization of the transmission line we have considered could be used to understand better the discretization of CMPS into MPS, if we can include the finite kinetic energy constraints that are present in this context. Further, the discretization of CMPS (or rather, field theories well described by CMPS) has recently been done with finite elements~\cite{shankar2026finiteelementmatrixproductstates} and Daubechies wavelets~\cite{kaplan2026waveletmatrixproductstates}. These correspond to particular (overlapping) bin shapes which may be useful in the context of continuous quantum measurement.

\begin{acknowledgments}
We thank Pierre Guilmin, Roberto Negrin, Rémi Robin, Pierre Rouchon, Christian Lantuéjoul, and Lev-Arcady Selem for helpful discussions. This work was partly supported by the European Union (ERC, QFT.zip project, Grant Agreement no. 101040260). Views and opinions expressed are however those of the authors only and do not necessarily reflect those of the European Union or the European Research Council Executive Agency. Neither the European Union nor the granting authority can be held responsible for them.
This work was also supported by the Plan France 2030 through the project ANR-22-PETQ-0006.
\end{acknowledgments}

\section*{Author contributions}
Hector Hutin proposed the method, developed the proof strategy and wrote the code used in the work. Antoine Tilloy extensively revised the manuscript, restructuring its narrative and improving its clarity and presentation. Large language models were used for proofreading, improving the clarity of some sentences and proofs, optimizing and formatting the original code for publication as supplementary material.

\bibliography{biblio.bib}

\newpage
\onecolumngrid
\appendix

\section{Robinet state for heterodyne readout} \label{app:heterodyne_robinet}
The heterodyne measurement can be built by introducing an independent auxiliary transmission line with field operator $\bhat(t)$, with $[\bhat(t), \bdag(u)] = \delta(t-u)$, in the vacuum state. One then performs two simultaneous homodyne measurements on the combined fields $c(t) = \frac{1}{\sqrt{2}}(\ahat(t) + \bhat(t))$ and $d(t) = \frac{1}{\sqrt{2}}(i\bhat(t)-i\ahat(t))$ \cite{wiseman1995}: the first measures the quadrature
\begin{equation} 
	\hat{X}_1(t) = \frac{1}{\sqrt{2}}\big(\ahat(t) + \bdag(t) + \adag(t) + \bhat(t)\big)
\end{equation}
yielding a real record $\frac{\drm y_1}{\drm t}(t)$, and the second measures the conjugate quadrature
\begin{equation}
	\hat{X}_2(t) = \frac{i}{\sqrt{2}}\big(\bhat(t) - \ahat(t) - \bdag(t) + \adag(t)\big)
\end{equation}
yielding $\frac{\drm y_2}{\drm t}(t)$. Note that $[\hat{X}_1(t), \hat{X}_2(u)] = 0$ for all $t, u$ 
, so both quadratures can be measured simultaneously. We note the complex measurement record $\frac{\drm y}{\drm t} = \frac{1}{\sqrt{2}}\left(\frac{\drm y_1}{\drm t} + i\frac{\drm y_2}{\drm t}\right)$. This can be viewed as the measurement of a complex operator $\hat{X}(t) = \frac{1}{\sqrt{2}}\left(\hat{X}_1(t) + i\hat{X}_2(t)\right) = \ahat(t) + \bdag(t)$. Integrating this measurement record against a complex function $f$ gives a complex outcome $I_f = \int_0^{\Delta t}  \drm tf^*(t)\frac{\drm y}{\drm t}$ of the measurement of the quadrature $\hat{X}[f] = \int_0^{\Delta t} f^*(t) \hat{X}(t) \drm t = \Ahat[f] + \Bdag[f^*]$ of the combined field $\ahat + \bdag$.
Separating $\hat{X}[f]$ into its hermitian and anti-hermitian parts:
\begin{align}
	\hat{X}_1[f] = \frac{\hat{X}[f] + \hat{X}^\dagger[f]}{2},\\
	\hat{X}_2[f] = \frac{\hat{X}[f] - \hat{X}^\dagger[f]}{2i},
\end{align}
we have $[\hat{X}_1[f], \hat{X}_2[f]] = 0$, so they can be measured simultaneously; their outcomes are the real and imaginary parts of $I_f$, respectively. The joint measurement of $\hat{X}_1[f]$ and $\hat{X}_2[f]$ can thus be viewed as the measurement of the complex operator $\hat{X}[f]$.
The back-action on the system is given by the projector $\Pi_{I_f}$ associated with this complex outcome $I_f$ is thus given by
\begin{align}\label{eq:heterodyne_projection1}
	\Pi_{I_f} :=& \delta\left(\Re( I_f)-\hat{X}_1[f]\right) \nonumber\\ \times&\delta\left(\Im( I_f)- \hat{X}_2[f]\right) \nonumber\\
	=
	&\delta^{(2)}\!\big( I_f - \Ahat[f] - \Bdag[f^*]\big).
\end{align}
The notation with a two-dimensional delta function can be used unambiguously, as the hermitian part and anti-hermitian part of $\Ahat[f] + \Bdag[f]$ commute: it is the product of the two delta functions of the two commuting operators. Using the Fourier representation of the delta function, we can write $\Pi_{I_f}$ as
\begin{align}\label{eq:heterodyne_projection2}
	\Pi_{I_f} &= \delta^{(2)}\!\big( I_f - \Ahat[f] - \Bdag[f^*]\big) \nonumber\\
	&= \frac{1}{\pi^2}\int \drm^2 p\, \e^{p^*  I_f - p  I_f^*} \nonumber\\
	&\quad \times \e^{\big(-p^*(\Ahat[f]+\Bdag[f^*])\,+\,p(\Adag[f]+\Bhat[f^*])\big)} \nonumber\\
	& = \frac{1}{\pi^2}\int \drm^2 p\, \e^{p^*  I_f - p  I_f^*}\,D^{\ahat}_{\Delta t}(p f)\,D^{\bhat}_{\Delta t}(-p^*f^*)
\end{align}
where the factorization of the exponential holds because $\Ahat[f]$ and $\Bhat[f^*]$ act on independent Hilbert spaces.
Since the auxiliary transmission line $\bhat$ is in the vacuum, we can trace it out, which gives
\begin{align}
	\tr_b\!\big[\Pi_{I_f}\, (\mathds{1} \otimes \vac_b \vacd_b)\big] & = \frac{1}{\pi^2}\int \drm^2 p\, \e^{p^*  I_f - p  I_f^*} \nonumber\\
	&\quad \times \e^{-\frac{\lVert f\rVert^2| p|^2}{2}}\,D^{\ahat}_{\Delta t}( p f)
\end{align}
where $\lVert f\rVert = \sqrt{\int_0^{\Delta t} |f(t)|^2 \drm t}$ is the norm of $f$ in $L^2([0,\Delta t],\mathbb{C})$.
For a normalized function $f$, this thus coincides with the anti-normalized operator-valued phase-space distribution of $\Ahat[f]$:
\begin{align}\label{eq:W_def}
	\hat{Q}_{f}( I_f) &= \frac{1}{\pi^2}\int \drm^2 p\, \e^{p^*  I_f - p  I_f^*}\,\e^{p^*\Ahat[f]} \e^{-p^*\Adag[f]},\\
	&= \frac{1}{\pi}D^{\ahat}_{\Delta t}( I_f f)\Pi_{0_f}D^{\ahat}_{\Delta t}(- I_f f) \nonumber
\end{align}
where we have introduced the projector on the zero-photon state $\Pi_{0_f} = \mathds{1}_S\otimes |0_f\rangle\langle 0_f|\otimes\mathds{1}_\text{rest}$ of the mode $\Ahat[f]$. The operator $\hat{Q}_f( I_f)$ is positive and verifies $\int_\mathbb{C}\drm^2  I_f\, \hat{Q}_f( I_f) = \mathds{1}$, and thus defines a POVM element for the measurement of the mode $\Ahat[f]$. We recover that when traced against a state $\rho$, $Q_f( I_f) = \tr[\rho\, \hat{Q}_{f}( I_f)]$ is the Husimi $Q$-function of the transmission line.

The Robinet state for heterodyne detection is thus given by a two-dimensional integral of tilted Liouvillians:
\begin{align}
	\tilde{\rho}_{\Delta t}(& I_f)\nonumber\\
	=&\int \frac{\drm^2 p}{\pi^2}\, \e^{p^*  I_f - p  I_f^*}\, \e^{-\frac{\lVert f\rVert^2|p|^2}{2}}\mathcal{T}\exp{\int_0^{\Delta t} \drm t \mathcal{L}_t^{p f}}\cdot \rho_0.
\end{align}

\section{Reference solver based on cascaded quantum systems}\label{sec:cascaded}
To obtain a reference solution for the binned stochastic master equation, we could make use of the Robinet method \cite{guilmin2025}, but we chose here to demonstrate an alternative method, based on cascaded quantum systems~\cite{kiilerich2019}. The idea is to append a cavity whose coupling to the line $\gamma(t)$ can be varied in time, after a circulator,  and which is initially in the vacuum state. Choosing the coupling appropriately, the cavity acquires the state of the outgoing mode defined by $g_1^{(1)}$. Measuring the cavity mode then gives the same back-action as measuring the zero mode of the line as we did in the main text.

\begin{figure}[h!]
	\centering
	\includegraphics[width=0.5\linewidth]{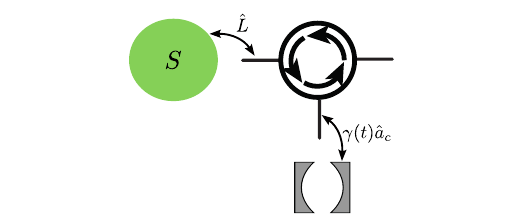}    \caption{Schematic of the cascaded system used to simulate the measurement back-action. The system $S$ is coupled to a waveguide, whose output is injected into a cavity via a circulator}
	\label{fig:cascaded_system}
\end{figure}
The output of the system is connected to the input of the cavity via a circulator, as shown in Fig.~\ref{fig:cascaded_system}. The system $S$ is described by the Hamiltonian $\Hhat_S$ and the jump operator $\Lhat$.
The total Hamiltonian reads
\begin{align}
	\Hhat = \Hhat_S + \frac{i\gamma(t)}{2}(\Lhat \adag_c - \Lhat^\dagger \ahat_c)
\end{align}
where $\ahat_c$ is the annihilation operator of the cavity mode. The total jump operator is simply $\Lhat_\mathrm{tot} = \gamma(t)\ahat_c + \Lhat$. The total density matrix $\rho$ of the system + cavity then follows the master equation
\begin{align}
	\partial_t\rho_t = -i[\Hhat, \rho] + \Dcal[\Lhat_\mathrm{tot}]\rho_t.
\end{align}
As it is detailed in \cite{kiilerich2019}, setting $\gamma(t) = -1/\sqrt{t}$ will make the cavity acquire the state of the outgoing mode defined by $g_1^{(1)}$. Up to the truncation of the cavity Hilbert space to a finite number of photons, this result is exact.
Finite quantum efficiency can be taken into account by setting
\begin{align}
	\Lhat^\eta_\mathrm{tot} = (\gamma(t)\ahat_c + \sqrt{\eta}\Lhat)
\end{align}
and adding an extra loss channel $\sqrt{1-\eta}\Lhat$:
\begin{align}
	\partial_t\rho_t = -i[\Hhat, \rho] + \Dcal[\Lhat^\eta_\mathrm{tot}]\rho_t + \Dcal[\sqrt{1-\eta}\Lhat]\rho_t.
\end{align}
We thus start with $\rho_{\mathrm{tot}}(0) = \rho_S(0)\otimes \ketbra{0}{0}$. The back-action associated with a measurement record $x$ over a time step $\Delta t$ gives the unnormalized state at $\Delta t$:
\begin{align}
	\tilde{\rho}_S(\Delta t) = \tr_\text{cav}((\mathds{1}\otimes\ketbra{x}{x})\rho_{\mathrm{tot}}(\Delta t)(\mathds{1}\otimes\ketbra{x}{x})).
\end{align}
Or more simply, using the cyclic property of the trace and the fact that $\mathds{1}\otimes\ketbra{x}{x}$ is a projector:
\begin{align}
	\rho_S(\Delta t) = \tr_\text{cav}((\mathds{1}\otimes\ketbra{x}{x})\rho_{\mathrm{tot}}(\Delta t)).
\end{align}
In practice, we approximate $\ket{x}$ by its projection on the Fock subspace, $\ket{x} \simeq \sum_{n=0}^{N_\mathrm{max}} h_n(x)\ket{n}$
where $h_n$ are the Hermite functions, and $N_\mathrm{max}$ is chosen such that the truncation error is negligible.

This method has the advantage of giving a physics intuition, making the measurement process transparent. It also gives an expansion of the probability of the measurement record:
\begin{align}
	p(x) = \tr_S((\mathds{1}\otimes\ketbra{x}{x})\rho_{\mathrm{tot}}(\Delta t)),
\end{align}
where $\tr_S$ denotes the partial trace over the system. This naturally expands $p$ in Hermite functions of $x$, and gives a physical justification as to why Hermite-Gauss quadrature is so well adapted in~\cite{guilmin2025}.

The cascaded system method still requires the simulation of a Lindblad equation in a larger Hilbert space (the system + the cavity mode) with a Hamiltonian exhibiting a singularity in 0, which is why it is probably less efficient than the Robinet method. On the other hand, it gives an analytic formula for the measurement record probability which is exact up to the truncation of the cavity Hilbert space, and can be used as a reference solver for the benchmarks in the main text (see~\cite{hutin2026robinetcalculus} for the implementation).

\section{Rigorous proof for infinite-mode states}
We collect here the rigorous argument behind the claims that some states cannot be represented with a finite number of modes. The object to consider here is the correlation function $\mathcal{G}_{\ket{\psi}}$, defined as the two-point correlation function of the state $\ket{\psi}$ of the line:
\begin{align}
	\mathcal{G}_{\ket{\psi}}(t_1,t_2) := \bra{\psi}\adag(t_1)\ahat(t_2)\ket{\psi}.
\end{align}
This correlation function is a kernel that can be diagonalized, in the form
\begin{align}
	\mathcal{G}_{\ket{\psi}}(t_1,t_2) = \sum_{n} n_\nu \phi_\nu(t_1)\phi^*_\nu(t_2)
\end{align}
with each of the eigenmodes $\phi_\nu$ normalized to 1, verifying for every $t_1 \in [0, \Delta t]$ the integral equation
\begin{align}
	\int_0^{\Delta t} \drm t_2 \mathcal{G}_{\ket{\psi}}(t_1,t_2)\phi_\nu(t_2) = n_\nu \phi_\nu(t_1).
\end{align}
Its eigenmodes $\phi_\nu$ give a minimal decomposition of the state into modes \cite{khanahmadi2023}. The eigenvalues $n_\nu$ give the population of these modes. Hence, $\ket{\psi}$ can be represented with a finite number of modes if and only if $\mathcal{G}_{\ket{\psi}}$ has only a finite number of non-zero eigenvalues \cite{khanahmadi2023}. In the two following cases, for $\ket{g_6'^{(2)}}$ and $\ket{\psi_\perp}$, we actually show that $\mathcal{G}$ has no zero eigenvalue, which means that every mode of the line is populated.

\subsection{State $\ket{g_6'^{(2)}}$}
\label{sec:appendix_g6_rigorous}
In the case of the state $\ket{g_6'^{(2)}}$, the correlation function is given by
\begin{align}
	\mathcal{G}_{\ket{g_6'^{(2)}}}(t_1,t_2)
	&:= \bra{g_6'^{(2)}}\adag(t_1)\ahat(t_2)\ket{g_6'^{(2)}},
	\qquad \tau := |t_1-t_2|, \\
	\mathcal{G}_{\ket{g_6'^{(2)}}}(t_1,t_2)
	&= \frac{1}{\Delta t^4}\Bigl(
	4\Delta t^3 - 6\Delta t^2(t_1+t_2)
	+ 12\Delta t\,(t_1 t_2-\tau^2) + 12\tau^3
	\Bigr) \, .
\end{align}
Let $K$ be the integral operator with kernel $K(t,s)=g_6'^{(2)}(t,s)$.
Then $\mathcal{G}=K^2$, so $\mathcal{G}$ and $K$ share the same eigenmodes.

We start from the eigenvalue equation for $K$:
\begin{equation}
\lambda\phi(t)=\int_0^{\Delta t}K(t,s)\phi(s)\,\drm s.
\end{equation}
Differentiating once, using $\partial_t K(t,s)=\frac{6}{\Delta t^2}\operatorname{sgn}(t-s)$, gives
\begin{equation}
\lambda\,\phi'(t)
	= \frac{6}{\Delta t^2}\int_0^{\Delta t}\operatorname{sgn}(t-s)\,\phi(s)\,\drm s
	= \frac{6}{\Delta t^2}\!\left(\int_0^t\phi(s)\,\drm s - \int_t^{\Delta t}\phi(s)\,\drm s\right).
\end{equation}
Differentiating a second time, using $\partial_t\operatorname{sgn}(t-s)=2\delta(t-s)$, gives
\begin{equation}
\lambda\,\phi''(t)=\frac{12}{\Delta t^2}\,\phi(t).
\end{equation}
If $\lambda=0$, this implies $\phi(t)=0$ for all $t$, i.e., only the trivial solution. Hence this kernel has no non-trivial null eigenvector, which means that every mode with a support in $[0, \Delta t]$ of the line is populated.

\subsection{State $\ket{\psi_\perp}$}\label{sec:appendix_psi_perp_multimode}
We provide here the rigorous proof that $\ket{\psi_\perp}$ is multimode. We look at the one-body correlation kernel of line $1$:

Starting from
\begin{equation}
	\ket{\psi_\perp} = \frac{1}{\sqrt{2}}\left( \ket{g_2^{1,2}} - \ket{g_2^{2,1}}\right),
\end{equation}
one gets
\begin{align}
	\hat{a}_1(t)\ket{\psi_\perp}
	&= \frac{1}{\Delta t}\int_0^{\Delta t}\drm t_1\,\big(\theta(t-t_1)-\theta(t_1-t)\big)\hat{a}_2^\dagger(t_1)\vac.
\end{align}
Hence, for $t,u\in[0,\Delta t]$, the one-body correlation kernel of line $1$ is
\begin{align}
	\mathcal{G}_{\ket{\psi_\perp}}^1(t,u)
	&:=\bra{\psi_\perp}\hat{a}_1^\dagger(t)\hat{a}_1(u)\ket{\psi_\perp}
	= \frac{\Delta t-2|t-u|}{\Delta t^2}.
\end{align}
The eigenvalue equation for this kernel reads
\begin{equation}
\lambda\phi(t)=\int_0^{\Delta t}\frac{\Delta t-2|t-s|}{\Delta t^2}\,\phi(s)\,\drm s.
\end{equation}
Differentiating twice in $t$ yields
\begin{equation}
\lambda\,\phi''(t)=-\frac{4}{\Delta t^2}\phi(t).
\end{equation}
Therefore, if $\lambda=0$, necessarily $\phi\equiv 0$: there is no non-trivial null eigenvector. As a consequence, this state populates every mode of the line $1$, and by symmetry all the modes of the line $2$ as well.

\section{Sampling a polynomial times a Gaussian}\label{sec:sampling_polynomial_gaussian}

We present here the algorithm used to sample the homodyne integrated measurement records.
The goal is to sample a random variable $\mathbf{X}$ according to a probability distribution of the form
\begin{align}
	p(x) = \sum_{k=0}^N c_k x^k e^{-x^2/\theta}
\end{align}
where $c_k \in \mathds{R}$ and $p(x) \geq 0$ for any $x \in \mathds{R}$. This is done with a rejection algorithm: we construct a probability distribution $q(x)$ such that $q(x) \geq p(x)$ for any $x\in \mathds{R}$, and sample $q(x)$ with a simple algorithm. We then keep the sampled value $X$ with probability $p(X)/q(X)$, otherwise we sample again. 

The key idea is that if $\mathbf{Y} \sim \Gamma(k, \theta)$ is a random variable following a Gamma distribution with shape $k$ and scale $\theta$, such that
\begin{align}
	f_{\mathbf{Y}}(y) = \frac{y^{k-1} e^{-y/\theta}}{\Gamma(k)\theta^k},
\end{align}
then $\mathbf{X} = \sqrt{\mathbf{Y}}$ follows a distribution with probability density distribution
\begin{align}
	f_{\mathbf{X}}(x) = \frac{2}{\Gamma(k)\theta^k} x^{2k-1} e^{-x^2/\theta}
\end{align}
for $x \geq 0$ and $f_{\mathbf{X}}(x) = 0$.
Knowing this, we can construct $q(x) = \sum_{k=0}^N R(c_k x^k)e^{-x^2/\theta}$ where $R(x) = \mathrm{max}(x, 0)$ is the ramp function. It is easy to see that this function verifies $q(x)\geq p(x)$ for any $x\in \mathds{R}$. Furthermore, $q(x)$ is the sum of positive functions. It is thus a statistical mixture. Then, given $\tilde{q}: x\rightarrow q(x)/\mathcal{N}$ the normalized version of $q$, we can sample $\tilde{q}$ with the following algorithm:
\begin{enumerate}
	\item Choose $k$ with probability $\frac{1}{\mathcal{N}} w_k$, where $w_k := \int_{-\infty}^\infty R(c_k x^k) e^{-x^2/\theta}\drm x$. If $k$ is even, $w_k = c_k\int_{-\infty}^\infty R(x^k) e^{-x^2/\theta}\drm x \times \theta(c_k)$: this vanishes whenever $c_k$ is negative, so such terms are simply dropped. If $k$ is odd, the two half-lines contribute equally by symmetry, and $w_k = |c_k|\int_{-\infty}^\infty R(x^k) e^{-x^2/\theta}\drm x$ regardless of the sign of $c_k$.
	\item Get a sample $X$ according to the distribution $R(\mathrm{sign}(c_k) x^k) e^{-x^2/\theta}$. If $k$ is even (and greater than 0 whose case is trivial), this is done by getting a sample $Y$ following a Gamma distribution with shape $(k+1)/2$ and scale $\theta$, and setting $X = \sqrt{Y}$ or $X = -\sqrt{Y}$ with probability $1/2$. If $k$ is odd, this is done by getting a sample $Y$ from a Gamma distribution with shape $(k+1)/2$ and scale $\theta$, and setting $X = \mathrm{sign}(c_k)\sqrt{Y}$: the sign must match $c_k$, since for odd $k$ the target density $R(\mathrm{sign}(c_k)x^k)e^{-x^2/\theta}$ is supported on the half-line whose sign is that of $c_k$, unlike the even case where it is symmetric.
\end{enumerate}
We can now accept the sample $X$ with probability $p(X)/q(X)$.

Note that a Gamma distribution with integer $k$ and scale $\theta$ can be sampled using uniform distributions as follows:

\begin{align}
	\mathbf{Y} = -\theta \sum_{i=1}^n\log( \mathbf{U}_i)
\end{align}
where $$\mathbf{U}_i \sim \mathcal{U}(0, 1)$$ are i.i.d uniform random variables between 0 and 1.

When $k$ is half-integer, we can use the fact that if $\mathbf{Z}$ is a standard normal variable, then $\mathbf{Z}^2$ follows a Gamma distribution with shape $1/2$ and scale $2$. Thus, if $k = n + 1/2$, we can sample $\mathbf{Y}$ with

\begin{align}
	\mathbf{Y} = \frac{\theta}{2}\mathbf{Z}^2  -\theta \sum_{i=1}^n\log( \mathbf{U}_i)
\end{align}

\section{Sampling the Husimi Q function}
Although not used in this work, we present here two methods to sample the Husimi Q function. First, writing $Q$ as $Q(x, y)$, the first one consists in sampling the marginal of the real part of Q as $Q_x(s) = \intinf Q(s, y)\drm y$, which gives an $x_0$ then $Q(x_0, y)$.

The second way consists in writing $Q$ in radial coordinates as $Q(r, \theta)$, sampling the marginal over $\theta$ as $Q_r(r) = \int_0^{2\pi} Q(r, \theta) r \drm \theta$, which gives $r_0$, then sampling $Q(r_0, \theta)$. The advantage of this method is that the marginal $Q_r(r)$ is easy to obtain, as the angular integration removes all the non-diagonal terms in the Fock basis expansion of $Q$:

\begin{align}
	Q_r(r) = 2 e^{-r^2} \sum_{n=0}^N  \frac{r^{2n+1}}{n!}\rho_{n, n}
\end{align}
which can be sampled without any rejection via the previous method, as this is a positive random variable and the $\rho_{n, n}$ are all positive.

Since the phase $\theta$ lives on $[-\pi, \pi]$, sampling $Q(r_0, \theta)$ can be done via numerical inversion of the cumulative distribution (typically with a Newton-Raphson method). The probability density function is given by
\begin{align}
	Q(r_0, \theta) = \frac{e^{-r_0^2}}{\pi} \sum_{m, n=0}^N \rho_{m, n} \frac{r_0^{m+n} e^{i(n-m)\theta}}{\sqrt{m! n!}}
\end{align}
and the cumulative distribution function by
\begin{align}
	C(\theta) = \int_0^\theta Q(r_0, \phi)\drm \phi = \frac{e^{-r_0^2}}{\pi} \sum_{m, n=0}^N \rho_{m, n} \frac{r_0^{m+n}}{\sqrt{m! n!}} \frac{e^{i(n-m)\theta} - 1}{i(n-m)}
\end{align}
with the convention that when $m=n$, the term is replaced by its limit value $\frac{e^{-r_0^2}}{\pi} \rho_{m, m} \frac{r_0^{2m}}{m!} \theta$.

\section{Explicit expansions}\label{app:explicit_expansions}
We give the explicit expansions to order $2N=8$ for the Lindblad equation and order $2N=6$ for the Robinet equation.

\subsection{Lindblad}
Here are the operators for the Lindblad equation at order $2N=8$:
\input{lindblad_8.tex}
\newpage
\subsection{Robinet}
Here are the operators for the integrated measurement at order $2N=6$:
\input{SME_6.tex}
\end{document}

%% file: lindblad_8.tex
{\fontsize{9pt}{12pt}\selectfont
\begin{flalign*}
    & M_{0} = \frac{\Delta t^{8} G^{8}}{40320} + \frac{\Delta t^{7} G^{7}}{5040} + \frac{\Delta t^{6} G^{6}}{720} + \frac{\Delta t^{5} G^{5}}{120} + \frac{\Delta t^{4} G^{4}}{24} && \\
    & \phantom{M_{0} =} + \frac{\Delta t^{3} G^{3}}{6} + \frac{\Delta t^{2} G^{2}}{2} + \Delta t G + 1 && \\
\end{flalign*}
\begin{flalign*}
    & M_{1} = \frac{\Delta t^{\frac{15}{2}} G L G^{6}}{40320} + \frac{\Delta t^{\frac{15}{2}} G^{2} L G^{5}}{40320} + \frac{\Delta t^{\frac{15}{2}} G^{3} L G^{4}}{40320} + \frac{\Delta t^{\frac{15}{2}} G^{4} L G^{3}}{40320} + \frac{\Delta t^{\frac{15}{2}} G^{5} L G^{2}}{40320} && \\
    & \phantom{M_{1} =} + \frac{\Delta t^{\frac{15}{2}} G^{6} L G}{40320} + \frac{\Delta t^{\frac{15}{2}} G^{7} L}{40320} + \frac{\Delta t^{\frac{15}{2}} L G^{7}}{40320} + \frac{\Delta t^{\frac{13}{2}} G L G^{5}}{5040} + \frac{\Delta t^{\frac{13}{2}} G^{2} L G^{4}}{5040} && \\
    & \phantom{M_{1} =} + \frac{\Delta t^{\frac{13}{2}} G^{3} L G^{3}}{5040} + \frac{\Delta t^{\frac{13}{2}} G^{4} L G^{2}}{5040} + \frac{\Delta t^{\frac{13}{2}} G^{5} L G}{5040} + \frac{\Delta t^{\frac{13}{2}} G^{6} L}{5040} + \frac{\Delta t^{\frac{13}{2}} L G^{6}}{5040} && \\
    & \phantom{M_{1} =} + \frac{\Delta t^{\frac{11}{2}} G L G^{4}}{720} + \frac{\Delta t^{\frac{11}{2}} G^{2} L G^{3}}{720} + \frac{\Delta t^{\frac{11}{2}} G^{3} L G^{2}}{720} + \frac{\Delta t^{\frac{11}{2}} G^{4} L G}{720} + \frac{\Delta t^{\frac{11}{2}} G^{5} L}{720} && \\
    & \phantom{M_{1} =} + \frac{\Delta t^{\frac{11}{2}} L G^{5}}{720} + \frac{\Delta t^{\frac{9}{2}} G L G^{3}}{120} + \frac{\Delta t^{\frac{9}{2}} G^{2} L G^{2}}{120} + \frac{\Delta t^{\frac{9}{2}} G^{3} L G}{120} + \frac{\Delta t^{\frac{9}{2}} G^{4} L}{120} && \\
    & \phantom{M_{1} =} + \frac{\Delta t^{\frac{9}{2}} L G^{4}}{120} + \frac{\Delta t^{\frac{7}{2}} G L G^{2}}{24} + \frac{\Delta t^{\frac{7}{2}} G^{2} L G}{24} + \frac{\Delta t^{\frac{7}{2}} G^{3} L}{24} + \frac{\Delta t^{\frac{7}{2}} L G^{3}}{24} && \\
    & \phantom{M_{1} =} + \frac{\Delta t^{\frac{5}{2}} G L G}{6} + \frac{\Delta t^{\frac{5}{2}} G^{2} L}{6} + \frac{\Delta t^{\frac{5}{2}} L G^{2}}{6} + \frac{\Delta t^{\frac{3}{2}} G L}{2} + \frac{\Delta t^{\frac{3}{2}} L G}{2} && \\
    & \phantom{M_{1} =} + \sqrt{\Delta t} L && \\
\end{flalign*}
\begin{flalign*}
    & M_{2} = \frac{\sqrt{2} \Delta t^{7} G L G L G^{4}}{40320} + \frac{\sqrt{2} \Delta t^{7} G L G^{2} L G^{3}}{40320} + \frac{\sqrt{2} \Delta t^{7} G L G^{3} L G^{2}}{40320} + \frac{\sqrt{2} \Delta t^{7} G L G^{4} L G}{40320} + \frac{\sqrt{2} \Delta t^{7} G L G^{5} L}{40320} && \\
    & \phantom{M_{2} =} + \frac{\sqrt{2} \Delta t^{7} G L^{2} G^{5}}{40320} + \frac{\sqrt{2} \Delta t^{7} G^{2} L G L G^{3}}{40320} + \frac{\sqrt{2} \Delta t^{7} G^{2} L G^{2} L G^{2}}{40320} + \frac{\sqrt{2} \Delta t^{7} G^{2} L G^{3} L G}{40320} + \frac{\sqrt{2} \Delta t^{7} G^{2} L G^{4} L}{40320} && \\
    & \phantom{M_{2} =} + \frac{\sqrt{2} \Delta t^{7} G^{2} L^{2} G^{4}}{40320} + \frac{\sqrt{2} \Delta t^{7} G^{3} L G L G^{2}}{40320} + \frac{\sqrt{2} \Delta t^{7} G^{3} L G^{2} L G}{40320} + \frac{\sqrt{2} \Delta t^{7} G^{3} L G^{3} L}{40320} + \frac{\sqrt{2} \Delta t^{7} G^{3} L^{2} G^{3}}{40320} && \\
    & \phantom{M_{2} =} + \frac{\sqrt{2} \Delta t^{7} G^{4} L G L G}{40320} + \frac{\sqrt{2} \Delta t^{7} G^{4} L G^{2} L}{40320} + \frac{\sqrt{2} \Delta t^{7} G^{4} L^{2} G^{2}}{40320} + \frac{\sqrt{2} \Delta t^{7} G^{5} L G L}{40320} + \frac{\sqrt{2} \Delta t^{7} G^{5} L^{2} G}{40320} && \\
    & \phantom{M_{2} =} + \frac{\sqrt{2} \Delta t^{7} G^{6} L^{2}}{40320} + \frac{\sqrt{2} \Delta t^{7} L G L G^{5}}{40320} + \frac{\sqrt{2} \Delta t^{7} L G^{2} L G^{4}}{40320} + \frac{\sqrt{2} \Delta t^{7} L G^{3} L G^{3}}{40320} + \frac{\sqrt{2} \Delta t^{7} L G^{4} L G^{2}}{40320} && \\
    & \phantom{M_{2} =} + \frac{\sqrt{2} \Delta t^{7} L G^{5} L G}{40320} + \frac{\sqrt{2} \Delta t^{7} L G^{6} L}{40320} + \frac{\sqrt{2} \Delta t^{7} L^{2} G^{6}}{40320} + \frac{\sqrt{2} \Delta t^{6} G L G L G^{3}}{5040} + \frac{\sqrt{2} \Delta t^{6} G L G^{2} L G^{2}}{5040} && \\
    & \phantom{M_{2} =} + \frac{\sqrt{2} \Delta t^{6} G L G^{3} L G}{5040} + \frac{\sqrt{2} \Delta t^{6} G L G^{4} L}{5040} + \frac{\sqrt{2} \Delta t^{6} G L^{2} G^{4}}{5040} + \frac{\sqrt{2} \Delta t^{6} G^{2} L G L G^{2}}{5040} + \frac{\sqrt{2} \Delta t^{6} G^{2} L G^{2} L G}{5040} && \\
    & \phantom{M_{2} =} + \frac{\sqrt{2} \Delta t^{6} G^{2} L G^{3} L}{5040} + \frac{\sqrt{2} \Delta t^{6} G^{2} L^{2} G^{3}}{5040} + \frac{\sqrt{2} \Delta t^{6} G^{3} L G L G}{5040} + \frac{\sqrt{2} \Delta t^{6} G^{3} L G^{2} L}{5040} + \frac{\sqrt{2} \Delta t^{6} G^{3} L^{2} G^{2}}{5040} && \\
    & \phantom{M_{2} =} + \frac{\sqrt{2} \Delta t^{6} G^{4} L G L}{5040} + \frac{\sqrt{2} \Delta t^{6} G^{4} L^{2} G}{5040} + \frac{\sqrt{2} \Delta t^{6} G^{5} L^{2}}{5040} + \frac{\sqrt{2} \Delta t^{6} L G L G^{4}}{5040} + \frac{\sqrt{2} \Delta t^{6} L G^{2} L G^{3}}{5040} && \\
    & \phantom{M_{2} =} + \frac{\sqrt{2} \Delta t^{6} L G^{3} L G^{2}}{5040} + \frac{\sqrt{2} \Delta t^{6} L G^{4} L G}{5040} + \frac{\sqrt{2} \Delta t^{6} L G^{5} L}{5040} + \frac{\sqrt{2} \Delta t^{6} L^{2} G^{5}}{5040} + \frac{\sqrt{2} \Delta t^{5} G L G L G^{2}}{720} && \\
    & \phantom{M_{2} =} + \frac{\sqrt{2} \Delta t^{5} G L G^{2} L G}{720} + \frac{\sqrt{2} \Delta t^{5} G L G^{3} L}{720} + \frac{\sqrt{2} \Delta t^{5} G L^{2} G^{3}}{720} + \frac{\sqrt{2} \Delta t^{5} G^{2} L G L G}{720} + \frac{\sqrt{2} \Delta t^{5} G^{2} L G^{2} L}{720} && \\
    & \phantom{M_{2} =} + \frac{\sqrt{2} \Delta t^{5} G^{2} L^{2} G^{2}}{720} + \frac{\sqrt{2} \Delta t^{5} G^{3} L G L}{720} + \frac{\sqrt{2} \Delta t^{5} G^{3} L^{2} G}{720} + \frac{\sqrt{2} \Delta t^{5} G^{4} L^{2}}{720} + \frac{\sqrt{2} \Delta t^{5} L G L G^{3}}{720} && \\
    & \phantom{M_{2} =} + \frac{\sqrt{2} \Delta t^{5} L G^{2} L G^{2}}{720} + \frac{\sqrt{2} \Delta t^{5} L G^{3} L G}{720} + \frac{\sqrt{2} \Delta t^{5} L G^{4} L}{720} + \frac{\sqrt{2} \Delta t^{5} L^{2} G^{4}}{720} + \frac{\sqrt{2} \Delta t^{4} G L G L G}{120} && \\
    & \phantom{M_{2} =} + \frac{\sqrt{2} \Delta t^{4} G L G^{2} L}{120} + \frac{\sqrt{2} \Delta t^{4} G L^{2} G^{2}}{120} + \frac{\sqrt{2} \Delta t^{4} G^{2} L G L}{120} + \frac{\sqrt{2} \Delta t^{4} G^{2} L^{2} G}{120} + \frac{\sqrt{2} \Delta t^{4} G^{3} L^{2}}{120} && \\
    & \phantom{M_{2} =} + \frac{\sqrt{2} \Delta t^{4} L G L G^{2}}{120} + \frac{\sqrt{2} \Delta t^{4} L G^{2} L G}{120} + \frac{\sqrt{2} \Delta t^{4} L G^{3} L}{120} + \frac{\sqrt{2} \Delta t^{4} L^{2} G^{3}}{120} + \frac{\sqrt{2} \Delta t^{3} G L G L}{24} && \\
    & \phantom{M_{2} =} + \frac{\sqrt{2} \Delta t^{3} G L^{2} G}{24} + \frac{\sqrt{2} \Delta t^{3} G^{2} L^{2}}{24} + \frac{\sqrt{2} \Delta t^{3} L G L G}{24} + \frac{\sqrt{2} \Delta t^{3} L G^{2} L}{24} + \frac{\sqrt{2} \Delta t^{3} L^{2} G^{2}}{24} && \\
    & \phantom{M_{2} =} + \frac{\sqrt{2} \Delta t^{2} G L^{2}}{6} + \frac{\sqrt{2} \Delta t^{2} L G L}{6} + \frac{\sqrt{2} \Delta t^{2} L^{2} G}{6} + \frac{\sqrt{2} \Delta t L^{2}}{2} && \\
\end{flalign*}
\begin{flalign*}
    & M_{3} = \frac{\sqrt{3} \Delta t^{\frac{13}{2}} G L G^{5}}{10080} + \frac{\sqrt{3} \Delta t^{\frac{13}{2}} G^{2} L G^{4}}{20160} - \frac{\sqrt{3} \Delta t^{\frac{13}{2}} G^{4} L G^{2}}{20160} - \frac{\sqrt{3} \Delta t^{\frac{13}{2}} G^{5} L G}{10080} - \frac{\sqrt{3} \Delta t^{\frac{13}{2}} G^{6} L}{6720} && \\
    & \phantom{M_{3} =} + \frac{\sqrt{3} \Delta t^{\frac{13}{2}} L G^{6}}{6720} + \frac{\sqrt{3} \Delta t^{\frac{11}{2}} G L G^{4}}{1680} + \frac{\sqrt{3} \Delta t^{\frac{11}{2}} G^{2} L G^{3}}{5040} - \frac{\sqrt{3} \Delta t^{\frac{11}{2}} G^{3} L G^{2}}{5040} - \frac{\sqrt{3} \Delta t^{\frac{11}{2}} G^{4} L G}{1680} && \\
    & \phantom{M_{3} =} - \frac{\sqrt{3} \Delta t^{\frac{11}{2}} G^{5} L}{1008} + \frac{\sqrt{3} \Delta t^{\frac{11}{2}} L G^{5}}{1008} + \frac{\sqrt{3} \Delta t^{\frac{9}{2}} G L G^{3}}{360} - \frac{\sqrt{3} \Delta t^{\frac{9}{2}} G^{3} L G}{360} - \frac{\sqrt{3} \Delta t^{\frac{9}{2}} G^{4} L}{180} && \\
    & \phantom{M_{3} =} + \frac{\sqrt{3} \Delta t^{\frac{9}{2}} L G^{4}}{180} + \frac{\sqrt{3} \Delta t^{\frac{7}{2}} G L G^{2}}{120} - \frac{\sqrt{3} \Delta t^{\frac{7}{2}} G^{2} L G}{120} - \frac{\sqrt{3} \Delta t^{\frac{7}{2}} G^{3} L}{40} + \frac{\sqrt{3} \Delta t^{\frac{7}{2}} L G^{3}}{40} && \\
    & \phantom{M_{3} =} - \frac{\sqrt{3} \Delta t^{\frac{5}{2}} G^{2} L}{12} + \frac{\sqrt{3} \Delta t^{\frac{5}{2}} L G^{2}}{12} - \frac{\sqrt{3} \Delta t^{\frac{3}{2}} G L}{6} + \frac{\sqrt{3} \Delta t^{\frac{3}{2}} L G}{6} && \\
\end{flalign*}
\begin{flalign*}
    & M_{4} = \frac{\sqrt{6} \Delta t^{\frac{13}{2}} G L G L G L G^{2}}{40320} + \frac{\sqrt{6} \Delta t^{\frac{13}{2}} G L G L G^{2} L G}{40320} + \frac{\sqrt{6} \Delta t^{\frac{13}{2}} G L G L G^{3} L}{40320} + \frac{\sqrt{6} \Delta t^{\frac{13}{2}} G L G L^{2} G^{3}}{40320} + \frac{\sqrt{6} \Delta t^{\frac{13}{2}} G L G^{2} L G L G}{40320} && \\
    & \phantom{M_{4} =} + \frac{\sqrt{6} \Delta t^{\frac{13}{2}} G L G^{2} L G^{2} L}{40320} + \frac{\sqrt{6} \Delta t^{\frac{13}{2}} G L G^{2} L^{2} G^{2}}{40320} + \frac{\sqrt{6} \Delta t^{\frac{13}{2}} G L G^{3} L G L}{40320} + \frac{\sqrt{6} \Delta t^{\frac{13}{2}} G L G^{3} L^{2} G}{40320} + \frac{\sqrt{6} \Delta t^{\frac{13}{2}} G L G^{4} L^{2}}{40320} && \\
    & \phantom{M_{4} =} + \frac{\sqrt{6} \Delta t^{\frac{13}{2}} G L^{2} G L G^{3}}{40320} + \frac{\sqrt{6} \Delta t^{\frac{13}{2}} G L^{2} G^{2} L G^{2}}{40320} + \frac{\sqrt{6} \Delta t^{\frac{13}{2}} G L^{2} G^{3} L G}{40320} + \frac{\sqrt{6} \Delta t^{\frac{13}{2}} G L^{2} G^{4} L}{40320} + \frac{\sqrt{6} \Delta t^{\frac{13}{2}} G L^{3} G^{4}}{40320} && \\
    & \phantom{M_{4} =} + \frac{\sqrt{6} \Delta t^{\frac{13}{2}} G^{2} L G L G L G}{40320} + \frac{\sqrt{6} \Delta t^{\frac{13}{2}} G^{2} L G L G^{2} L}{40320} + \frac{\sqrt{6} \Delta t^{\frac{13}{2}} G^{2} L G L^{2} G^{2}}{40320} + \frac{\sqrt{6} \Delta t^{\frac{13}{2}} G^{2} L G^{2} L G L}{40320} + \frac{\sqrt{6} \Delta t^{\frac{13}{2}} G^{2} L G^{2} L^{2} G}{40320} && \\
    & \phantom{M_{4} =} + \frac{\sqrt{6} \Delta t^{\frac{13}{2}} G^{2} L G^{3} L^{2}}{40320} + \frac{\sqrt{6} \Delta t^{\frac{13}{2}} G^{2} L^{2} G L G^{2}}{40320} + \frac{\sqrt{6} \Delta t^{\frac{13}{2}} G^{2} L^{2} G^{2} L G}{40320} + \frac{\sqrt{6} \Delta t^{\frac{13}{2}} G^{2} L^{2} G^{3} L}{40320} + \frac{\sqrt{6} \Delta t^{\frac{13}{2}} G^{2} L^{3} G^{3}}{40320} && \\
    & \phantom{M_{4} =} + \frac{\sqrt{6} \Delta t^{\frac{13}{2}} G^{3} L G L G L}{40320} + \frac{\sqrt{6} \Delta t^{\frac{13}{2}} G^{3} L G L^{2} G}{40320} + \frac{\sqrt{6} \Delta t^{\frac{13}{2}} G^{3} L G^{2} L^{2}}{40320} + \frac{\sqrt{6} \Delta t^{\frac{13}{2}} G^{3} L^{2} G L G}{40320} + \frac{\sqrt{6} \Delta t^{\frac{13}{2}} G^{3} L^{2} G^{2} L}{40320} && \\
    & \phantom{M_{4} =} + \frac{\sqrt{6} \Delta t^{\frac{13}{2}} G^{3} L^{3} G^{2}}{40320} + \frac{\sqrt{6} \Delta t^{\frac{13}{2}} G^{4} L G L^{2}}{40320} + \frac{\sqrt{6} \Delta t^{\frac{13}{2}} G^{4} L^{2} G L}{40320} + \frac{\sqrt{6} \Delta t^{\frac{13}{2}} G^{4} L^{3} G}{40320} + \frac{\sqrt{6} \Delta t^{\frac{13}{2}} G^{5} L^{3}}{40320} && \\
    & \phantom{M_{4} =} + \frac{\sqrt{6} \Delta t^{\frac{13}{2}} L G L G L G^{3}}{40320} + \frac{\sqrt{6} \Delta t^{\frac{13}{2}} L G L G^{2} L G^{2}}{40320} + \frac{\sqrt{6} \Delta t^{\frac{13}{2}} L G L G^{3} L G}{40320} + \frac{\sqrt{6} \Delta t^{\frac{13}{2}} L G L G^{4} L}{40320} + \frac{\sqrt{6} \Delta t^{\frac{13}{2}} L G L^{2} G^{4}}{40320} && \\
    & \phantom{M_{4} =} + \frac{\sqrt{6} \Delta t^{\frac{13}{2}} L G^{2} L G L G^{2}}{40320} + \frac{\sqrt{6} \Delta t^{\frac{13}{2}} L G^{2} L G^{2} L G}{40320} + \frac{\sqrt{6} \Delta t^{\frac{13}{2}} L G^{2} L G^{3} L}{40320} + \frac{\sqrt{6} \Delta t^{\frac{13}{2}} L G^{2} L^{2} G^{3}}{40320} + \frac{\sqrt{6} \Delta t^{\frac{13}{2}} L G^{3} L G L G}{40320} && \\
    & \phantom{M_{4} =} + \frac{\sqrt{6} \Delta t^{\frac{13}{2}} L G^{3} L G^{2} L}{40320} + \frac{\sqrt{6} \Delta t^{\frac{13}{2}} L G^{3} L^{2} G^{2}}{40320} + \frac{\sqrt{6} \Delta t^{\frac{13}{2}} L G^{4} L G L}{40320} + \frac{\sqrt{6} \Delta t^{\frac{13}{2}} L G^{4} L^{2} G}{40320} + \frac{\sqrt{6} \Delta t^{\frac{13}{2}} L G^{5} L^{2}}{40320} && \\
    & \phantom{M_{4} =} + \frac{\sqrt{6} \Delta t^{\frac{13}{2}} L^{2} G L G^{4}}{40320} + \frac{\sqrt{6} \Delta t^{\frac{13}{2}} L^{2} G^{2} L G^{3}}{40320} + \frac{\sqrt{6} \Delta t^{\frac{13}{2}} L^{2} G^{3} L G^{2}}{40320} + \frac{\sqrt{6} \Delta t^{\frac{13}{2}} L^{2} G^{4} L G}{40320} + \frac{\sqrt{6} \Delta t^{\frac{13}{2}} L^{2} G^{5} L}{40320} && \\
    & \phantom{M_{4} =} + \frac{\sqrt{6} \Delta t^{\frac{13}{2}} L^{3} G^{5}}{40320} + \frac{\sqrt{6} \Delta t^{\frac{11}{2}} G L G L G L G}{5040} + \frac{\sqrt{6} \Delta t^{\frac{11}{2}} G L G L G^{2} L}{5040} + \frac{\sqrt{6} \Delta t^{\frac{11}{2}} G L G L^{2} G^{2}}{5040} + \frac{\sqrt{6} \Delta t^{\frac{11}{2}} G L G^{2} L G L}{5040} && \\
    & \phantom{M_{4} =} + \frac{\sqrt{6} \Delta t^{\frac{11}{2}} G L G^{2} L^{2} G}{5040} + \frac{\sqrt{6} \Delta t^{\frac{11}{2}} G L G^{3} L^{2}}{5040} + \frac{\sqrt{6} \Delta t^{\frac{11}{2}} G L^{2} G L G^{2}}{5040} + \frac{\sqrt{6} \Delta t^{\frac{11}{2}} G L^{2} G^{2} L G}{5040} + \frac{\sqrt{6} \Delta t^{\frac{11}{2}} G L^{2} G^{3} L}{5040} && \\
    & \phantom{M_{4} =} + \frac{\sqrt{6} \Delta t^{\frac{11}{2}} G L^{3} G^{3}}{5040} + \frac{\sqrt{6} \Delta t^{\frac{11}{2}} G^{2} L G L G L}{5040} + \frac{\sqrt{6} \Delta t^{\frac{11}{2}} G^{2} L G L^{2} G}{5040} + \frac{\sqrt{6} \Delta t^{\frac{11}{2}} G^{2} L G^{2} L^{2}}{5040} + \frac{\sqrt{6} \Delta t^{\frac{11}{2}} G^{2} L^{2} G L G}{5040} && \\
    & \phantom{M_{4} =} + \frac{\sqrt{6} \Delta t^{\frac{11}{2}} G^{2} L^{2} G^{2} L}{5040} + \frac{\sqrt{6} \Delta t^{\frac{11}{2}} G^{2} L^{3} G^{2}}{5040} + \frac{\sqrt{6} \Delta t^{\frac{11}{2}} G^{3} L G L^{2}}{5040} + \frac{\sqrt{6} \Delta t^{\frac{11}{2}} G^{3} L^{2} G L}{5040} + \frac{\sqrt{6} \Delta t^{\frac{11}{2}} G^{3} L^{3} G}{5040} && \\
    & \phantom{M_{4} =} + \frac{\sqrt{6} \Delta t^{\frac{11}{2}} G^{4} L^{3}}{5040} + \frac{\sqrt{6} \Delta t^{\frac{11}{2}} L G L G L G^{2}}{5040} + \frac{\sqrt{6} \Delta t^{\frac{11}{2}} L G L G^{2} L G}{5040} + \frac{\sqrt{6} \Delta t^{\frac{11}{2}} L G L G^{3} L}{5040} + \frac{\sqrt{6} \Delta t^{\frac{11}{2}} L G L^{2} G^{3}}{5040} && \\
    & \phantom{M_{4} =} + \frac{\sqrt{6} \Delta t^{\frac{11}{2}} L G^{2} L G L G}{5040} + \frac{\sqrt{6} \Delta t^{\frac{11}{2}} L G^{2} L G^{2} L}{5040} + \frac{\sqrt{6} \Delta t^{\frac{11}{2}} L G^{2} L^{2} G^{2}}{5040} + \frac{\sqrt{6} \Delta t^{\frac{11}{2}} L G^{3} L G L}{5040} + \frac{\sqrt{6} \Delta t^{\frac{11}{2}} L G^{3} L^{2} G}{5040} && \\
    & \phantom{M_{4} =} + \frac{\sqrt{6} \Delta t^{\frac{11}{2}} L G^{4} L^{2}}{5040} + \frac{\sqrt{6} \Delta t^{\frac{11}{2}} L^{2} G L G^{3}}{5040} + \frac{\sqrt{6} \Delta t^{\frac{11}{2}} L^{2} G^{2} L G^{2}}{5040} + \frac{\sqrt{6} \Delta t^{\frac{11}{2}} L^{2} G^{3} L G}{5040} + \frac{\sqrt{6} \Delta t^{\frac{11}{2}} L^{2} G^{4} L}{5040} && \\
    & \phantom{M_{4} =} + \frac{\sqrt{6} \Delta t^{\frac{11}{2}} L^{3} G^{4}}{5040} + \frac{\sqrt{6} \Delta t^{\frac{9}{2}} G L G L G L}{720} + \frac{\sqrt{6} \Delta t^{\frac{9}{2}} G L G L^{2} G}{720} + \frac{\sqrt{6} \Delta t^{\frac{9}{2}} G L G^{2} L^{2}}{720} + \frac{\sqrt{6} \Delta t^{\frac{9}{2}} G L^{2} G L G}{720} && \\
    & \phantom{M_{4} =} + \frac{\sqrt{6} \Delta t^{\frac{9}{2}} G L^{2} G^{2} L}{720} + \frac{\sqrt{6} \Delta t^{\frac{9}{2}} G L^{3} G^{2}}{720} + \frac{\sqrt{6} \Delta t^{\frac{9}{2}} G^{2} L G L^{2}}{720} + \frac{\sqrt{6} \Delta t^{\frac{9}{2}} G^{2} L^{2} G L}{720} + \frac{\sqrt{6} \Delta t^{\frac{9}{2}} G^{2} L^{3} G}{720} && \\
    & \phantom{M_{4} =} + \frac{\sqrt{6} \Delta t^{\frac{9}{2}} G^{3} L^{3}}{720} + \frac{\sqrt{6} \Delta t^{\frac{9}{2}} L G L G L G}{720} + \frac{\sqrt{6} \Delta t^{\frac{9}{2}} L G L G^{2} L}{720} + \frac{\sqrt{6} \Delta t^{\frac{9}{2}} L G L^{2} G^{2}}{720} + \frac{\sqrt{6} \Delta t^{\frac{9}{2}} L G^{2} L G L}{720} && \\
    & \phantom{M_{4} =} + \frac{\sqrt{6} \Delta t^{\frac{9}{2}} L G^{2} L^{2} G}{720} + \frac{\sqrt{6} \Delta t^{\frac{9}{2}} L G^{3} L^{2}}{720} + \frac{\sqrt{6} \Delta t^{\frac{9}{2}} L^{2} G L G^{2}}{720} + \frac{\sqrt{6} \Delta t^{\frac{9}{2}} L^{2} G^{2} L G}{720} + \frac{\sqrt{6} \Delta t^{\frac{9}{2}} L^{2} G^{3} L}{720} && \\
    & \phantom{M_{4} =} + \frac{\sqrt{6} \Delta t^{\frac{9}{2}} L^{3} G^{3}}{720} + \frac{\sqrt{6} \Delta t^{\frac{7}{2}} G L G L^{2}}{120} + \frac{\sqrt{6} \Delta t^{\frac{7}{2}} G L^{2} G L}{120} + \frac{\sqrt{6} \Delta t^{\frac{7}{2}} G L^{3} G}{120} + \frac{\sqrt{6} \Delta t^{\frac{7}{2}} G^{2} L^{3}}{120} && \\
    & \phantom{M_{4} =} + \frac{\sqrt{6} \Delta t^{\frac{7}{2}} L G L G L}{120} + \frac{\sqrt{6} \Delta t^{\frac{7}{2}} L G L^{2} G}{120} + \frac{\sqrt{6} \Delta t^{\frac{7}{2}} L G^{2} L^{2}}{120} + \frac{\sqrt{6} \Delta t^{\frac{7}{2}} L^{2} G L G}{120} + \frac{\sqrt{6} \Delta t^{\frac{7}{2}} L^{2} G^{2} L}{120} && \\
    & \phantom{M_{4} =} + \frac{\sqrt{6} \Delta t^{\frac{7}{2}} L^{3} G^{2}}{120} + \frac{\sqrt{6} \Delta t^{\frac{5}{2}} G L^{3}}{24} + \frac{\sqrt{6} \Delta t^{\frac{5}{2}} L G L^{2}}{24} + \frac{\sqrt{6} \Delta t^{\frac{5}{2}} L^{2} G L}{24} + \frac{\sqrt{6} \Delta t^{\frac{5}{2}} L^{3} G}{24} && \\
    & \phantom{M_{4} =} + \frac{\sqrt{6} \Delta t^{\frac{3}{2}} L^{3}}{6} && \\
\end{flalign*}
\begin{flalign*}
    & M_{5} = \frac{\Delta t^{6} G L G L G^{3}}{5040} + \frac{\Delta t^{6} G L G^{2} L G^{2}}{20160} - \frac{\Delta t^{6} G L G^{3} L G}{10080} - \frac{\Delta t^{6} G L G^{4} L}{4032} + \frac{\Delta t^{6} G L^{2} G^{4}}{2880} && \\
    & \phantom{M_{5} =} + \frac{\Delta t^{6} G^{2} L G L G^{2}}{20160} - \frac{\Delta t^{6} G^{2} L G^{2} L G}{10080} - \frac{\Delta t^{6} G^{2} L G^{3} L}{4032} + \frac{\Delta t^{6} G^{2} L^{2} G^{3}}{5040} - \frac{\Delta t^{6} G^{3} L G L G}{10080} && \\
    & \phantom{M_{5} =} - \frac{\Delta t^{6} G^{3} L G^{2} L}{4032} + \frac{\Delta t^{6} G^{3} L^{2} G^{2}}{20160} - \frac{\Delta t^{6} G^{4} L G L}{4032} - \frac{\Delta t^{6} G^{4} L^{2} G}{10080} - \frac{\Delta t^{6} G^{5} L^{2}}{4032} && \\
    & \phantom{M_{5} =} + \frac{\Delta t^{6} L G L G^{4}}{2880} + \frac{\Delta t^{6} L G^{2} L G^{3}}{5040} + \frac{\Delta t^{6} L G^{3} L G^{2}}{20160} - \frac{\Delta t^{6} L G^{4} L G}{10080} - \frac{\Delta t^{6} L G^{5} L}{4032} && \\
    & \phantom{M_{5} =} + \frac{\Delta t^{6} L^{2} G^{5}}{2016} + \frac{\Delta t^{5} G L G L G^{2}}{1260} - \frac{\Delta t^{5} G L G^{2} L G}{2520} - \frac{\Delta t^{5} G L G^{3} L}{630} + \frac{\Delta t^{5} G L^{2} G^{3}}{504} && \\
    & \phantom{M_{5} =} - \frac{\Delta t^{5} G^{2} L G L G}{2520} - \frac{\Delta t^{5} G^{2} L G^{2} L}{630} + \frac{\Delta t^{5} G^{2} L^{2} G^{2}}{1260} - \frac{\Delta t^{5} G^{3} L G L}{630} - \frac{\Delta t^{5} G^{3} L^{2} G}{2520} && \\
    & \phantom{M_{5} =} - \frac{\Delta t^{5} G^{4} L^{2}}{630} + \frac{\Delta t^{5} L G L G^{3}}{504} + \frac{\Delta t^{5} L G^{2} L G^{2}}{1260} - \frac{\Delta t^{5} L G^{3} L G}{2520} - \frac{\Delta t^{5} L G^{4} L}{630} && \\
    & \phantom{M_{5} =} + \frac{\Delta t^{5} L^{2} G^{4}}{315} - \frac{\Delta t^{4} G L G^{2} L}{120} + \frac{\Delta t^{4} G L^{2} G^{2}}{120} - \frac{\Delta t^{4} G^{2} L G L}{120} - \frac{\Delta t^{4} G^{3} L^{2}}{120} && \\
    & \phantom{M_{5} =} + \frac{\Delta t^{4} L G L G^{2}}{120} - \frac{\Delta t^{4} L G^{3} L}{120} + \frac{\Delta t^{4} L^{2} G^{3}}{60} - \frac{\Delta t^{3} G L G L}{30} + \frac{\Delta t^{3} G L^{2} G}{60} && \\
    & \phantom{M_{5} =} - \frac{\Delta t^{3} G^{2} L^{2}}{30} + \frac{\Delta t^{3} L G L G}{60} - \frac{\Delta t^{3} L G^{2} L}{30} + \frac{\Delta t^{3} L^{2} G^{2}}{15} - \frac{\Delta t^{2} G L^{2}}{12} && \\
    & \phantom{M_{5} =} - \frac{\Delta t^{2} L G L}{12} + \frac{\Delta t^{2} L^{2} G}{6} && \\
\end{flalign*}
\begin{flalign*}
    & M_{6} = \frac{\sqrt{3} \Delta t^{6} G L G^{2} L G^{2}}{20160} + \frac{\sqrt{3} \Delta t^{6} G L G^{3} L G}{10080} + \frac{\sqrt{3} \Delta t^{6} G L G^{4} L}{6720} - \frac{\sqrt{3} \Delta t^{6} G L^{2} G^{4}}{20160} - \frac{\sqrt{3} \Delta t^{6} G^{2} L G L G^{2}}{20160} && \\
    & \phantom{M_{6} =} + \frac{\sqrt{3} \Delta t^{6} G^{2} L G^{3} L}{20160} - \frac{\sqrt{3} \Delta t^{6} G^{2} L^{2} G^{3}}{10080} - \frac{\sqrt{3} \Delta t^{6} G^{3} L G L G}{10080} - \frac{\sqrt{3} \Delta t^{6} G^{3} L G^{2} L}{20160} - \frac{\sqrt{3} \Delta t^{6} G^{3} L^{2} G^{2}}{6720} && \\
    & \phantom{M_{6} =} - \frac{\sqrt{3} \Delta t^{6} G^{4} L G L}{6720} - \frac{\sqrt{3} \Delta t^{6} G^{4} L^{2} G}{5040} - \frac{\sqrt{3} \Delta t^{6} G^{5} L^{2}}{4032} + \frac{\sqrt{3} \Delta t^{6} L G L G^{4}}{20160} + \frac{\sqrt{3} \Delta t^{6} L G^{2} L G^{3}}{10080} && \\
    & \phantom{M_{6} =} + \frac{\sqrt{3} \Delta t^{6} L G^{3} L G^{2}}{6720} + \frac{\sqrt{3} \Delta t^{6} L G^{4} L G}{5040} + \frac{\sqrt{3} \Delta t^{6} L G^{5} L}{4032} + \frac{\sqrt{3} \Delta t^{5} G L G^{2} L G}{2520} + \frac{\sqrt{3} \Delta t^{5} G L G^{3} L}{1260} && \\
    & \phantom{M_{6} =} - \frac{\sqrt{3} \Delta t^{5} G L^{2} G^{3}}{2520} - \frac{\sqrt{3} \Delta t^{5} G^{2} L G L G}{2520} - \frac{\sqrt{3} \Delta t^{5} G^{2} L^{2} G^{2}}{1260} - \frac{\sqrt{3} \Delta t^{5} G^{3} L G L}{1260} - \frac{\sqrt{3} \Delta t^{5} G^{3} L^{2} G}{840} && \\
    & \phantom{M_{6} =} - \frac{\sqrt{3} \Delta t^{5} G^{4} L^{2}}{630} + \frac{\sqrt{3} \Delta t^{5} L G L G^{3}}{2520} + \frac{\sqrt{3} \Delta t^{5} L G^{2} L G^{2}}{1260} + \frac{\sqrt{3} \Delta t^{5} L G^{3} L G}{840} + \frac{\sqrt{3} \Delta t^{5} L G^{4} L}{630} && \\
    & \phantom{M_{6} =} + \frac{\sqrt{3} \Delta t^{4} G L G^{2} L}{360} - \frac{\sqrt{3} \Delta t^{4} G L^{2} G^{2}}{360} - \frac{\sqrt{3} \Delta t^{4} G^{2} L G L}{360} - \frac{\sqrt{3} \Delta t^{4} G^{2} L^{2} G}{180} - \frac{\sqrt{3} \Delta t^{4} G^{3} L^{2}}{120} && \\
    & \phantom{M_{6} =} + \frac{\sqrt{3} \Delta t^{4} L G L G^{2}}{360} + \frac{\sqrt{3} \Delta t^{4} L G^{2} L G}{180} + \frac{\sqrt{3} \Delta t^{4} L G^{3} L}{120} - \frac{\sqrt{3} \Delta t^{3} G L^{2} G}{60} - \frac{\sqrt{3} \Delta t^{3} G^{2} L^{2}}{30} && \\
    & \phantom{M_{6} =} + \frac{\sqrt{3} \Delta t^{3} L G L G}{60} + \frac{\sqrt{3} \Delta t^{3} L G^{2} L}{30} - \frac{\sqrt{3} \Delta t^{2} G L^{2}}{12} + \frac{\sqrt{3} \Delta t^{2} L G L}{12} && \\
\end{flalign*}
\begin{flalign*}
    & M_{7} = \frac{\sqrt{6} \Delta t^{6} G L G L G L G L}{20160} + \frac{\sqrt{6} \Delta t^{6} G L G L G L^{2} G}{20160} + \frac{\sqrt{6} \Delta t^{6} G L G L G^{2} L^{2}}{20160} + \frac{\sqrt{6} \Delta t^{6} G L G L^{2} G L G}{20160} + \frac{\sqrt{6} \Delta t^{6} G L G L^{2} G^{2} L}{20160} && \\
    & \phantom{M_{7} =} + \frac{\sqrt{6} \Delta t^{6} G L G L^{3} G^{2}}{20160} + \frac{\sqrt{6} \Delta t^{6} G L G^{2} L G L^{2}}{20160} + \frac{\sqrt{6} \Delta t^{6} G L G^{2} L^{2} G L}{20160} + \frac{\sqrt{6} \Delta t^{6} G L G^{2} L^{3} G}{20160} + \frac{\sqrt{6} \Delta t^{6} G L G^{3} L^{3}}{20160} && \\
    & \phantom{M_{7} =} + \frac{\sqrt{6} \Delta t^{6} G L^{2} G L G L G}{20160} + \frac{\sqrt{6} \Delta t^{6} G L^{2} G L G^{2} L}{20160} + \frac{\sqrt{6} \Delta t^{6} G L^{2} G L^{2} G^{2}}{20160} + \frac{\sqrt{6} \Delta t^{6} G L^{2} G^{2} L G L}{20160} + \frac{\sqrt{6} \Delta t^{6} G L^{2} G^{2} L^{2} G}{20160} && \\
    & \phantom{M_{7} =} + \frac{\sqrt{6} \Delta t^{6} G L^{2} G^{3} L^{2}}{20160} + \frac{\sqrt{6} \Delta t^{6} G L^{3} G L G^{2}}{20160} + \frac{\sqrt{6} \Delta t^{6} G L^{3} G^{2} L G}{20160} + \frac{\sqrt{6} \Delta t^{6} G L^{3} G^{3} L}{20160} + \frac{\sqrt{6} \Delta t^{6} G L^{4} G^{3}}{20160} && \\
    & \phantom{M_{7} =} + \frac{\sqrt{6} \Delta t^{6} G^{2} L G L G L^{2}}{20160} + \frac{\sqrt{6} \Delta t^{6} G^{2} L G L^{2} G L}{20160} + \frac{\sqrt{6} \Delta t^{6} G^{2} L G L^{3} G}{20160} + \frac{\sqrt{6} \Delta t^{6} G^{2} L G^{2} L^{3}}{20160} + \frac{\sqrt{6} \Delta t^{6} G^{2} L^{2} G L G L}{20160} && \\
    & \phantom{M_{7} =} + \frac{\sqrt{6} \Delta t^{6} G^{2} L^{2} G L^{2} G}{20160} + \frac{\sqrt{6} \Delta t^{6} G^{2} L^{2} G^{2} L^{2}}{20160} + \frac{\sqrt{6} \Delta t^{6} G^{2} L^{3} G L G}{20160} + \frac{\sqrt{6} \Delta t^{6} G^{2} L^{3} G^{2} L}{20160} + \frac{\sqrt{6} \Delta t^{6} G^{2} L^{4} G^{2}}{20160} && \\
    & \phantom{M_{7} =} + \frac{\sqrt{6} \Delta t^{6} G^{3} L G L^{3}}{20160} + \frac{\sqrt{6} \Delta t^{6} G^{3} L^{2} G L^{2}}{20160} + \frac{\sqrt{6} \Delta t^{6} G^{3} L^{3} G L}{20160} + \frac{\sqrt{6} \Delta t^{6} G^{3} L^{4} G}{20160} + \frac{\sqrt{6} \Delta t^{6} G^{4} L^{4}}{20160} && \\
    & \phantom{M_{7} =} + \frac{\sqrt{6} \Delta t^{6} L G L G L G L G}{20160} + \frac{\sqrt{6} \Delta t^{6} L G L G L G^{2} L}{20160} + \frac{\sqrt{6} \Delta t^{6} L G L G L^{2} G^{2}}{20160} + \frac{\sqrt{6} \Delta t^{6} L G L G^{2} L G L}{20160} + \frac{\sqrt{6} \Delta t^{6} L G L G^{2} L^{2} G}{20160} && \\
    & \phantom{M_{7} =} + \frac{\sqrt{6} \Delta t^{6} L G L G^{3} L^{2}}{20160} + \frac{\sqrt{6} \Delta t^{6} L G L^{2} G L G^{2}}{20160} + \frac{\sqrt{6} \Delta t^{6} L G L^{2} G^{2} L G}{20160} + \frac{\sqrt{6} \Delta t^{6} L G L^{2} G^{3} L}{20160} + \frac{\sqrt{6} \Delta t^{6} L G L^{3} G^{3}}{20160} && \\
    & \phantom{M_{7} =} + \frac{\sqrt{6} \Delta t^{6} L G^{2} L G L G L}{20160} + \frac{\sqrt{6} \Delta t^{6} L G^{2} L G L^{2} G}{20160} + \frac{\sqrt{6} \Delta t^{6} L G^{2} L G^{2} L^{2}}{20160} + \frac{\sqrt{6} \Delta t^{6} L G^{2} L^{2} G L G}{20160} + \frac{\sqrt{6} \Delta t^{6} L G^{2} L^{2} G^{2} L}{20160} && \\
    & \phantom{M_{7} =} + \frac{\sqrt{6} \Delta t^{6} L G^{2} L^{3} G^{2}}{20160} + \frac{\sqrt{6} \Delta t^{6} L G^{3} L G L^{2}}{20160} + \frac{\sqrt{6} \Delta t^{6} L G^{3} L^{2} G L}{20160} + \frac{\sqrt{6} \Delta t^{6} L G^{3} L^{3} G}{20160} + \frac{\sqrt{6} \Delta t^{6} L G^{4} L^{3}}{20160} && \\
    & \phantom{M_{7} =} + \frac{\sqrt{6} \Delta t^{6} L^{2} G L G L G^{2}}{20160} + \frac{\sqrt{6} \Delta t^{6} L^{2} G L G^{2} L G}{20160} + \frac{\sqrt{6} \Delta t^{6} L^{2} G L G^{3} L}{20160} + \frac{\sqrt{6} \Delta t^{6} L^{2} G L^{2} G^{3}}{20160} + \frac{\sqrt{6} \Delta t^{6} L^{2} G^{2} L G L G}{20160} && \\
    & \phantom{M_{7} =} + \frac{\sqrt{6} \Delta t^{6} L^{2} G^{2} L G^{2} L}{20160} + \frac{\sqrt{6} \Delta t^{6} L^{2} G^{2} L^{2} G^{2}}{20160} + \frac{\sqrt{6} \Delta t^{6} L^{2} G^{3} L G L}{20160} + \frac{\sqrt{6} \Delta t^{6} L^{2} G^{3} L^{2} G}{20160} + \frac{\sqrt{6} \Delta t^{6} L^{2} G^{4} L^{2}}{20160} && \\
    & \phantom{M_{7} =} + \frac{\sqrt{6} \Delta t^{6} L^{3} G L G^{3}}{20160} + \frac{\sqrt{6} \Delta t^{6} L^{3} G^{2} L G^{2}}{20160} + \frac{\sqrt{6} \Delta t^{6} L^{3} G^{3} L G}{20160} + \frac{\sqrt{6} \Delta t^{6} L^{3} G^{4} L}{20160} + \frac{\sqrt{6} \Delta t^{6} L^{4} G^{4}}{20160} && \\
    & \phantom{M_{7} =} + \frac{\sqrt{6} \Delta t^{5} G L G L G L^{2}}{2520} + \frac{\sqrt{6} \Delta t^{5} G L G L^{2} G L}{2520} + \frac{\sqrt{6} \Delta t^{5} G L G L^{3} G}{2520} + \frac{\sqrt{6} \Delta t^{5} G L G^{2} L^{3}}{2520} + \frac{\sqrt{6} \Delta t^{5} G L^{2} G L G L}{2520} && \\
    & \phantom{M_{7} =} + \frac{\sqrt{6} \Delta t^{5} G L^{2} G L^{2} G}{2520} + \frac{\sqrt{6} \Delta t^{5} G L^{2} G^{2} L^{2}}{2520} + \frac{\sqrt{6} \Delta t^{5} G L^{3} G L G}{2520} + \frac{\sqrt{6} \Delta t^{5} G L^{3} G^{2} L}{2520} + \frac{\sqrt{6} \Delta t^{5} G L^{4} G^{2}}{2520} && \\
    & \phantom{M_{7} =} + \frac{\sqrt{6} \Delta t^{5} G^{2} L G L^{3}}{2520} + \frac{\sqrt{6} \Delta t^{5} G^{2} L^{2} G L^{2}}{2520} + \frac{\sqrt{6} \Delta t^{5} G^{2} L^{3} G L}{2520} + \frac{\sqrt{6} \Delta t^{5} G^{2} L^{4} G}{2520} + \frac{\sqrt{6} \Delta t^{5} G^{3} L^{4}}{2520} && \\
    & \phantom{M_{7} =} + \frac{\sqrt{6} \Delta t^{5} L G L G L G L}{2520} + \frac{\sqrt{6} \Delta t^{5} L G L G L^{2} G}{2520} + \frac{\sqrt{6} \Delta t^{5} L G L G^{2} L^{2}}{2520} + \frac{\sqrt{6} \Delta t^{5} L G L^{2} G L G}{2520} + \frac{\sqrt{6} \Delta t^{5} L G L^{2} G^{2} L}{2520} && \\
    & \phantom{M_{7} =} + \frac{\sqrt{6} \Delta t^{5} L G L^{3} G^{2}}{2520} + \frac{\sqrt{6} \Delta t^{5} L G^{2} L G L^{2}}{2520} + \frac{\sqrt{6} \Delta t^{5} L G^{2} L^{2} G L}{2520} + \frac{\sqrt{6} \Delta t^{5} L G^{2} L^{3} G}{2520} + \frac{\sqrt{6} \Delta t^{5} L G^{3} L^{3}}{2520} && \\
    & \phantom{M_{7} =} + \frac{\sqrt{6} \Delta t^{5} L^{2} G L G L G}{2520} + \frac{\sqrt{6} \Delta t^{5} L^{2} G L G^{2} L}{2520} + \frac{\sqrt{6} \Delta t^{5} L^{2} G L^{2} G^{2}}{2520} + \frac{\sqrt{6} \Delta t^{5} L^{2} G^{2} L G L}{2520} + \frac{\sqrt{6} \Delta t^{5} L^{2} G^{2} L^{2} G}{2520} && \\
    & \phantom{M_{7} =} + \frac{\sqrt{6} \Delta t^{5} L^{2} G^{3} L^{2}}{2520} + \frac{\sqrt{6} \Delta t^{5} L^{3} G L G^{2}}{2520} + \frac{\sqrt{6} \Delta t^{5} L^{3} G^{2} L G}{2520} + \frac{\sqrt{6} \Delta t^{5} L^{3} G^{3} L}{2520} + \frac{\sqrt{6} \Delta t^{5} L^{4} G^{3}}{2520} && \\
    & \phantom{M_{7} =} + \frac{\sqrt{6} \Delta t^{4} G L G L^{3}}{360} + \frac{\sqrt{6} \Delta t^{4} G L^{2} G L^{2}}{360} + \frac{\sqrt{6} \Delta t^{4} G L^{3} G L}{360} + \frac{\sqrt{6} \Delta t^{4} G L^{4} G}{360} + \frac{\sqrt{6} \Delta t^{4} G^{2} L^{4}}{360} && \\
    & \phantom{M_{7} =} + \frac{\sqrt{6} \Delta t^{4} L G L G L^{2}}{360} + \frac{\sqrt{6} \Delta t^{4} L G L^{2} G L}{360} + \frac{\sqrt{6} \Delta t^{4} L G L^{3} G}{360} + \frac{\sqrt{6} \Delta t^{4} L G^{2} L^{3}}{360} + \frac{\sqrt{6} \Delta t^{4} L^{2} G L G L}{360} && \\
    & \phantom{M_{7} =} + \frac{\sqrt{6} \Delta t^{4} L^{2} G L^{2} G}{360} + \frac{\sqrt{6} \Delta t^{4} L^{2} G^{2} L^{2}}{360} + \frac{\sqrt{6} \Delta t^{4} L^{3} G L G}{360} + \frac{\sqrt{6} \Delta t^{4} L^{3} G^{2} L}{360} + \frac{\sqrt{6} \Delta t^{4} L^{4} G^{2}}{360} && \\
    & \phantom{M_{7} =} + \frac{\sqrt{6} \Delta t^{3} G L^{4}}{60} + \frac{\sqrt{6} \Delta t^{3} L G L^{3}}{60} + \frac{\sqrt{6} \Delta t^{3} L^{2} G L^{2}}{60} + \frac{\sqrt{6} \Delta t^{3} L^{3} G L}{60} + \frac{\sqrt{6} \Delta t^{3} L^{4} G}{60} && \\
    & \phantom{M_{7} =} + \frac{\sqrt{6} \Delta t^{2} L^{4}}{12} && \\
\end{flalign*}
\begin{flalign*}
    & M_{8} = - \frac{\sqrt{5} \Delta t^{\frac{11}{2}} G L G^{4}}{10080} - \frac{\sqrt{5} \Delta t^{\frac{11}{2}} G^{2} L G^{3}}{2520} - \frac{\sqrt{5} \Delta t^{\frac{11}{2}} G^{3} L G^{2}}{2520} - \frac{\sqrt{5} \Delta t^{\frac{11}{2}} G^{4} L G}{10080} + \frac{\sqrt{5} \Delta t^{\frac{11}{2}} G^{5} L}{2016} && \\
    & \phantom{M_{8} =} + \frac{\sqrt{5} \Delta t^{\frac{11}{2}} L G^{5}}{2016} - \frac{\sqrt{5} \Delta t^{\frac{9}{2}} G L G^{3}}{840} - \frac{\sqrt{5} \Delta t^{\frac{9}{2}} G^{2} L G^{2}}{420} - \frac{\sqrt{5} \Delta t^{\frac{9}{2}} G^{3} L G}{840} + \frac{\sqrt{5} \Delta t^{\frac{9}{2}} G^{4} L}{420} && \\
    & \phantom{M_{8} =} + \frac{\sqrt{5} \Delta t^{\frac{9}{2}} L G^{4}}{420} - \frac{\sqrt{5} \Delta t^{\frac{7}{2}} G L G^{2}}{120} - \frac{\sqrt{5} \Delta t^{\frac{7}{2}} G^{2} L G}{120} + \frac{\sqrt{5} \Delta t^{\frac{7}{2}} G^{3} L}{120} + \frac{\sqrt{5} \Delta t^{\frac{7}{2}} L G^{3}}{120} && \\
    & \phantom{M_{8} =} - \frac{\sqrt{5} \Delta t^{\frac{5}{2}} G L G}{30} + \frac{\sqrt{5} \Delta t^{\frac{5}{2}} G^{2} L}{60} + \frac{\sqrt{5} \Delta t^{\frac{5}{2}} L G^{2}}{60} && \\
\end{flalign*}
\begin{flalign*}
    & M_{9} = - \frac{\sqrt{10} \Delta t^{\frac{11}{2}} G L G L G^{2} L}{10080} + \frac{\sqrt{10} \Delta t^{\frac{11}{2}} G L G L^{2} G^{2}}{10080} - \frac{\sqrt{10} \Delta t^{\frac{11}{2}} G L G^{2} L G L}{10080} - \frac{\sqrt{10} \Delta t^{\frac{11}{2}} G L G^{3} L^{2}}{10080} + \frac{\sqrt{10} \Delta t^{\frac{11}{2}} G L^{2} G L G^{2}}{10080} && \\
    & \phantom{M_{9} =} - \frac{\sqrt{10} \Delta t^{\frac{11}{2}} G L^{2} G^{3} L}{10080} + \frac{\sqrt{10} \Delta t^{\frac{11}{2}} G L^{3} G^{3}}{5040} - \frac{\sqrt{10} \Delta t^{\frac{11}{2}} G^{2} L G L G L}{10080} - \frac{\sqrt{10} \Delta t^{\frac{11}{2}} G^{2} L G^{2} L^{2}}{10080} - \frac{\sqrt{10} \Delta t^{\frac{11}{2}} G^{2} L^{2} G^{2} L}{10080} && \\
    & \phantom{M_{9} =} + \frac{\sqrt{10} \Delta t^{\frac{11}{2}} G^{2} L^{3} G^{2}}{10080} - \frac{\sqrt{10} \Delta t^{\frac{11}{2}} G^{3} L G L^{2}}{10080} - \frac{\sqrt{10} \Delta t^{\frac{11}{2}} G^{3} L^{2} G L}{10080} - \frac{\sqrt{10} \Delta t^{\frac{11}{2}} G^{4} L^{3}}{10080} + \frac{\sqrt{10} \Delta t^{\frac{11}{2}} L G L G L G^{2}}{10080} && \\
    & \phantom{M_{9} =} - \frac{\sqrt{10} \Delta t^{\frac{11}{2}} L G L G^{3} L}{10080} + \frac{\sqrt{10} \Delta t^{\frac{11}{2}} L G L^{2} G^{3}}{5040} - \frac{\sqrt{10} \Delta t^{\frac{11}{2}} L G^{2} L G^{2} L}{10080} + \frac{\sqrt{10} \Delta t^{\frac{11}{2}} L G^{2} L^{2} G^{2}}{10080} - \frac{\sqrt{10} \Delta t^{\frac{11}{2}} L G^{3} L G L}{10080} && \\
    & \phantom{M_{9} =} - \frac{\sqrt{10} \Delta t^{\frac{11}{2}} L G^{4} L^{2}}{10080} + \frac{\sqrt{10} \Delta t^{\frac{11}{2}} L^{2} G L G^{3}}{5040} + \frac{\sqrt{10} \Delta t^{\frac{11}{2}} L^{2} G^{2} L G^{2}}{10080} - \frac{\sqrt{10} \Delta t^{\frac{11}{2}} L^{2} G^{4} L}{10080} + \frac{\sqrt{10} \Delta t^{\frac{11}{2}} L^{3} G^{4}}{3360} && \\
    & \phantom{M_{9} =} - \frac{\sqrt{10} \Delta t^{\frac{9}{2}} G L G L G L}{1680} + \frac{\sqrt{10} \Delta t^{\frac{9}{2}} G L G L^{2} G}{5040} - \frac{\sqrt{10} \Delta t^{\frac{9}{2}} G L G^{2} L^{2}}{1680} + \frac{\sqrt{10} \Delta t^{\frac{9}{2}} G L^{2} G L G}{5040} - \frac{\sqrt{10} \Delta t^{\frac{9}{2}} G L^{2} G^{2} L}{1680} && \\
    & \phantom{M_{9} =} + \frac{\sqrt{10} \Delta t^{\frac{9}{2}} G L^{3} G^{2}}{1008} - \frac{\sqrt{10} \Delta t^{\frac{9}{2}} G^{2} L G L^{2}}{1680} - \frac{\sqrt{10} \Delta t^{\frac{9}{2}} G^{2} L^{2} G L}{1680} + \frac{\sqrt{10} \Delta t^{\frac{9}{2}} G^{2} L^{3} G}{5040} - \frac{\sqrt{10} \Delta t^{\frac{9}{2}} G^{3} L^{3}}{1680} && \\
    & \phantom{M_{9} =} + \frac{\sqrt{10} \Delta t^{\frac{9}{2}} L G L G L G}{5040} - \frac{\sqrt{10} \Delta t^{\frac{9}{2}} L G L G^{2} L}{1680} + \frac{\sqrt{10} \Delta t^{\frac{9}{2}} L G L^{2} G^{2}}{1008} - \frac{\sqrt{10} \Delta t^{\frac{9}{2}} L G^{2} L G L}{1680} + \frac{\sqrt{10} \Delta t^{\frac{9}{2}} L G^{2} L^{2} G}{5040} && \\
    & \phantom{M_{9} =} - \frac{\sqrt{10} \Delta t^{\frac{9}{2}} L G^{3} L^{2}}{1680} + \frac{\sqrt{10} \Delta t^{\frac{9}{2}} L^{2} G L G^{2}}{1008} + \frac{\sqrt{10} \Delta t^{\frac{9}{2}} L^{2} G^{2} L G}{5040} - \frac{\sqrt{10} \Delta t^{\frac{9}{2}} L^{2} G^{3} L}{1680} + \frac{\sqrt{10} \Delta t^{\frac{9}{2}} L^{3} G^{3}}{560} && \\
    & \phantom{M_{9} =} - \frac{\sqrt{10} \Delta t^{\frac{7}{2}} G L G L^{2}}{360} - \frac{\sqrt{10} \Delta t^{\frac{7}{2}} G L^{2} G L}{360} + \frac{\sqrt{10} \Delta t^{\frac{7}{2}} G L^{3} G}{360} - \frac{\sqrt{10} \Delta t^{\frac{7}{2}} G^{2} L^{3}}{360} - \frac{\sqrt{10} \Delta t^{\frac{7}{2}} L G L G L}{360} && \\
    & \phantom{M_{9} =} + \frac{\sqrt{10} \Delta t^{\frac{7}{2}} L G L^{2} G}{360} - \frac{\sqrt{10} \Delta t^{\frac{7}{2}} L G^{2} L^{2}}{360} + \frac{\sqrt{10} \Delta t^{\frac{7}{2}} L^{2} G L G}{360} - \frac{\sqrt{10} \Delta t^{\frac{7}{2}} L^{2} G^{2} L}{360} + \frac{\sqrt{10} \Delta t^{\frac{7}{2}} L^{3} G^{2}}{120} && \\
    & \phantom{M_{9} =} - \frac{\sqrt{10} \Delta t^{\frac{5}{2}} G L^{3}}{120} - \frac{\sqrt{10} \Delta t^{\frac{5}{2}} L G L^{2}}{120} - \frac{\sqrt{10} \Delta t^{\frac{5}{2}} L^{2} G L}{120} + \frac{\sqrt{10} \Delta t^{\frac{5}{2}} L^{3} G}{40} && \\
\end{flalign*}
\begin{flalign*}
    & M_{10} = \frac{\sqrt{5} \Delta t^{\frac{11}{2}} G L G L G^{2} L}{10080} - \frac{\sqrt{5} \Delta t^{\frac{11}{2}} G L G L^{2} G^{2}}{10080} - \frac{\sqrt{5} \Delta t^{\frac{11}{2}} G L G^{2} L G L}{20160} - \frac{\sqrt{5} \Delta t^{\frac{11}{2}} G L G^{2} L^{2} G}{6720} - \frac{\sqrt{5} \Delta t^{\frac{11}{2}} G L G^{3} L^{2}}{5040} && \\
    & \phantom{M_{10} =} + \frac{\sqrt{5} \Delta t^{\frac{11}{2}} G L^{2} G L G^{2}}{20160} + \frac{\sqrt{5} \Delta t^{\frac{11}{2}} G L^{2} G^{2} L G}{6720} + \frac{\sqrt{5} \Delta t^{\frac{11}{2}} G L^{2} G^{3} L}{4032} - \frac{\sqrt{5} \Delta t^{\frac{11}{2}} G L^{3} G^{3}}{20160} - \frac{\sqrt{5} \Delta t^{\frac{11}{2}} G^{2} L G L G L}{20160} && \\
    & \phantom{M_{10} =} - \frac{\sqrt{5} \Delta t^{\frac{11}{2}} G^{2} L G L^{2} G}{6720} - \frac{\sqrt{5} \Delta t^{\frac{11}{2}} G^{2} L G^{2} L^{2}}{5040} + \frac{\sqrt{5} \Delta t^{\frac{11}{2}} G^{2} L^{2} G^{2} L}{10080} - \frac{\sqrt{5} \Delta t^{\frac{11}{2}} G^{2} L^{3} G^{2}}{10080} - \frac{\sqrt{5} \Delta t^{\frac{11}{2}} G^{3} L G L^{2}}{5040} && \\
    & \phantom{M_{10} =} - \frac{\sqrt{5} \Delta t^{\frac{11}{2}} G^{3} L^{2} G L}{20160} - \frac{\sqrt{5} \Delta t^{\frac{11}{2}} G^{3} L^{3} G}{6720} - \frac{\sqrt{5} \Delta t^{\frac{11}{2}} G^{4} L^{3}}{5040} + \frac{\sqrt{5} \Delta t^{\frac{11}{2}} L G L G L G^{2}}{20160} + \frac{\sqrt{5} \Delta t^{\frac{11}{2}} L G L G^{2} L G}{6720} && \\
    & \phantom{M_{10} =} + \frac{\sqrt{5} \Delta t^{\frac{11}{2}} L G L G^{3} L}{4032} - \frac{\sqrt{5} \Delta t^{\frac{11}{2}} L G L^{2} G^{3}}{20160} + \frac{\sqrt{5} \Delta t^{\frac{11}{2}} L G^{2} L G^{2} L}{10080} - \frac{\sqrt{5} \Delta t^{\frac{11}{2}} L G^{2} L^{2} G^{2}}{10080} - \frac{\sqrt{5} \Delta t^{\frac{11}{2}} L G^{3} L G L}{20160} && \\
    & \phantom{M_{10} =} - \frac{\sqrt{5} \Delta t^{\frac{11}{2}} L G^{3} L^{2} G}{6720} - \frac{\sqrt{5} \Delta t^{\frac{11}{2}} L G^{4} L^{2}}{5040} + \frac{\sqrt{5} \Delta t^{\frac{11}{2}} L^{2} G L G^{3}}{10080} + \frac{\sqrt{5} \Delta t^{\frac{11}{2}} L^{2} G^{2} L G^{2}}{5040} + \frac{\sqrt{5} \Delta t^{\frac{11}{2}} L^{2} G^{3} L G}{3360} && \\
    & \phantom{M_{10} =} + \frac{\sqrt{5} \Delta t^{\frac{11}{2}} L^{2} G^{4} L}{2520} - \frac{\sqrt{5} \Delta t^{\frac{9}{2}} G L G L^{2} G}{1260} - \frac{\sqrt{5} \Delta t^{\frac{9}{2}} G L G^{2} L^{2}}{840} + \frac{\sqrt{5} \Delta t^{\frac{9}{2}} G L^{2} G L G}{2520} + \frac{\sqrt{5} \Delta t^{\frac{9}{2}} G L^{2} G^{2} L}{840} && \\
    & \phantom{M_{10} =} - \frac{\sqrt{5} \Delta t^{\frac{9}{2}} G L^{3} G^{2}}{2520} - \frac{\sqrt{5} \Delta t^{\frac{9}{2}} G^{2} L G L^{2}}{840} - \frac{\sqrt{5} \Delta t^{\frac{9}{2}} G^{2} L^{3} G}{1260} - \frac{\sqrt{5} \Delta t^{\frac{9}{2}} G^{3} L^{3}}{840} + \frac{\sqrt{5} \Delta t^{\frac{9}{2}} L G L G L G}{2520} && \\
    & \phantom{M_{10} =} + \frac{\sqrt{5} \Delta t^{\frac{9}{2}} L G L G^{2} L}{840} - \frac{\sqrt{5} \Delta t^{\frac{9}{2}} L G L^{2} G^{2}}{2520} - \frac{\sqrt{5} \Delta t^{\frac{9}{2}} L G^{2} L^{2} G}{1260} - \frac{\sqrt{5} \Delta t^{\frac{9}{2}} L G^{3} L^{2}}{840} + \frac{\sqrt{5} \Delta t^{\frac{9}{2}} L^{2} G L G^{2}}{1260} && \\
    & \phantom{M_{10} =} + \frac{\sqrt{5} \Delta t^{\frac{9}{2}} L^{2} G^{2} L G}{630} + \frac{\sqrt{5} \Delta t^{\frac{9}{2}} L^{2} G^{3} L}{420} - \frac{\sqrt{5} \Delta t^{\frac{7}{2}} G L G L^{2}}{180} + \frac{\sqrt{5} \Delta t^{\frac{7}{2}} G L^{2} G L}{360} - \frac{\sqrt{5} \Delta t^{\frac{7}{2}} G L^{3} G}{360} && \\
    & \phantom{M_{10} =} - \frac{\sqrt{5} \Delta t^{\frac{7}{2}} G^{2} L^{3}}{180} + \frac{\sqrt{5} \Delta t^{\frac{7}{2}} L G L G L}{360} - \frac{\sqrt{5} \Delta t^{\frac{7}{2}} L G L^{2} G}{360} - \frac{\sqrt{5} \Delta t^{\frac{7}{2}} L G^{2} L^{2}}{180} + \frac{\sqrt{5} \Delta t^{\frac{7}{2}} L^{2} G L G}{180} && \\
    & \phantom{M_{10} =} + \frac{\sqrt{5} \Delta t^{\frac{7}{2}} L^{2} G^{2} L}{90} - \frac{\sqrt{5} \Delta t^{\frac{5}{2}} G L^{3}}{60} - \frac{\sqrt{5} \Delta t^{\frac{5}{2}} L G L^{2}}{60} + \frac{\sqrt{5} \Delta t^{\frac{5}{2}} L^{2} G L}{30} && \\
\end{flalign*}
\begin{flalign*}
    & M_{11} = \frac{\sqrt{15} \Delta t^{\frac{11}{2}} G L G^{2} L G L}{20160} + \frac{\sqrt{15} \Delta t^{\frac{11}{2}} G L G^{2} L^{2} G}{20160} + \frac{\sqrt{15} \Delta t^{\frac{11}{2}} G L G^{3} L^{2}}{10080} - \frac{\sqrt{15} \Delta t^{\frac{11}{2}} G L^{2} G L G^{2}}{20160} - \frac{\sqrt{15} \Delta t^{\frac{11}{2}} G L^{2} G^{2} L G}{20160} && \\
    & \phantom{M_{11} =} - \frac{\sqrt{15} \Delta t^{\frac{11}{2}} G L^{2} G^{3} L}{20160} - \frac{\sqrt{15} \Delta t^{\frac{11}{2}} G L^{3} G^{3}}{20160} - \frac{\sqrt{15} \Delta t^{\frac{11}{2}} G^{2} L G L G L}{20160} - \frac{\sqrt{15} \Delta t^{\frac{11}{2}} G^{2} L G L^{2} G}{20160} - \frac{\sqrt{15} \Delta t^{\frac{11}{2}} G^{2} L^{2} G L G}{10080} && \\
    & \phantom{M_{11} =} - \frac{\sqrt{15} \Delta t^{\frac{11}{2}} G^{2} L^{2} G^{2} L}{10080} - \frac{\sqrt{15} \Delta t^{\frac{11}{2}} G^{2} L^{3} G^{2}}{10080} - \frac{\sqrt{15} \Delta t^{\frac{11}{2}} G^{3} L G L^{2}}{10080} - \frac{\sqrt{15} \Delta t^{\frac{11}{2}} G^{3} L^{2} G L}{6720} - \frac{\sqrt{15} \Delta t^{\frac{11}{2}} G^{3} L^{3} G}{6720} && \\
    & \phantom{M_{11} =} - \frac{\sqrt{15} \Delta t^{\frac{11}{2}} G^{4} L^{3}}{5040} + \frac{\sqrt{15} \Delta t^{\frac{11}{2}} L G L G L G^{2}}{20160} + \frac{\sqrt{15} \Delta t^{\frac{11}{2}} L G L G^{2} L G}{20160} + \frac{\sqrt{15} \Delta t^{\frac{11}{2}} L G L G^{3} L}{20160} + \frac{\sqrt{15} \Delta t^{\frac{11}{2}} L G L^{2} G^{3}}{20160} && \\
    & \phantom{M_{11} =} + \frac{\sqrt{15} \Delta t^{\frac{11}{2}} L G^{2} L G L G}{10080} + \frac{\sqrt{15} \Delta t^{\frac{11}{2}} L G^{2} L G^{2} L}{10080} + \frac{\sqrt{15} \Delta t^{\frac{11}{2}} L G^{2} L^{2} G^{2}}{10080} + \frac{\sqrt{15} \Delta t^{\frac{11}{2}} L G^{3} L G L}{6720} + \frac{\sqrt{15} \Delta t^{\frac{11}{2}} L G^{3} L^{2} G}{6720} && \\
    & \phantom{M_{11} =} + \frac{\sqrt{15} \Delta t^{\frac{11}{2}} L G^{4} L^{2}}{5040} + \frac{\sqrt{15} \Delta t^{\frac{9}{2}} G L G^{2} L^{2}}{2520} - \frac{\sqrt{15} \Delta t^{\frac{9}{2}} G L^{2} G L G}{2520} - \frac{\sqrt{15} \Delta t^{\frac{9}{2}} G L^{2} G^{2} L}{2520} - \frac{\sqrt{15} \Delta t^{\frac{9}{2}} G L^{3} G^{2}}{2520} && \\
    & \phantom{M_{11} =} - \frac{\sqrt{15} \Delta t^{\frac{9}{2}} G^{2} L G L^{2}}{2520} - \frac{\sqrt{15} \Delta t^{\frac{9}{2}} G^{2} L^{2} G L}{1260} - \frac{\sqrt{15} \Delta t^{\frac{9}{2}} G^{2} L^{3} G}{1260} - \frac{\sqrt{15} \Delta t^{\frac{9}{2}} G^{3} L^{3}}{840} + \frac{\sqrt{15} \Delta t^{\frac{9}{2}} L G L G L G}{2520} && \\
    & \phantom{M_{11} =} + \frac{\sqrt{15} \Delta t^{\frac{9}{2}} L G L G^{2} L}{2520} + \frac{\sqrt{15} \Delta t^{\frac{9}{2}} L G L^{2} G^{2}}{2520} + \frac{\sqrt{15} \Delta t^{\frac{9}{2}} L G^{2} L G L}{1260} + \frac{\sqrt{15} \Delta t^{\frac{9}{2}} L G^{2} L^{2} G}{1260} + \frac{\sqrt{15} \Delta t^{\frac{9}{2}} L G^{3} L^{2}}{840} && \\
    & \phantom{M_{11} =} - \frac{\sqrt{15} \Delta t^{\frac{7}{2}} G L^{2} G L}{360} - \frac{\sqrt{15} \Delta t^{\frac{7}{2}} G L^{3} G}{360} - \frac{\sqrt{15} \Delta t^{\frac{7}{2}} G^{2} L^{3}}{180} + \frac{\sqrt{15} \Delta t^{\frac{7}{2}} L G L G L}{360} + \frac{\sqrt{15} \Delta t^{\frac{7}{2}} L G L^{2} G}{360} && \\
    & \phantom{M_{11} =} + \frac{\sqrt{15} \Delta t^{\frac{7}{2}} L G^{2} L^{2}}{180} - \frac{\sqrt{15} \Delta t^{\frac{5}{2}} G L^{3}}{60} + \frac{\sqrt{15} \Delta t^{\frac{5}{2}} L G L^{2}}{60} && \\
\end{flalign*}
\begin{flalign*}
    & M_{12} = \frac{\sqrt{30} \Delta t^{\frac{11}{2}} G L G L G L^{3}}{20160} + \frac{\sqrt{30} \Delta t^{\frac{11}{2}} G L G L^{2} G L^{2}}{20160} + \frac{\sqrt{30} \Delta t^{\frac{11}{2}} G L G L^{3} G L}{20160} + \frac{\sqrt{30} \Delta t^{\frac{11}{2}} G L G L^{4} G}{20160} + \frac{\sqrt{30} \Delta t^{\frac{11}{2}} G L G^{2} L^{4}}{20160} && \\
    & \phantom{M_{12} =} + \frac{\sqrt{30} \Delta t^{\frac{11}{2}} G L^{2} G L G L^{2}}{20160} + \frac{\sqrt{30} \Delta t^{\frac{11}{2}} G L^{2} G L^{2} G L}{20160} + \frac{\sqrt{30} \Delta t^{\frac{11}{2}} G L^{2} G L^{3} G}{20160} + \frac{\sqrt{30} \Delta t^{\frac{11}{2}} G L^{2} G^{2} L^{3}}{20160} + \frac{\sqrt{30} \Delta t^{\frac{11}{2}} G L^{3} G L G L}{20160} && \\
    & \phantom{M_{12} =} + \frac{\sqrt{30} \Delta t^{\frac{11}{2}} G L^{3} G L^{2} G}{20160} + \frac{\sqrt{30} \Delta t^{\frac{11}{2}} G L^{3} G^{2} L^{2}}{20160} + \frac{\sqrt{30} \Delta t^{\frac{11}{2}} G L^{4} G L G}{20160} + \frac{\sqrt{30} \Delta t^{\frac{11}{2}} G L^{4} G^{2} L}{20160} + \frac{\sqrt{30} \Delta t^{\frac{11}{2}} G L^{5} G^{2}}{20160} && \\
    & \phantom{M_{12} =} + \frac{\sqrt{30} \Delta t^{\frac{11}{2}} G^{2} L G L^{4}}{20160} + \frac{\sqrt{30} \Delta t^{\frac{11}{2}} G^{2} L^{2} G L^{3}}{20160} + \frac{\sqrt{30} \Delta t^{\frac{11}{2}} G^{2} L^{3} G L^{2}}{20160} + \frac{\sqrt{30} \Delta t^{\frac{11}{2}} G^{2} L^{4} G L}{20160} + \frac{\sqrt{30} \Delta t^{\frac{11}{2}} G^{2} L^{5} G}{20160} && \\
    & \phantom{M_{12} =} + \frac{\sqrt{30} \Delta t^{\frac{11}{2}} G^{3} L^{5}}{20160} + \frac{\sqrt{30} \Delta t^{\frac{11}{2}} L G L G L G L^{2}}{20160} + \frac{\sqrt{30} \Delta t^{\frac{11}{2}} L G L G L^{2} G L}{20160} + \frac{\sqrt{30} \Delta t^{\frac{11}{2}} L G L G L^{3} G}{20160} + \frac{\sqrt{30} \Delta t^{\frac{11}{2}} L G L G^{2} L^{3}}{20160} && \\
    & \phantom{M_{12} =} + \frac{\sqrt{30} \Delta t^{\frac{11}{2}} L G L^{2} G L G L}{20160} + \frac{\sqrt{30} \Delta t^{\frac{11}{2}} L G L^{2} G L^{2} G}{20160} + \frac{\sqrt{30} \Delta t^{\frac{11}{2}} L G L^{2} G^{2} L^{2}}{20160} + \frac{\sqrt{30} \Delta t^{\frac{11}{2}} L G L^{3} G L G}{20160} + \frac{\sqrt{30} \Delta t^{\frac{11}{2}} L G L^{3} G^{2} L}{20160} && \\
    & \phantom{M_{12} =} + \frac{\sqrt{30} \Delta t^{\frac{11}{2}} L G L^{4} G^{2}}{20160} + \frac{\sqrt{30} \Delta t^{\frac{11}{2}} L G^{2} L G L^{3}}{20160} + \frac{\sqrt{30} \Delta t^{\frac{11}{2}} L G^{2} L^{2} G L^{2}}{20160} + \frac{\sqrt{30} \Delta t^{\frac{11}{2}} L G^{2} L^{3} G L}{20160} + \frac{\sqrt{30} \Delta t^{\frac{11}{2}} L G^{2} L^{4} G}{20160} && \\
    & \phantom{M_{12} =} + \frac{\sqrt{30} \Delta t^{\frac{11}{2}} L G^{3} L^{4}}{20160} + \frac{\sqrt{30} \Delta t^{\frac{11}{2}} L^{2} G L G L G L}{20160} + \frac{\sqrt{30} \Delta t^{\frac{11}{2}} L^{2} G L G L^{2} G}{20160} + \frac{\sqrt{30} \Delta t^{\frac{11}{2}} L^{2} G L G^{2} L^{2}}{20160} + \frac{\sqrt{30} \Delta t^{\frac{11}{2}} L^{2} G L^{2} G L G}{20160} && \\
    & \phantom{M_{12} =} + \frac{\sqrt{30} \Delta t^{\frac{11}{2}} L^{2} G L^{2} G^{2} L}{20160} + \frac{\sqrt{30} \Delta t^{\frac{11}{2}} L^{2} G L^{3} G^{2}}{20160} + \frac{\sqrt{30} \Delta t^{\frac{11}{2}} L^{2} G^{2} L G L^{2}}{20160} + \frac{\sqrt{30} \Delta t^{\frac{11}{2}} L^{2} G^{2} L^{2} G L}{20160} + \frac{\sqrt{30} \Delta t^{\frac{11}{2}} L^{2} G^{2} L^{3} G}{20160} && \\
    & \phantom{M_{12} =} + \frac{\sqrt{30} \Delta t^{\frac{11}{2}} L^{2} G^{3} L^{3}}{20160} + \frac{\sqrt{30} \Delta t^{\frac{11}{2}} L^{3} G L G L G}{20160} + \frac{\sqrt{30} \Delta t^{\frac{11}{2}} L^{3} G L G^{2} L}{20160} + \frac{\sqrt{30} \Delta t^{\frac{11}{2}} L^{3} G L^{2} G^{2}}{20160} + \frac{\sqrt{30} \Delta t^{\frac{11}{2}} L^{3} G^{2} L G L}{20160} && \\
    & \phantom{M_{12} =} + \frac{\sqrt{30} \Delta t^{\frac{11}{2}} L^{3} G^{2} L^{2} G}{20160} + \frac{\sqrt{30} \Delta t^{\frac{11}{2}} L^{3} G^{3} L^{2}}{20160} + \frac{\sqrt{30} \Delta t^{\frac{11}{2}} L^{4} G L G^{2}}{20160} + \frac{\sqrt{30} \Delta t^{\frac{11}{2}} L^{4} G^{2} L G}{20160} + \frac{\sqrt{30} \Delta t^{\frac{11}{2}} L^{4} G^{3} L}{20160} && \\
    & \phantom{M_{12} =} + \frac{\sqrt{30} \Delta t^{\frac{11}{2}} L^{5} G^{3}}{20160} + \frac{\sqrt{30} \Delta t^{\frac{9}{2}} G L G L^{4}}{2520} + \frac{\sqrt{30} \Delta t^{\frac{9}{2}} G L^{2} G L^{3}}{2520} + \frac{\sqrt{30} \Delta t^{\frac{9}{2}} G L^{3} G L^{2}}{2520} + \frac{\sqrt{30} \Delta t^{\frac{9}{2}} G L^{4} G L}{2520} && \\
    & \phantom{M_{12} =} + \frac{\sqrt{30} \Delta t^{\frac{9}{2}} G L^{5} G}{2520} + \frac{\sqrt{30} \Delta t^{\frac{9}{2}} G^{2} L^{5}}{2520} + \frac{\sqrt{30} \Delta t^{\frac{9}{2}} L G L G L^{3}}{2520} + \frac{\sqrt{30} \Delta t^{\frac{9}{2}} L G L^{2} G L^{2}}{2520} + \frac{\sqrt{30} \Delta t^{\frac{9}{2}} L G L^{3} G L}{2520} && \\
    & \phantom{M_{12} =} + \frac{\sqrt{30} \Delta t^{\frac{9}{2}} L G L^{4} G}{2520} + \frac{\sqrt{30} \Delta t^{\frac{9}{2}} L G^{2} L^{4}}{2520} + \frac{\sqrt{30} \Delta t^{\frac{9}{2}} L^{2} G L G L^{2}}{2520} + \frac{\sqrt{30} \Delta t^{\frac{9}{2}} L^{2} G L^{2} G L}{2520} + \frac{\sqrt{30} \Delta t^{\frac{9}{2}} L^{2} G L^{3} G}{2520} && \\
    & \phantom{M_{12} =} + \frac{\sqrt{30} \Delta t^{\frac{9}{2}} L^{2} G^{2} L^{3}}{2520} + \frac{\sqrt{30} \Delta t^{\frac{9}{2}} L^{3} G L G L}{2520} + \frac{\sqrt{30} \Delta t^{\frac{9}{2}} L^{3} G L^{2} G}{2520} + \frac{\sqrt{30} \Delta t^{\frac{9}{2}} L^{3} G^{2} L^{2}}{2520} + \frac{\sqrt{30} \Delta t^{\frac{9}{2}} L^{4} G L G}{2520} && \\
    & \phantom{M_{12} =} + \frac{\sqrt{30} \Delta t^{\frac{9}{2}} L^{4} G^{2} L}{2520} + \frac{\sqrt{30} \Delta t^{\frac{9}{2}} L^{5} G^{2}}{2520} + \frac{\sqrt{30} \Delta t^{\frac{7}{2}} G L^{5}}{360} + \frac{\sqrt{30} \Delta t^{\frac{7}{2}} L G L^{4}}{360} + \frac{\sqrt{30} \Delta t^{\frac{7}{2}} L^{2} G L^{3}}{360} && \\
    & \phantom{M_{12} =} + \frac{\sqrt{30} \Delta t^{\frac{7}{2}} L^{3} G L^{2}}{360} + \frac{\sqrt{30} \Delta t^{\frac{7}{2}} L^{4} G L}{360} + \frac{\sqrt{30} \Delta t^{\frac{7}{2}} L^{5} G}{360} + \frac{\sqrt{30} \Delta t^{\frac{5}{2}} L^{5}}{60} && \\
\end{flalign*}
\begin{flalign*}
    & M_{13} = - \frac{\sqrt{6} \Delta t^{5} G L G L G^{2}}{2520} - \frac{\sqrt{6} \Delta t^{5} G L G^{2} L G}{3360} + \frac{\sqrt{6} \Delta t^{5} G L G^{3} L}{3360} - \frac{\sqrt{6} \Delta t^{5} G^{2} L G L G}{3360} + \frac{\sqrt{6} \Delta t^{5} G^{2} L G^{2} L}{3360} && \\
    & \phantom{M_{13} =} - \frac{\sqrt{6} \Delta t^{5} G^{2} L^{2} G^{2}}{2520} + \frac{\sqrt{6} \Delta t^{5} G^{3} L G L}{3360} - \frac{\sqrt{6} \Delta t^{5} G^{3} L^{2} G}{3360} + \frac{\sqrt{6} \Delta t^{5} G^{4} L^{2}}{3360} - \frac{\sqrt{6} \Delta t^{5} L G^{2} L G^{2}}{2520} && \\
    & \phantom{M_{13} =} - \frac{\sqrt{6} \Delta t^{5} L G^{3} L G}{3360} + \frac{\sqrt{6} \Delta t^{5} L G^{4} L}{3360} + \frac{\sqrt{6} \Delta t^{5} L^{2} G^{4}}{1120} - \frac{\sqrt{6} \Delta t^{4} G L G L G}{504} + \frac{\sqrt{6} \Delta t^{4} G L G^{2} L}{840} && \\
    & \phantom{M_{13} =} - \frac{\sqrt{6} \Delta t^{4} G L^{2} G^{2}}{840} + \frac{\sqrt{6} \Delta t^{4} G^{2} L G L}{840} - \frac{\sqrt{6} \Delta t^{4} G^{2} L^{2} G}{504} + \frac{\sqrt{6} \Delta t^{4} G^{3} L^{2}}{840} - \frac{\sqrt{6} \Delta t^{4} L G L G^{2}}{840} && \\
    & \phantom{M_{13} =} - \frac{\sqrt{6} \Delta t^{4} L G^{2} L G}{504} + \frac{\sqrt{6} \Delta t^{4} L G^{3} L}{840} + \frac{\sqrt{6} \Delta t^{4} L^{2} G^{3}}{280} + \frac{\sqrt{6} \Delta t^{3} G L G L}{360} - \frac{\sqrt{6} \Delta t^{3} G L^{2} G}{120} && \\
    & \phantom{M_{13} =} + \frac{\sqrt{6} \Delta t^{3} G^{2} L^{2}}{360} - \frac{\sqrt{6} \Delta t^{3} L G L G}{120} + \frac{\sqrt{6} \Delta t^{3} L G^{2} L}{360} + \frac{\sqrt{6} \Delta t^{3} L^{2} G^{2}}{120} && \\
\end{flalign*}
\begin{flalign*}
    & M_{14} = \frac{\sqrt{2} \Delta t^{5} G L G^{2} L G}{6720} - \frac{\sqrt{2} \Delta t^{5} G L G^{3} L}{2240} - \frac{\sqrt{2} \Delta t^{5} G L^{2} G^{3}}{1120} - \frac{\sqrt{2} \Delta t^{5} G^{2} L G L G}{6720} - \frac{\sqrt{2} \Delta t^{5} G^{2} L^{2} G^{2}}{960} && \\
    & \phantom{M_{14} =} + \frac{\sqrt{2} \Delta t^{5} G^{3} L G L}{2240} - \frac{\sqrt{2} \Delta t^{5} G^{3} L^{2} G}{2240} + \frac{\sqrt{2} \Delta t^{5} G^{4} L^{2}}{1120} + \frac{\sqrt{2} \Delta t^{5} L G L G^{3}}{1120} + \frac{\sqrt{2} \Delta t^{5} L G^{2} L G^{2}}{960} && \\
    & \phantom{M_{14} =} + \frac{\sqrt{2} \Delta t^{5} L G^{3} L G}{2240} - \frac{\sqrt{2} \Delta t^{5} L G^{4} L}{1120} - \frac{\sqrt{2} \Delta t^{4} G L G^{2} L}{840} - \frac{\sqrt{2} \Delta t^{4} G L^{2} G^{2}}{210} + \frac{\sqrt{2} \Delta t^{4} G^{2} L G L}{840} && \\
    & \phantom{M_{14} =} - \frac{\sqrt{2} \Delta t^{4} G^{2} L^{2} G}{280} + \frac{\sqrt{2} \Delta t^{4} G^{3} L^{2}}{280} + \frac{\sqrt{2} \Delta t^{4} L G L G^{2}}{210} + \frac{\sqrt{2} \Delta t^{4} L G^{2} L G}{280} - \frac{\sqrt{2} \Delta t^{4} L G^{3} L}{280} && \\
    & \phantom{M_{14} =} - \frac{\sqrt{2} \Delta t^{3} G L^{2} G}{60} + \frac{\sqrt{2} \Delta t^{3} G^{2} L^{2}}{120} + \frac{\sqrt{2} \Delta t^{3} L G L G}{60} - \frac{\sqrt{2} \Delta t^{3} L G^{2} L}{120} && \\
\end{flalign*}
\begin{flalign*}
    & M_{15} = - \frac{\sqrt{30} \Delta t^{5} G L G L G^{2}}{10080} - \frac{\sqrt{30} \Delta t^{5} G L G^{2} L G}{6720} - \frac{\sqrt{30} \Delta t^{5} G L G^{3} L}{6720} - \frac{\sqrt{30} \Delta t^{5} G^{2} L G L G}{6720} - \frac{\sqrt{30} \Delta t^{5} G^{2} L G^{2} L}{3360} && \\
    & \phantom{M_{15} =} + \frac{\sqrt{30} \Delta t^{5} G^{2} L^{2} G^{2}}{20160} - \frac{\sqrt{30} \Delta t^{5} G^{3} L G L}{6720} + \frac{\sqrt{30} \Delta t^{5} G^{3} L^{2} G}{6720} + \frac{\sqrt{30} \Delta t^{5} G^{4} L^{2}}{3360} + \frac{\sqrt{30} \Delta t^{5} L G^{2} L G^{2}}{20160} && \\
    & \phantom{M_{15} =} + \frac{\sqrt{30} \Delta t^{5} L G^{3} L G}{6720} + \frac{\sqrt{30} \Delta t^{5} L G^{4} L}{3360} - \frac{\sqrt{30} \Delta t^{4} G L G L G}{1260} - \frac{\sqrt{30} \Delta t^{4} G L G^{2} L}{840} - \frac{\sqrt{30} \Delta t^{4} G^{2} L G L}{840} && \\
    & \phantom{M_{15} =} + \frac{\sqrt{30} \Delta t^{4} G^{2} L^{2} G}{2520} + \frac{\sqrt{30} \Delta t^{4} G^{3} L^{2}}{840} + \frac{\sqrt{30} \Delta t^{4} L G^{2} L G}{2520} + \frac{\sqrt{30} \Delta t^{4} L G^{3} L}{840} - \frac{\sqrt{30} \Delta t^{3} G L G L}{180} && \\
    & \phantom{M_{15} =} + \frac{\sqrt{30} \Delta t^{3} G^{2} L^{2}}{360} + \frac{\sqrt{30} \Delta t^{3} L G^{2} L}{360} && \\
\end{flalign*}
\begin{flalign*}
    & M_{16} = - \frac{\Delta t^{5} G L G L G L^{2}}{2240} - \frac{\Delta t^{5} G L G L^{2} G L}{2240} + \frac{\Delta t^{5} G L G L^{3} G}{3360} - \frac{\Delta t^{5} G L G^{2} L^{3}}{2240} - \frac{\Delta t^{5} G L^{2} G L G L}{2240} && \\
    & \phantom{M_{16} =} + \frac{\Delta t^{5} G L^{2} G L^{2} G}{3360} - \frac{\Delta t^{5} G L^{2} G^{2} L^{2}}{2240} + \frac{\Delta t^{5} G L^{3} G L G}{3360} - \frac{\Delta t^{5} G L^{3} G^{2} L}{2240} + \frac{\Delta t^{5} G L^{4} G^{2}}{960} && \\
    & \phantom{M_{16} =} - \frac{\Delta t^{5} G^{2} L G L^{3}}{2240} - \frac{\Delta t^{5} G^{2} L^{2} G L^{2}}{2240} - \frac{\Delta t^{5} G^{2} L^{3} G L}{2240} + \frac{\Delta t^{5} G^{2} L^{4} G}{3360} - \frac{\Delta t^{5} G^{3} L^{4}}{2240} && \\
    & \phantom{M_{16} =} - \frac{\Delta t^{5} L G L G L G L}{2240} + \frac{\Delta t^{5} L G L G L^{2} G}{3360} - \frac{\Delta t^{5} L G L G^{2} L^{2}}{2240} + \frac{\Delta t^{5} L G L^{2} G L G}{3360} - \frac{\Delta t^{5} L G L^{2} G^{2} L}{2240} && \\
    & \phantom{M_{16} =} + \frac{\Delta t^{5} L G L^{3} G^{2}}{960} - \frac{\Delta t^{5} L G^{2} L G L^{2}}{2240} - \frac{\Delta t^{5} L G^{2} L^{2} G L}{2240} + \frac{\Delta t^{5} L G^{2} L^{3} G}{3360} - \frac{\Delta t^{5} L G^{3} L^{3}}{2240} && \\
    & \phantom{M_{16} =} + \frac{\Delta t^{5} L^{2} G L G L G}{3360} - \frac{\Delta t^{5} L^{2} G L G^{2} L}{2240} + \frac{\Delta t^{5} L^{2} G L^{2} G^{2}}{960} - \frac{\Delta t^{5} L^{2} G^{2} L G L}{2240} + \frac{\Delta t^{5} L^{2} G^{2} L^{2} G}{3360} && \\
    & \phantom{M_{16} =} - \frac{\Delta t^{5} L^{2} G^{3} L^{2}}{2240} + \frac{\Delta t^{5} L^{3} G L G^{2}}{960} + \frac{\Delta t^{5} L^{3} G^{2} L G}{3360} - \frac{\Delta t^{5} L^{3} G^{3} L}{2240} + \frac{\Delta t^{5} L^{4} G^{3}}{560} && \\
    & \phantom{M_{16} =} - \frac{\Delta t^{4} G L G L^{3}}{420} - \frac{\Delta t^{4} G L^{2} G L^{2}}{420} - \frac{\Delta t^{4} G L^{3} G L}{420} + \frac{\Delta t^{4} G L^{4} G}{280} - \frac{\Delta t^{4} G^{2} L^{4}}{420} && \\
    & \phantom{M_{16} =} - \frac{\Delta t^{4} L G L G L^{2}}{420} - \frac{\Delta t^{4} L G L^{2} G L}{420} + \frac{\Delta t^{4} L G L^{3} G}{280} - \frac{\Delta t^{4} L G^{2} L^{3}}{420} - \frac{\Delta t^{4} L^{2} G L G L}{420} && \\
    & \phantom{M_{16} =} + \frac{\Delta t^{4} L^{2} G L^{2} G}{280} - \frac{\Delta t^{4} L^{2} G^{2} L^{2}}{420} + \frac{\Delta t^{4} L^{3} G L G}{280} - \frac{\Delta t^{4} L^{3} G^{2} L}{420} + \frac{\Delta t^{4} L^{4} G^{2}}{105} && \\
    & \phantom{M_{16} =} - \frac{\Delta t^{3} G L^{4}}{120} - \frac{\Delta t^{3} L G L^{3}}{120} - \frac{\Delta t^{3} L^{2} G L^{2}}{120} - \frac{\Delta t^{3} L^{3} G L}{120} + \frac{\Delta t^{3} L^{4} G}{30} && \\
\end{flalign*}
\begin{flalign*}
    & M_{17} = - \frac{\sqrt{15} \Delta t^{5} G L G L G L^{2}}{6720} + \frac{\sqrt{15} \Delta t^{5} G L G L^{2} G L}{20160} - \frac{\sqrt{15} \Delta t^{5} G L G L^{3} G}{10080} - \frac{\sqrt{15} \Delta t^{5} G L G^{2} L^{3}}{6720} + \frac{\sqrt{15} \Delta t^{5} G L^{2} G L G L}{20160} && \\
    & \phantom{M_{17} =} - \frac{\sqrt{15} \Delta t^{5} G L^{2} G L^{2} G}{10080} - \frac{\sqrt{15} \Delta t^{5} G L^{2} G^{2} L^{2}}{6720} + \frac{\sqrt{15} \Delta t^{5} G L^{3} G L G}{10080} + \frac{\sqrt{15} \Delta t^{5} G L^{3} G^{2} L}{4032} - \frac{\sqrt{15} \Delta t^{5} G L^{4} G^{2}}{20160} && \\
    & \phantom{M_{17} =} - \frac{\sqrt{15} \Delta t^{5} G^{2} L G L^{3}}{6720} - \frac{\sqrt{15} \Delta t^{5} G^{2} L^{2} G L^{2}}{6720} + \frac{\sqrt{15} \Delta t^{5} G^{2} L^{3} G L}{20160} - \frac{\sqrt{15} \Delta t^{5} G^{2} L^{4} G}{10080} - \frac{\sqrt{15} \Delta t^{5} G^{3} L^{4}}{6720} && \\
    & \phantom{M_{17} =} + \frac{\sqrt{15} \Delta t^{5} L G L G L G L}{20160} - \frac{\sqrt{15} \Delta t^{5} L G L G L^{2} G}{10080} - \frac{\sqrt{15} \Delta t^{5} L G L G^{2} L^{2}}{6720} + \frac{\sqrt{15} \Delta t^{5} L G L^{2} G L G}{10080} + \frac{\sqrt{15} \Delta t^{5} L G L^{2} G^{2} L}{4032} && \\
    & \phantom{M_{17} =} - \frac{\sqrt{15} \Delta t^{5} L G L^{3} G^{2}}{20160} - \frac{\sqrt{15} \Delta t^{5} L G^{2} L G L^{2}}{6720} + \frac{\sqrt{15} \Delta t^{5} L G^{2} L^{2} G L}{20160} - \frac{\sqrt{15} \Delta t^{5} L G^{2} L^{3} G}{10080} - \frac{\sqrt{15} \Delta t^{5} L G^{3} L^{3}}{6720} && \\
    & \phantom{M_{17} =} + \frac{\sqrt{15} \Delta t^{5} L^{2} G L G L G}{10080} + \frac{\sqrt{15} \Delta t^{5} L^{2} G L G^{2} L}{4032} - \frac{\sqrt{15} \Delta t^{5} L^{2} G L^{2} G^{2}}{20160} + \frac{\sqrt{15} \Delta t^{5} L^{2} G^{2} L G L}{20160} - \frac{\sqrt{15} \Delta t^{5} L^{2} G^{2} L^{2} G}{10080} && \\
    & \phantom{M_{17} =} - \frac{\sqrt{15} \Delta t^{5} L^{2} G^{3} L^{2}}{6720} + \frac{\sqrt{15} \Delta t^{5} L^{3} G L G^{2}}{6720} + \frac{\sqrt{15} \Delta t^{5} L^{3} G^{2} L G}{3360} + \frac{\sqrt{15} \Delta t^{5} L^{3} G^{3} L}{2240} - \frac{\sqrt{15} \Delta t^{4} G L G L^{3}}{1260} && \\
    & \phantom{M_{17} =} - \frac{\sqrt{15} \Delta t^{4} G L^{2} G L^{2}}{1260} + \frac{\sqrt{15} \Delta t^{4} G L^{3} G L}{1260} - \frac{\sqrt{15} \Delta t^{4} G L^{4} G}{2520} - \frac{\sqrt{15} \Delta t^{4} G^{2} L^{4}}{1260} - \frac{\sqrt{15} \Delta t^{4} L G L G L^{2}}{1260} && \\
    & \phantom{M_{17} =} + \frac{\sqrt{15} \Delta t^{4} L G L^{2} G L}{1260} - \frac{\sqrt{15} \Delta t^{4} L G L^{3} G}{2520} - \frac{\sqrt{15} \Delta t^{4} L G^{2} L^{3}}{1260} + \frac{\sqrt{15} \Delta t^{4} L^{2} G L G L}{1260} - \frac{\sqrt{15} \Delta t^{4} L^{2} G L^{2} G}{2520} && \\
    & \phantom{M_{17} =} - \frac{\sqrt{15} \Delta t^{4} L^{2} G^{2} L^{2}}{1260} + \frac{\sqrt{15} \Delta t^{4} L^{3} G L G}{840} + \frac{\sqrt{15} \Delta t^{4} L^{3} G^{2} L}{420} - \frac{\sqrt{15} \Delta t^{3} G L^{4}}{360} - \frac{\sqrt{15} \Delta t^{3} L G L^{3}}{360} && \\
    & \phantom{M_{17} =} - \frac{\sqrt{15} \Delta t^{3} L^{2} G L^{2}}{360} + \frac{\sqrt{15} \Delta t^{3} L^{3} G L}{120} && \\
\end{flalign*}
\begin{flalign*}
    & M_{18} = - \frac{\sqrt{30} \Delta t^{5} G L G L^{2} G L}{10080} - \frac{\sqrt{30} \Delta t^{5} G L G L^{3} G}{10080} - \frac{\sqrt{30} \Delta t^{5} G L G^{2} L^{3}}{6720} + \frac{\sqrt{30} \Delta t^{5} G L^{2} G L G L}{20160} + \frac{\sqrt{30} \Delta t^{5} G L^{2} G L^{2} G}{20160} && \\
    & \phantom{M_{18} =} + \frac{\sqrt{30} \Delta t^{5} G L^{2} G^{2} L^{2}}{6720} - \frac{\sqrt{30} \Delta t^{5} G L^{3} G L G}{20160} - \frac{\sqrt{30} \Delta t^{5} G L^{3} G^{2} L}{20160} - \frac{\sqrt{30} \Delta t^{5} G L^{4} G^{2}}{20160} - \frac{\sqrt{30} \Delta t^{5} G^{2} L G L^{3}}{6720} && \\
    & \phantom{M_{18} =} - \frac{\sqrt{30} \Delta t^{5} G^{2} L^{3} G L}{10080} - \frac{\sqrt{30} \Delta t^{5} G^{2} L^{4} G}{10080} - \frac{\sqrt{30} \Delta t^{5} G^{3} L^{4}}{6720} + \frac{\sqrt{30} \Delta t^{5} L G L G L G L}{20160} + \frac{\sqrt{30} \Delta t^{5} L G L G L^{2} G}{20160} && \\
    & \phantom{M_{18} =} + \frac{\sqrt{30} \Delta t^{5} L G L G^{2} L^{2}}{6720} - \frac{\sqrt{30} \Delta t^{5} L G L^{2} G L G}{20160} - \frac{\sqrt{30} \Delta t^{5} L G L^{2} G^{2} L}{20160} - \frac{\sqrt{30} \Delta t^{5} L G L^{3} G^{2}}{20160} - \frac{\sqrt{30} \Delta t^{5} L G^{2} L^{2} G L}{10080} && \\
    & \phantom{M_{18} =} - \frac{\sqrt{30} \Delta t^{5} L G^{2} L^{3} G}{10080} - \frac{\sqrt{30} \Delta t^{5} L G^{3} L^{3}}{6720} + \frac{\sqrt{30} \Delta t^{5} L^{2} G L G L G}{10080} + \frac{\sqrt{30} \Delta t^{5} L^{2} G L G^{2} L}{10080} + \frac{\sqrt{30} \Delta t^{5} L^{2} G L^{2} G^{2}}{10080} && \\
    & \phantom{M_{18} =} + \frac{\sqrt{30} \Delta t^{5} L^{2} G^{2} L G L}{5040} + \frac{\sqrt{30} \Delta t^{5} L^{2} G^{2} L^{2} G}{5040} + \frac{\sqrt{30} \Delta t^{5} L^{2} G^{3} L^{2}}{3360} - \frac{\sqrt{30} \Delta t^{4} G L G L^{3}}{1260} + \frac{\sqrt{30} \Delta t^{4} G L^{2} G L^{2}}{2520} && \\
    & \phantom{M_{18} =} - \frac{\sqrt{30} \Delta t^{4} G L^{3} G L}{2520} - \frac{\sqrt{30} \Delta t^{4} G L^{4} G}{2520} - \frac{\sqrt{30} \Delta t^{4} G^{2} L^{4}}{1260} + \frac{\sqrt{30} \Delta t^{4} L G L G L^{2}}{2520} - \frac{\sqrt{30} \Delta t^{4} L G L^{2} G L}{2520} && \\
    & \phantom{M_{18} =} - \frac{\sqrt{30} \Delta t^{4} L G L^{3} G}{2520} - \frac{\sqrt{30} \Delta t^{4} L G^{2} L^{3}}{1260} + \frac{\sqrt{30} \Delta t^{4} L^{2} G L G L}{1260} + \frac{\sqrt{30} \Delta t^{4} L^{2} G L^{2} G}{1260} + \frac{\sqrt{30} \Delta t^{4} L^{2} G^{2} L^{2}}{630} && \\
    & \phantom{M_{18} =} - \frac{\sqrt{30} \Delta t^{3} G L^{4}}{360} - \frac{\sqrt{30} \Delta t^{3} L G L^{3}}{360} + \frac{\sqrt{30} \Delta t^{3} L^{2} G L^{2}}{180} && \\
\end{flalign*}
\begin{flalign*}
    & M_{19} = \frac{\sqrt{10} \Delta t^{5} G L G^{2} L^{3}}{6720} - \frac{\sqrt{10} \Delta t^{5} G L^{2} G L G L}{6720} - \frac{\sqrt{10} \Delta t^{5} G L^{2} G L^{2} G}{6720} - \frac{\sqrt{10} \Delta t^{5} G L^{2} G^{2} L^{2}}{6720} - \frac{\sqrt{10} \Delta t^{5} G L^{3} G L G}{6720} && \\
    & \phantom{M_{19} =} - \frac{\sqrt{10} \Delta t^{5} G L^{3} G^{2} L}{6720} - \frac{\sqrt{10} \Delta t^{5} G L^{4} G^{2}}{6720} - \frac{\sqrt{10} \Delta t^{5} G^{2} L G L^{3}}{6720} - \frac{\sqrt{10} \Delta t^{5} G^{2} L^{2} G L^{2}}{3360} - \frac{\sqrt{10} \Delta t^{5} G^{2} L^{3} G L}{3360} && \\
    & \phantom{M_{19} =} - \frac{\sqrt{10} \Delta t^{5} G^{2} L^{4} G}{3360} - \frac{\sqrt{10} \Delta t^{5} G^{3} L^{4}}{2240} + \frac{\sqrt{10} \Delta t^{5} L G L G L G L}{6720} + \frac{\sqrt{10} \Delta t^{5} L G L G L^{2} G}{6720} + \frac{\sqrt{10} \Delta t^{5} L G L G^{2} L^{2}}{6720} && \\
    & \phantom{M_{19} =} + \frac{\sqrt{10} \Delta t^{5} L G L^{2} G L G}{6720} + \frac{\sqrt{10} \Delta t^{5} L G L^{2} G^{2} L}{6720} + \frac{\sqrt{10} \Delta t^{5} L G L^{3} G^{2}}{6720} + \frac{\sqrt{10} \Delta t^{5} L G^{2} L G L^{2}}{3360} + \frac{\sqrt{10} \Delta t^{5} L G^{2} L^{2} G L}{3360} && \\
    & \phantom{M_{19} =} + \frac{\sqrt{10} \Delta t^{5} L G^{2} L^{3} G}{3360} + \frac{\sqrt{10} \Delta t^{5} L G^{3} L^{3}}{2240} - \frac{\sqrt{10} \Delta t^{4} G L^{2} G L^{2}}{840} - \frac{\sqrt{10} \Delta t^{4} G L^{3} G L}{840} - \frac{\sqrt{10} \Delta t^{4} G L^{4} G}{840} && \\
    & \phantom{M_{19} =} - \frac{\sqrt{10} \Delta t^{4} G^{2} L^{4}}{420} + \frac{\sqrt{10} \Delta t^{4} L G L G L^{2}}{840} + \frac{\sqrt{10} \Delta t^{4} L G L^{2} G L}{840} + \frac{\sqrt{10} \Delta t^{4} L G L^{3} G}{840} + \frac{\sqrt{10} \Delta t^{4} L G^{2} L^{3}}{420} && \\
    & \phantom{M_{19} =} - \frac{\sqrt{10} \Delta t^{3} G L^{4}}{120} + \frac{\sqrt{10} \Delta t^{3} L G L^{3}}{120} && \\
\end{flalign*}
\begin{flalign*}
    & M_{20} = \frac{\sqrt{5} \Delta t^{5} G L G L^{5}}{3360} + \frac{\sqrt{5} \Delta t^{5} G L^{2} G L^{4}}{3360} + \frac{\sqrt{5} \Delta t^{5} G L^{3} G L^{3}}{3360} + \frac{\sqrt{5} \Delta t^{5} G L^{4} G L^{2}}{3360} + \frac{\sqrt{5} \Delta t^{5} G L^{5} G L}{3360} && \\
    & \phantom{M_{20} =} + \frac{\sqrt{5} \Delta t^{5} G L^{6} G}{3360} + \frac{\sqrt{5} \Delta t^{5} G^{2} L^{6}}{3360} + \frac{\sqrt{5} \Delta t^{5} L G L G L^{4}}{3360} + \frac{\sqrt{5} \Delta t^{5} L G L^{2} G L^{3}}{3360} + \frac{\sqrt{5} \Delta t^{5} L G L^{3} G L^{2}}{3360} && \\
    & \phantom{M_{20} =} + \frac{\sqrt{5} \Delta t^{5} L G L^{4} G L}{3360} + \frac{\sqrt{5} \Delta t^{5} L G L^{5} G}{3360} + \frac{\sqrt{5} \Delta t^{5} L G^{2} L^{5}}{3360} + \frac{\sqrt{5} \Delta t^{5} L^{2} G L G L^{3}}{3360} + \frac{\sqrt{5} \Delta t^{5} L^{2} G L^{2} G L^{2}}{3360} && \\
    & \phantom{M_{20} =} + \frac{\sqrt{5} \Delta t^{5} L^{2} G L^{3} G L}{3360} + \frac{\sqrt{5} \Delta t^{5} L^{2} G L^{4} G}{3360} + \frac{\sqrt{5} \Delta t^{5} L^{2} G^{2} L^{4}}{3360} + \frac{\sqrt{5} \Delta t^{5} L^{3} G L G L^{2}}{3360} + \frac{\sqrt{5} \Delta t^{5} L^{3} G L^{2} G L}{3360} && \\
    & \phantom{M_{20} =} + \frac{\sqrt{5} \Delta t^{5} L^{3} G L^{3} G}{3360} + \frac{\sqrt{5} \Delta t^{5} L^{3} G^{2} L^{3}}{3360} + \frac{\sqrt{5} \Delta t^{5} L^{4} G L G L}{3360} + \frac{\sqrt{5} \Delta t^{5} L^{4} G L^{2} G}{3360} + \frac{\sqrt{5} \Delta t^{5} L^{4} G^{2} L^{2}}{3360} && \\
    & \phantom{M_{20} =} + \frac{\sqrt{5} \Delta t^{5} L^{5} G L G}{3360} + \frac{\sqrt{5} \Delta t^{5} L^{5} G^{2} L}{3360} + \frac{\sqrt{5} \Delta t^{5} L^{6} G^{2}}{3360} + \frac{\sqrt{5} \Delta t^{4} G L^{6}}{420} + \frac{\sqrt{5} \Delta t^{4} L G L^{5}}{420} && \\
    & \phantom{M_{20} =} + \frac{\sqrt{5} \Delta t^{4} L^{2} G L^{4}}{420} + \frac{\sqrt{5} \Delta t^{4} L^{3} G L^{3}}{420} + \frac{\sqrt{5} \Delta t^{4} L^{4} G L^{2}}{420} + \frac{\sqrt{5} \Delta t^{4} L^{5} G L}{420} + \frac{\sqrt{5} \Delta t^{4} L^{6} G}{420} && \\
    & \phantom{M_{20} =} + \frac{\sqrt{5} \Delta t^{3} L^{6}}{60} && \\
\end{flalign*}
\begin{flalign*}
    & M_{21} = - \frac{\sqrt{7} \Delta t^{\frac{9}{2}} G L G^{3}}{840} + \frac{\sqrt{7} \Delta t^{\frac{9}{2}} G^{3} L G}{840} - \frac{\sqrt{7} \Delta t^{\frac{9}{2}} G^{4} L}{1680} + \frac{\sqrt{7} \Delta t^{\frac{9}{2}} L G^{4}}{1680} - \frac{\sqrt{7} \Delta t^{\frac{7}{2}} G L G^{2}}{280} && \\
    & \phantom{M_{21} =} + \frac{\sqrt{7} \Delta t^{\frac{7}{2}} G^{2} L G}{280} - \frac{\sqrt{7} \Delta t^{\frac{7}{2}} G^{3} L}{840} + \frac{\sqrt{7} \Delta t^{\frac{7}{2}} L G^{3}}{840} && \\
\end{flalign*}
\begin{flalign*}
    & M_{22} = \frac{\sqrt{14} \Delta t^{\frac{9}{2}} G L G L G L}{6720} - \frac{\sqrt{14} \Delta t^{\frac{9}{2}} G L G L^{2} G}{2880} + \frac{\sqrt{14} \Delta t^{\frac{9}{2}} G L G^{2} L^{2}}{6720} - \frac{\sqrt{14} \Delta t^{\frac{9}{2}} G L^{2} G L G}{2880} + \frac{\sqrt{14} \Delta t^{\frac{9}{2}} G L^{2} G^{2} L}{6720} && \\
    & \phantom{M_{22} =} - \frac{\sqrt{14} \Delta t^{\frac{9}{2}} G L^{3} G^{2}}{10080} + \frac{\sqrt{14} \Delta t^{\frac{9}{2}} G^{2} L G L^{2}}{6720} + \frac{\sqrt{14} \Delta t^{\frac{9}{2}} G^{2} L^{2} G L}{6720} - \frac{\sqrt{14} \Delta t^{\frac{9}{2}} G^{2} L^{3} G}{2880} + \frac{\sqrt{14} \Delta t^{\frac{9}{2}} G^{3} L^{3}}{6720} && \\
    & \phantom{M_{22} =} - \frac{\sqrt{14} \Delta t^{\frac{9}{2}} L G L G L G}{2880} + \frac{\sqrt{14} \Delta t^{\frac{9}{2}} L G L G^{2} L}{6720} - \frac{\sqrt{14} \Delta t^{\frac{9}{2}} L G L^{2} G^{2}}{10080} + \frac{\sqrt{14} \Delta t^{\frac{9}{2}} L G^{2} L G L}{6720} - \frac{\sqrt{14} \Delta t^{\frac{9}{2}} L G^{2} L^{2} G}{2880} && \\
    & \phantom{M_{22} =} + \frac{\sqrt{14} \Delta t^{\frac{9}{2}} L G^{3} L^{2}}{6720} - \frac{\sqrt{14} \Delta t^{\frac{9}{2}} L^{2} G L G^{2}}{10080} - \frac{\sqrt{14} \Delta t^{\frac{9}{2}} L^{2} G^{2} L G}{2880} + \frac{\sqrt{14} \Delta t^{\frac{9}{2}} L^{2} G^{3} L}{6720} + \frac{\sqrt{14} \Delta t^{\frac{9}{2}} L^{3} G^{3}}{1120} && \\
    & \phantom{M_{22} =} + \frac{\sqrt{14} \Delta t^{\frac{7}{2}} G L G L^{2}}{2520} + \frac{\sqrt{14} \Delta t^{\frac{7}{2}} G L^{2} G L}{2520} - \frac{\sqrt{14} \Delta t^{\frac{7}{2}} G L^{3} G}{630} + \frac{\sqrt{14} \Delta t^{\frac{7}{2}} G^{2} L^{3}}{2520} + \frac{\sqrt{14} \Delta t^{\frac{7}{2}} L G L G L}{2520} && \\
    & \phantom{M_{22} =} - \frac{\sqrt{14} \Delta t^{\frac{7}{2}} L G L^{2} G}{630} + \frac{\sqrt{14} \Delta t^{\frac{7}{2}} L G^{2} L^{2}}{2520} - \frac{\sqrt{14} \Delta t^{\frac{7}{2}} L^{2} G L G}{630} + \frac{\sqrt{14} \Delta t^{\frac{7}{2}} L^{2} G^{2} L}{2520} + \frac{\sqrt{14} \Delta t^{\frac{7}{2}} L^{3} G^{2}}{420} && \\
\end{flalign*}
\begin{flalign*}
    & M_{23} = - \frac{\sqrt{7} \Delta t^{\frac{9}{2}} G L G L^{2} G}{2520} + \frac{\sqrt{7} \Delta t^{\frac{9}{2}} G L G^{2} L^{2}}{3360} + \frac{\sqrt{7} \Delta t^{\frac{9}{2}} G L^{2} G L G}{5040} - \frac{\sqrt{7} \Delta t^{\frac{9}{2}} G L^{2} G^{2} L}{3360} - \frac{\sqrt{7} \Delta t^{\frac{9}{2}} G L^{3} G^{2}}{2016} && \\
    & \phantom{M_{23} =} + \frac{\sqrt{7} \Delta t^{\frac{9}{2}} G^{2} L G L^{2}}{3360} - \frac{\sqrt{7} \Delta t^{\frac{9}{2}} G^{2} L^{3} G}{2520} + \frac{\sqrt{7} \Delta t^{\frac{9}{2}} G^{3} L^{3}}{3360} + \frac{\sqrt{7} \Delta t^{\frac{9}{2}} L G L G L G}{5040} - \frac{\sqrt{7} \Delta t^{\frac{9}{2}} L G L G^{2} L}{3360} && \\
    & \phantom{M_{23} =} - \frac{\sqrt{7} \Delta t^{\frac{9}{2}} L G L^{2} G^{2}}{2016} - \frac{\sqrt{7} \Delta t^{\frac{9}{2}} L G^{2} L^{2} G}{2520} + \frac{\sqrt{7} \Delta t^{\frac{9}{2}} L G^{3} L^{2}}{3360} + \frac{\sqrt{7} \Delta t^{\frac{9}{2}} L^{2} G L G^{2}}{1008} + \frac{\sqrt{7} \Delta t^{\frac{9}{2}} L^{2} G^{2} L G}{1260} && \\
    & \phantom{M_{23} =} - \frac{\sqrt{7} \Delta t^{\frac{9}{2}} L^{2} G^{3} L}{1680} + \frac{\sqrt{7} \Delta t^{\frac{7}{2}} G L G L^{2}}{1260} - \frac{\sqrt{7} \Delta t^{\frac{7}{2}} G L^{2} G L}{2520} - \frac{\sqrt{7} \Delta t^{\frac{7}{2}} G L^{3} G}{504} + \frac{\sqrt{7} \Delta t^{\frac{7}{2}} G^{2} L^{3}}{1260} && \\
    & \phantom{M_{23} =} - \frac{\sqrt{7} \Delta t^{\frac{7}{2}} L G L G L}{2520} - \frac{\sqrt{7} \Delta t^{\frac{7}{2}} L G L^{2} G}{504} + \frac{\sqrt{7} \Delta t^{\frac{7}{2}} L G^{2} L^{2}}{1260} + \frac{\sqrt{7} \Delta t^{\frac{7}{2}} L^{2} G L G}{252} - \frac{\sqrt{7} \Delta t^{\frac{7}{2}} L^{2} G^{2} L}{630} && \\
\end{flalign*}
\begin{flalign*}
    & M_{24} = - \frac{\sqrt{42} \Delta t^{\frac{9}{2}} G L G L G L}{4032} + \frac{\sqrt{42} \Delta t^{\frac{9}{2}} G L G L^{2} G}{20160} + \frac{\sqrt{42} \Delta t^{\frac{9}{2}} G L G^{2} L^{2}}{6720} - \frac{\sqrt{42} \Delta t^{\frac{9}{2}} G L^{2} G L G}{6720} - \frac{\sqrt{42} \Delta t^{\frac{9}{2}} G L^{2} G^{2} L}{6720} && \\
    & \phantom{M_{24} =} + \frac{\sqrt{42} \Delta t^{\frac{9}{2}} G^{2} L G L^{2}}{6720} - \frac{\sqrt{42} \Delta t^{\frac{9}{2}} G^{2} L^{2} G L}{4032} + \frac{\sqrt{42} \Delta t^{\frac{9}{2}} G^{2} L^{3} G}{20160} + \frac{\sqrt{42} \Delta t^{\frac{9}{2}} G^{3} L^{3}}{6720} - \frac{\sqrt{42} \Delta t^{\frac{9}{2}} L G L G L G}{6720} && \\
    & \phantom{M_{24} =} - \frac{\sqrt{42} \Delta t^{\frac{9}{2}} L G L G^{2} L}{6720} - \frac{\sqrt{42} \Delta t^{\frac{9}{2}} L G^{2} L G L}{4032} + \frac{\sqrt{42} \Delta t^{\frac{9}{2}} L G^{2} L^{2} G}{20160} + \frac{\sqrt{42} \Delta t^{\frac{9}{2}} L G^{3} L^{2}}{6720} + \frac{\sqrt{42} \Delta t^{\frac{9}{2}} L^{2} G^{2} L G}{6720} && \\
    & \phantom{M_{24} =} + \frac{\sqrt{42} \Delta t^{\frac{9}{2}} L^{2} G^{3} L}{2240} + \frac{\sqrt{42} \Delta t^{\frac{7}{2}} G L G L^{2}}{2520} - \frac{\sqrt{42} \Delta t^{\frac{7}{2}} G L^{2} G L}{840} + \frac{\sqrt{42} \Delta t^{\frac{7}{2}} G^{2} L^{3}}{2520} - \frac{\sqrt{42} \Delta t^{\frac{7}{2}} L G L G L}{840} && \\
    & \phantom{M_{24} =} + \frac{\sqrt{42} \Delta t^{\frac{7}{2}} L G^{2} L^{2}}{2520} + \frac{\sqrt{42} \Delta t^{\frac{7}{2}} L^{2} G^{2} L}{840} && \\
\end{flalign*}
\begin{flalign*}
    & M_{25} = - \frac{\sqrt{21} \Delta t^{\frac{9}{2}} G L G^{2} L^{2}}{10080} - \frac{\sqrt{21} \Delta t^{\frac{9}{2}} G L^{2} G L G}{5040} + \frac{\sqrt{21} \Delta t^{\frac{9}{2}} G L^{2} G^{2} L}{10080} - \frac{\sqrt{21} \Delta t^{\frac{9}{2}} G L^{3} G^{2}}{2016} + \frac{\sqrt{21} \Delta t^{\frac{9}{2}} G^{2} L G L^{2}}{10080} && \\
    & \phantom{M_{25} =} + \frac{\sqrt{21} \Delta t^{\frac{9}{2}} G^{2} L^{2} G L}{5040} - \frac{\sqrt{21} \Delta t^{\frac{9}{2}} G^{2} L^{3} G}{2520} + \frac{\sqrt{21} \Delta t^{\frac{9}{2}} G^{3} L^{3}}{3360} + \frac{\sqrt{21} \Delta t^{\frac{9}{2}} L G L G L G}{5040} - \frac{\sqrt{21} \Delta t^{\frac{9}{2}} L G L G^{2} L}{10080} && \\
    & \phantom{M_{25} =} + \frac{\sqrt{21} \Delta t^{\frac{9}{2}} L G L^{2} G^{2}}{2016} - \frac{\sqrt{21} \Delta t^{\frac{9}{2}} L G^{2} L G L}{5040} + \frac{\sqrt{21} \Delta t^{\frac{9}{2}} L G^{2} L^{2} G}{2520} - \frac{\sqrt{21} \Delta t^{\frac{9}{2}} L G^{3} L^{2}}{3360} + \frac{\sqrt{21} \Delta t^{\frac{7}{2}} G L^{2} G L}{2520} && \\
    & \phantom{M_{25} =} - \frac{\sqrt{21} \Delta t^{\frac{7}{2}} G L^{3} G}{504} + \frac{\sqrt{21} \Delta t^{\frac{7}{2}} G^{2} L^{3}}{1260} - \frac{\sqrt{21} \Delta t^{\frac{7}{2}} L G L G L}{2520} + \frac{\sqrt{21} \Delta t^{\frac{7}{2}} L G L^{2} G}{504} - \frac{\sqrt{21} \Delta t^{\frac{7}{2}} L G^{2} L^{2}}{1260} && \\
\end{flalign*}
\begin{flalign*}
    & M_{26} = - \frac{\sqrt{14} \Delta t^{\frac{9}{2}} G L G^{2} L^{2}}{6720} - \frac{\sqrt{14} \Delta t^{\frac{9}{2}} G L^{2} G L G}{3360} - \frac{\sqrt{14} \Delta t^{\frac{9}{2}} G L^{2} G^{2} L}{1680} + \frac{\sqrt{14} \Delta t^{\frac{9}{2}} G^{2} L G L^{2}}{6720} - \frac{\sqrt{14} \Delta t^{\frac{9}{2}} G^{2} L^{2} G L}{2240} && \\
    & \phantom{M_{26} =} + \frac{\sqrt{14} \Delta t^{\frac{9}{2}} G^{2} L^{3} G}{6720} + \frac{\sqrt{14} \Delta t^{\frac{9}{2}} G^{3} L^{3}}{2240} + \frac{\sqrt{14} \Delta t^{\frac{9}{2}} L G L G L G}{3360} + \frac{\sqrt{14} \Delta t^{\frac{9}{2}} L G L G^{2} L}{1680} + \frac{\sqrt{14} \Delta t^{\frac{9}{2}} L G^{2} L G L}{2240} && \\
    & \phantom{M_{26} =} - \frac{\sqrt{14} \Delta t^{\frac{9}{2}} L G^{2} L^{2} G}{6720} - \frac{\sqrt{14} \Delta t^{\frac{9}{2}} L G^{3} L^{2}}{2240} - \frac{\sqrt{14} \Delta t^{\frac{7}{2}} G L^{2} G L}{420} + \frac{\sqrt{14} \Delta t^{\frac{7}{2}} G^{2} L^{3}}{840} + \frac{\sqrt{14} \Delta t^{\frac{7}{2}} L G L G L}{420} && \\
    & \phantom{M_{26} =} - \frac{\sqrt{14} \Delta t^{\frac{7}{2}} L G^{2} L^{2}}{840} && \\
\end{flalign*}
\begin{flalign*}
    & M_{27} = - \frac{\sqrt{210} \Delta t^{\frac{9}{2}} G L G L G L}{10080} - \frac{\sqrt{210} \Delta t^{\frac{9}{2}} G L G L^{2} G}{10080} - \frac{\sqrt{210} \Delta t^{\frac{9}{2}} G L G^{2} L^{2}}{6720} - \frac{\sqrt{210} \Delta t^{\frac{9}{2}} G^{2} L G L^{2}}{6720} + \frac{\sqrt{210} \Delta t^{\frac{9}{2}} G^{2} L^{2} G L}{20160} && \\
    & \phantom{M_{27} =} + \frac{\sqrt{210} \Delta t^{\frac{9}{2}} G^{2} L^{3} G}{20160} + \frac{\sqrt{210} \Delta t^{\frac{9}{2}} G^{3} L^{3}}{6720} + \frac{\sqrt{210} \Delta t^{\frac{9}{2}} L G^{2} L G L}{20160} + \frac{\sqrt{210} \Delta t^{\frac{9}{2}} L G^{2} L^{2} G}{20160} + \frac{\sqrt{210} \Delta t^{\frac{9}{2}} L G^{3} L^{2}}{6720} && \\
    & \phantom{M_{27} =} - \frac{\sqrt{210} \Delta t^{\frac{7}{2}} G L G L^{2}}{1260} + \frac{\sqrt{210} \Delta t^{\frac{7}{2}} G^{2} L^{3}}{2520} + \frac{\sqrt{210} \Delta t^{\frac{7}{2}} L G^{2} L^{2}}{2520} && \\
\end{flalign*}
\begin{flalign*}
    & M_{28} = - \frac{\sqrt{42} \Delta t^{\frac{9}{2}} G L G L^{4}}{10080} - \frac{\sqrt{42} \Delta t^{\frac{9}{2}} G L^{2} G L^{3}}{10080} - \frac{\sqrt{42} \Delta t^{\frac{9}{2}} G L^{3} G L^{2}}{10080} - \frac{\sqrt{42} \Delta t^{\frac{9}{2}} G L^{4} G L}{10080} + \frac{\sqrt{42} \Delta t^{\frac{9}{2}} G L^{5} G}{5040} && \\
    & \phantom{M_{28} =} - \frac{\sqrt{42} \Delta t^{\frac{9}{2}} G^{2} L^{5}}{10080} - \frac{\sqrt{42} \Delta t^{\frac{9}{2}} L G L G L^{3}}{10080} - \frac{\sqrt{42} \Delta t^{\frac{9}{2}} L G L^{2} G L^{2}}{10080} - \frac{\sqrt{42} \Delta t^{\frac{9}{2}} L G L^{3} G L}{10080} + \frac{\sqrt{42} \Delta t^{\frac{9}{2}} L G L^{4} G}{5040} && \\
    & \phantom{M_{28} =} - \frac{\sqrt{42} \Delta t^{\frac{9}{2}} L G^{2} L^{4}}{10080} - \frac{\sqrt{42} \Delta t^{\frac{9}{2}} L^{2} G L G L^{2}}{10080} - \frac{\sqrt{42} \Delta t^{\frac{9}{2}} L^{2} G L^{2} G L}{10080} + \frac{\sqrt{42} \Delta t^{\frac{9}{2}} L^{2} G L^{3} G}{5040} - \frac{\sqrt{42} \Delta t^{\frac{9}{2}} L^{2} G^{2} L^{3}}{10080} && \\
    & \phantom{M_{28} =} - \frac{\sqrt{42} \Delta t^{\frac{9}{2}} L^{3} G L G L}{10080} + \frac{\sqrt{42} \Delta t^{\frac{9}{2}} L^{3} G L^{2} G}{5040} - \frac{\sqrt{42} \Delta t^{\frac{9}{2}} L^{3} G^{2} L^{2}}{10080} + \frac{\sqrt{42} \Delta t^{\frac{9}{2}} L^{4} G L G}{5040} - \frac{\sqrt{42} \Delta t^{\frac{9}{2}} L^{4} G^{2} L}{10080} && \\
    & \phantom{M_{28} =} + \frac{\sqrt{42} \Delta t^{\frac{9}{2}} L^{5} G^{2}}{2016} - \frac{\sqrt{42} \Delta t^{\frac{7}{2}} G L^{5}}{2520} - \frac{\sqrt{42} \Delta t^{\frac{7}{2}} L G L^{4}}{2520} - \frac{\sqrt{42} \Delta t^{\frac{7}{2}} L^{2} G L^{3}}{2520} - \frac{\sqrt{42} \Delta t^{\frac{7}{2}} L^{3} G L^{2}}{2520} && \\
    & \phantom{M_{28} =} - \frac{\sqrt{42} \Delta t^{\frac{7}{2}} L^{4} G L}{2520} + \frac{\sqrt{42} \Delta t^{\frac{7}{2}} L^{5} G}{504} && \\
\end{flalign*}
\begin{flalign*}
    & M_{29} = - \frac{\sqrt{7} \Delta t^{\frac{9}{2}} G L G L^{4}}{3360} - \frac{\sqrt{7} \Delta t^{\frac{9}{2}} G L^{2} G L^{3}}{3360} - \frac{\sqrt{7} \Delta t^{\frac{9}{2}} G L^{3} G L^{2}}{3360} + \frac{\sqrt{7} \Delta t^{\frac{9}{2}} G L^{4} G L}{2240} - \frac{\sqrt{7} \Delta t^{\frac{9}{2}} G L^{5} G}{6720} && \\
    & \phantom{M_{29} =} - \frac{\sqrt{7} \Delta t^{\frac{9}{2}} G^{2} L^{5}}{3360} - \frac{\sqrt{7} \Delta t^{\frac{9}{2}} L G L G L^{3}}{3360} - \frac{\sqrt{7} \Delta t^{\frac{9}{2}} L G L^{2} G L^{2}}{3360} + \frac{\sqrt{7} \Delta t^{\frac{9}{2}} L G L^{3} G L}{2240} - \frac{\sqrt{7} \Delta t^{\frac{9}{2}} L G L^{4} G}{6720} && \\
    & \phantom{M_{29} =} - \frac{\sqrt{7} \Delta t^{\frac{9}{2}} L G^{2} L^{4}}{3360} - \frac{\sqrt{7} \Delta t^{\frac{9}{2}} L^{2} G L G L^{2}}{3360} + \frac{\sqrt{7} \Delta t^{\frac{9}{2}} L^{2} G L^{2} G L}{2240} - \frac{\sqrt{7} \Delta t^{\frac{9}{2}} L^{2} G L^{3} G}{6720} - \frac{\sqrt{7} \Delta t^{\frac{9}{2}} L^{2} G^{2} L^{3}}{3360} && \\
    & \phantom{M_{29} =} + \frac{\sqrt{7} \Delta t^{\frac{9}{2}} L^{3} G L G L}{2240} - \frac{\sqrt{7} \Delta t^{\frac{9}{2}} L^{3} G L^{2} G}{6720} - \frac{\sqrt{7} \Delta t^{\frac{9}{2}} L^{3} G^{2} L^{2}}{3360} + \frac{\sqrt{7} \Delta t^{\frac{9}{2}} L^{4} G L G}{1680} + \frac{\sqrt{7} \Delta t^{\frac{9}{2}} L^{4} G^{2} L}{840} && \\
    & \phantom{M_{29} =} - \frac{\sqrt{7} \Delta t^{\frac{7}{2}} G L^{5}}{840} - \frac{\sqrt{7} \Delta t^{\frac{7}{2}} L G L^{4}}{840} - \frac{\sqrt{7} \Delta t^{\frac{7}{2}} L^{2} G L^{3}}{840} - \frac{\sqrt{7} \Delta t^{\frac{7}{2}} L^{3} G L^{2}}{840} + \frac{\sqrt{7} \Delta t^{\frac{7}{2}} L^{4} G L}{210} && \\
\end{flalign*}
\begin{flalign*}
    & M_{30} = - \frac{\sqrt{105} \Delta t^{\frac{9}{2}} G L G L^{4}}{10080} - \frac{\sqrt{105} \Delta t^{\frac{9}{2}} G L^{2} G L^{3}}{10080} + \frac{\sqrt{105} \Delta t^{\frac{9}{2}} G L^{3} G L^{2}}{10080} - \frac{\sqrt{105} \Delta t^{\frac{9}{2}} G L^{4} G L}{20160} - \frac{\sqrt{105} \Delta t^{\frac{9}{2}} G L^{5} G}{20160} && \\
    & \phantom{M_{30} =} - \frac{\sqrt{105} \Delta t^{\frac{9}{2}} G^{2} L^{5}}{10080} - \frac{\sqrt{105} \Delta t^{\frac{9}{2}} L G L G L^{3}}{10080} + \frac{\sqrt{105} \Delta t^{\frac{9}{2}} L G L^{2} G L^{2}}{10080} - \frac{\sqrt{105} \Delta t^{\frac{9}{2}} L G L^{3} G L}{20160} - \frac{\sqrt{105} \Delta t^{\frac{9}{2}} L G L^{4} G}{20160} && \\
    & \phantom{M_{30} =} - \frac{\sqrt{105} \Delta t^{\frac{9}{2}} L G^{2} L^{4}}{10080} + \frac{\sqrt{105} \Delta t^{\frac{9}{2}} L^{2} G L G L^{2}}{10080} - \frac{\sqrt{105} \Delta t^{\frac{9}{2}} L^{2} G L^{2} G L}{20160} - \frac{\sqrt{105} \Delta t^{\frac{9}{2}} L^{2} G L^{3} G}{20160} - \frac{\sqrt{105} \Delta t^{\frac{9}{2}} L^{2} G^{2} L^{3}}{10080} && \\
    & \phantom{M_{30} =} + \frac{\sqrt{105} \Delta t^{\frac{9}{2}} L^{3} G L G L}{6720} + \frac{\sqrt{105} \Delta t^{\frac{9}{2}} L^{3} G L^{2} G}{6720} + \frac{\sqrt{105} \Delta t^{\frac{9}{2}} L^{3} G^{2} L^{2}}{3360} - \frac{\sqrt{105} \Delta t^{\frac{7}{2}} G L^{5}}{2520} - \frac{\sqrt{105} \Delta t^{\frac{7}{2}} L G L^{4}}{2520} && \\
    & \phantom{M_{30} =} - \frac{\sqrt{105} \Delta t^{\frac{7}{2}} L^{2} G L^{3}}{2520} + \frac{\sqrt{105} \Delta t^{\frac{7}{2}} L^{3} G L^{2}}{840} && \\
\end{flalign*}
\begin{flalign*}
    & M_{31} = - \frac{\sqrt{210} \Delta t^{\frac{9}{2}} G L G L^{4}}{10080} + \frac{\sqrt{210} \Delta t^{\frac{9}{2}} G L^{2} G L^{3}}{20160} - \frac{\sqrt{210} \Delta t^{\frac{9}{2}} G L^{3} G L^{2}}{20160} - \frac{\sqrt{210} \Delta t^{\frac{9}{2}} G L^{4} G L}{20160} - \frac{\sqrt{210} \Delta t^{\frac{9}{2}} G L^{5} G}{20160} && \\
    & \phantom{M_{31} =} - \frac{\sqrt{210} \Delta t^{\frac{9}{2}} G^{2} L^{5}}{10080} + \frac{\sqrt{210} \Delta t^{\frac{9}{2}} L G L G L^{3}}{20160} - \frac{\sqrt{210} \Delta t^{\frac{9}{2}} L G L^{2} G L^{2}}{20160} - \frac{\sqrt{210} \Delta t^{\frac{9}{2}} L G L^{3} G L}{20160} - \frac{\sqrt{210} \Delta t^{\frac{9}{2}} L G L^{4} G}{20160} && \\
    & \phantom{M_{31} =} - \frac{\sqrt{210} \Delta t^{\frac{9}{2}} L G^{2} L^{4}}{10080} + \frac{\sqrt{210} \Delta t^{\frac{9}{2}} L^{2} G L G L^{2}}{10080} + \frac{\sqrt{210} \Delta t^{\frac{9}{2}} L^{2} G L^{2} G L}{10080} + \frac{\sqrt{210} \Delta t^{\frac{9}{2}} L^{2} G L^{3} G}{10080} + \frac{\sqrt{210} \Delta t^{\frac{9}{2}} L^{2} G^{2} L^{3}}{5040} && \\
    & \phantom{M_{31} =} - \frac{\sqrt{210} \Delta t^{\frac{7}{2}} G L^{5}}{2520} - \frac{\sqrt{210} \Delta t^{\frac{7}{2}} L G L^{4}}{2520} + \frac{\sqrt{210} \Delta t^{\frac{7}{2}} L^{2} G L^{3}}{1260} && \\
\end{flalign*}
\begin{flalign*}
    & M_{32} = - \frac{\sqrt{70} \Delta t^{\frac{9}{2}} G L^{2} G L^{3}}{6720} - \frac{\sqrt{70} \Delta t^{\frac{9}{2}} G L^{3} G L^{2}}{6720} - \frac{\sqrt{70} \Delta t^{\frac{9}{2}} G L^{4} G L}{6720} - \frac{\sqrt{70} \Delta t^{\frac{9}{2}} G L^{5} G}{6720} - \frac{\sqrt{70} \Delta t^{\frac{9}{2}} G^{2} L^{5}}{3360} && \\
    & \phantom{M_{32} =} + \frac{\sqrt{70} \Delta t^{\frac{9}{2}} L G L G L^{3}}{6720} + \frac{\sqrt{70} \Delta t^{\frac{9}{2}} L G L^{2} G L^{2}}{6720} + \frac{\sqrt{70} \Delta t^{\frac{9}{2}} L G L^{3} G L}{6720} + \frac{\sqrt{70} \Delta t^{\frac{9}{2}} L G L^{4} G}{6720} + \frac{\sqrt{70} \Delta t^{\frac{9}{2}} L G^{2} L^{4}}{3360} && \\
    & \phantom{M_{32} =} - \frac{\sqrt{70} \Delta t^{\frac{7}{2}} G L^{5}}{840} + \frac{\sqrt{70} \Delta t^{\frac{7}{2}} L G L^{4}}{840} && \\
\end{flalign*}
\begin{flalign*}
    & M_{33} = \frac{\sqrt{35} \Delta t^{\frac{9}{2}} G L^{7}}{3360} + \frac{\sqrt{35} \Delta t^{\frac{9}{2}} L G L^{6}}{3360} + \frac{\sqrt{35} \Delta t^{\frac{9}{2}} L^{2} G L^{5}}{3360} + \frac{\sqrt{35} \Delta t^{\frac{9}{2}} L^{3} G L^{4}}{3360} + \frac{\sqrt{35} \Delta t^{\frac{9}{2}} L^{4} G L^{3}}{3360} && \\
    & \phantom{M_{33} =} + \frac{\sqrt{35} \Delta t^{\frac{9}{2}} L^{5} G L^{2}}{3360} + \frac{\sqrt{35} \Delta t^{\frac{9}{2}} L^{6} G L}{3360} + \frac{\sqrt{35} \Delta t^{\frac{9}{2}} L^{7} G}{3360} + \frac{\sqrt{35} \Delta t^{\frac{7}{2}} L^{7}}{420} && \\
\end{flalign*}
\begin{flalign*}
    & M_{34} = \frac{\sqrt{2} \Delta t^{4} G L G L G}{840} - \frac{\sqrt{2} \Delta t^{4} G L G^{2} L}{3360} - \frac{\sqrt{2} \Delta t^{4} G L^{2} G^{2}}{560} - \frac{\sqrt{2} \Delta t^{4} G^{2} L G L}{3360} + \frac{\sqrt{2} \Delta t^{4} G^{2} L^{2} G}{840} && \\
    & \phantom{M_{34} =} - \frac{\sqrt{2} \Delta t^{4} G^{3} L^{2}}{3360} - \frac{\sqrt{2} \Delta t^{4} L G L G^{2}}{560} + \frac{\sqrt{2} \Delta t^{4} L G^{2} L G}{840} - \frac{\sqrt{2} \Delta t^{4} L G^{3} L}{3360} + \frac{\sqrt{2} \Delta t^{4} L^{2} G^{3}}{840} && \\
\end{flalign*}
\begin{flalign*}
    & M_{35} = \frac{\sqrt{6} \Delta t^{4} G L G^{2} L}{10080} - \frac{\sqrt{6} \Delta t^{4} G L^{2} G^{2}}{1008} - \frac{\sqrt{6} \Delta t^{4} G^{2} L G L}{10080} + \frac{\sqrt{6} \Delta t^{4} G^{2} L^{2} G}{1008} - \frac{\sqrt{6} \Delta t^{4} G^{3} L^{2}}{3360} && \\
    & \phantom{M_{35} =} + \frac{\sqrt{6} \Delta t^{4} L G L G^{2}}{1008} - \frac{\sqrt{6} \Delta t^{4} L G^{2} L G}{1008} + \frac{\sqrt{6} \Delta t^{4} L G^{3} L}{3360} && \\
\end{flalign*}
\begin{flalign*}
    & M_{36} = - \frac{\sqrt{10} \Delta t^{4} G L G L G}{840} + \frac{\sqrt{10} \Delta t^{4} G L G^{2} L}{3360} + \frac{\sqrt{10} \Delta t^{4} G^{2} L G L}{3360} + \frac{\sqrt{10} \Delta t^{4} G^{2} L^{2} G}{1680} - \frac{\sqrt{10} \Delta t^{4} G^{3} L^{2}}{3360} && \\
    & \phantom{M_{36} =} + \frac{\sqrt{10} \Delta t^{4} L G^{2} L G}{1680} - \frac{\sqrt{10} \Delta t^{4} L G^{3} L}{3360} && \\
\end{flalign*}
\begin{flalign*}
    & M_{37} = - \frac{\sqrt{14} \Delta t^{4} G L G^{2} L}{1120} + \frac{\sqrt{14} \Delta t^{4} G^{2} L G L}{1120} - \frac{\sqrt{14} \Delta t^{4} G^{3} L^{2}}{3360} + \frac{\sqrt{14} \Delta t^{4} L G^{3} L}{3360} && \\
\end{flalign*}
\begin{flalign*}
    & M_{38} = \frac{\sqrt{3} \Delta t^{4} G L G L^{3}}{5040} + \frac{\sqrt{3} \Delta t^{4} G L^{2} G L^{2}}{5040} + \frac{\sqrt{3} \Delta t^{4} G L^{3} G L}{5040} - \frac{\sqrt{3} \Delta t^{4} G L^{4} G}{1008} + \frac{\sqrt{3} \Delta t^{4} G^{2} L^{4}}{5040} && \\
    & \phantom{M_{38} =} + \frac{\sqrt{3} \Delta t^{4} L G L G L^{2}}{5040} + \frac{\sqrt{3} \Delta t^{4} L G L^{2} G L}{5040} - \frac{\sqrt{3} \Delta t^{4} L G L^{3} G}{1008} + \frac{\sqrt{3} \Delta t^{4} L G^{2} L^{3}}{5040} + \frac{\sqrt{3} \Delta t^{4} L^{2} G L G L}{5040} && \\
    & \phantom{M_{38} =} - \frac{\sqrt{3} \Delta t^{4} L^{2} G L^{2} G}{1008} + \frac{\sqrt{3} \Delta t^{4} L^{2} G^{2} L^{2}}{5040} - \frac{\sqrt{3} \Delta t^{4} L^{3} G L G}{1008} + \frac{\sqrt{3} \Delta t^{4} L^{3} G^{2} L}{5040} + \frac{\sqrt{3} \Delta t^{4} L^{4} G^{2}}{504} && \\
\end{flalign*}
\begin{flalign*}
    & M_{39} = \frac{\sqrt{5} \Delta t^{4} G L G L^{3}}{5040} + \frac{\sqrt{5} \Delta t^{4} G L^{2} G L^{2}}{5040} - \frac{\sqrt{5} \Delta t^{4} G L^{3} G L}{5040} - \frac{\sqrt{5} \Delta t^{4} G L^{4} G}{1680} + \frac{\sqrt{5} \Delta t^{4} G^{2} L^{4}}{5040} && \\
    & \phantom{M_{39} =} + \frac{\sqrt{5} \Delta t^{4} L G L G L^{2}}{5040} - \frac{\sqrt{5} \Delta t^{4} L G L^{2} G L}{5040} - \frac{\sqrt{5} \Delta t^{4} L G L^{3} G}{1680} + \frac{\sqrt{5} \Delta t^{4} L G^{2} L^{3}}{5040} - \frac{\sqrt{5} \Delta t^{4} L^{2} G L G L}{5040} && \\
    & \phantom{M_{39} =} - \frac{\sqrt{5} \Delta t^{4} L^{2} G L^{2} G}{1680} + \frac{\sqrt{5} \Delta t^{4} L^{2} G^{2} L^{2}}{5040} + \frac{\sqrt{5} \Delta t^{4} L^{3} G L G}{560} - \frac{\sqrt{5} \Delta t^{4} L^{3} G^{2} L}{1680} && \\
\end{flalign*}
\begin{flalign*}
    & M_{40} = \frac{\sqrt{7} \Delta t^{4} G L G L^{3}}{5040} + \frac{\sqrt{7} \Delta t^{4} G L^{2} G L^{2}}{5040} - \frac{\sqrt{7} \Delta t^{4} G L^{3} G L}{1260} + \frac{\sqrt{7} \Delta t^{4} G^{2} L^{4}}{5040} + \frac{\sqrt{7} \Delta t^{4} L G L G L^{2}}{5040} && \\
    & \phantom{M_{40} =} - \frac{\sqrt{7} \Delta t^{4} L G L^{2} G L}{1260} + \frac{\sqrt{7} \Delta t^{4} L G^{2} L^{3}}{5040} - \frac{\sqrt{7} \Delta t^{4} L^{2} G L G L}{1260} + \frac{\sqrt{7} \Delta t^{4} L^{2} G^{2} L^{2}}{5040} + \frac{\sqrt{7} \Delta t^{4} L^{3} G^{2} L}{840} && \\
\end{flalign*}
\begin{flalign*}
    & M_{41} = \frac{\sqrt{10} \Delta t^{4} G L G L^{3}}{5040} - \frac{\sqrt{10} \Delta t^{4} G L^{2} G L^{2}}{10080} + \frac{\sqrt{10} \Delta t^{4} G L^{3} G L}{10080} - \frac{\sqrt{10} \Delta t^{4} G L^{4} G}{1680} + \frac{\sqrt{10} \Delta t^{4} G^{2} L^{4}}{5040} && \\
    & \phantom{M_{41} =} - \frac{\sqrt{10} \Delta t^{4} L G L G L^{2}}{10080} + \frac{\sqrt{10} \Delta t^{4} L G L^{2} G L}{10080} - \frac{\sqrt{10} \Delta t^{4} L G L^{3} G}{1680} + \frac{\sqrt{10} \Delta t^{4} L G^{2} L^{3}}{5040} - \frac{\sqrt{10} \Delta t^{4} L^{2} G L G L}{5040} && \\
    & \phantom{M_{41} =} + \frac{\sqrt{10} \Delta t^{4} L^{2} G L^{2} G}{840} - \frac{\sqrt{10} \Delta t^{4} L^{2} G^{2} L^{2}}{2520} && \\
\end{flalign*}
\begin{flalign*}
    & M_{42} = \frac{\sqrt{14} \Delta t^{4} G L G L^{3}}{5040} - \frac{\sqrt{14} \Delta t^{4} G L^{2} G L^{2}}{10080} - \frac{\sqrt{14} \Delta t^{4} G L^{3} G L}{2016} + \frac{\sqrt{14} \Delta t^{4} G^{2} L^{4}}{5040} - \frac{\sqrt{14} \Delta t^{4} L G L G L^{2}}{10080} && \\
    & \phantom{M_{42} =} - \frac{\sqrt{14} \Delta t^{4} L G L^{2} G L}{2016} + \frac{\sqrt{14} \Delta t^{4} L G^{2} L^{3}}{5040} + \frac{\sqrt{14} \Delta t^{4} L^{2} G L G L}{1008} - \frac{\sqrt{14} \Delta t^{4} L^{2} G^{2} L^{2}}{2520} && \\
\end{flalign*}
\begin{flalign*}
    & M_{43} = \frac{\sqrt{21} \Delta t^{4} G L G L^{3}}{5040} - \frac{\sqrt{21} \Delta t^{4} G L^{2} G L^{2}}{1680} + \frac{\sqrt{21} \Delta t^{4} G^{2} L^{4}}{5040} - \frac{\sqrt{21} \Delta t^{4} L G L G L^{2}}{1680} + \frac{\sqrt{21} \Delta t^{4} L G^{2} L^{3}}{5040} && \\
    & \phantom{M_{43} =} + \frac{\sqrt{21} \Delta t^{4} L^{2} G^{2} L^{2}}{1680} && \\
\end{flalign*}
\begin{flalign*}
    & M_{44} = \frac{\sqrt{30} \Delta t^{4} G L^{2} G L^{2}}{10080} + \frac{\sqrt{30} \Delta t^{4} G L^{3} G L}{10080} - \frac{\sqrt{30} \Delta t^{4} G L^{4} G}{1680} + \frac{\sqrt{30} \Delta t^{4} G^{2} L^{4}}{5040} - \frac{\sqrt{30} \Delta t^{4} L G L G L^{2}}{10080} && \\
    & \phantom{M_{44} =} - \frac{\sqrt{30} \Delta t^{4} L G L^{2} G L}{10080} + \frac{\sqrt{30} \Delta t^{4} L G L^{3} G}{1680} - \frac{\sqrt{30} \Delta t^{4} L G^{2} L^{3}}{5040} && \\
\end{flalign*}
\begin{flalign*}
    & M_{45} = \frac{\sqrt{42} \Delta t^{4} G L^{2} G L^{2}}{10080} - \frac{\sqrt{42} \Delta t^{4} G L^{3} G L}{2016} + \frac{\sqrt{42} \Delta t^{4} G^{2} L^{4}}{5040} - \frac{\sqrt{42} \Delta t^{4} L G L G L^{2}}{10080} + \frac{\sqrt{42} \Delta t^{4} L G L^{2} G L}{2016} && \\
    & \phantom{M_{45} =} - \frac{\sqrt{42} \Delta t^{4} L G^{2} L^{3}}{5040} && \\
\end{flalign*}
\begin{flalign*}
    & M_{46} = - \frac{\sqrt{7} \Delta t^{4} G L^{2} G L^{2}}{840} + \frac{\sqrt{7} \Delta t^{4} G^{2} L^{4}}{1680} + \frac{\sqrt{7} \Delta t^{4} L G L G L^{2}}{840} - \frac{\sqrt{7} \Delta t^{4} L G^{2} L^{3}}{1680} && \\
\end{flalign*}
\begin{flalign*}
    & M_{47} = - \frac{\sqrt{105} \Delta t^{4} G L G L^{3}}{2520} + \frac{\sqrt{105} \Delta t^{4} G^{2} L^{4}}{5040} + \frac{\sqrt{105} \Delta t^{4} L G^{2} L^{3}}{5040} && \\
\end{flalign*}
\begin{flalign*}
    & M_{48} = - \frac{\sqrt{15} \Delta t^{4} G L^{6}}{5040} - \frac{\sqrt{15} \Delta t^{4} L G L^{5}}{5040} - \frac{\sqrt{15} \Delta t^{4} L^{2} G L^{4}}{5040} - \frac{\sqrt{15} \Delta t^{4} L^{3} G L^{3}}{5040} - \frac{\sqrt{15} \Delta t^{4} L^{4} G L^{2}}{5040} && \\
    & \phantom{M_{48} =} - \frac{\sqrt{15} \Delta t^{4} L^{5} G L}{5040} + \frac{\sqrt{15} \Delta t^{4} L^{6} G}{840} && \\
\end{flalign*}
\begin{flalign*}
    & M_{49} = - \frac{\sqrt{21} \Delta t^{4} G L^{6}}{5040} - \frac{\sqrt{21} \Delta t^{4} L G L^{5}}{5040} - \frac{\sqrt{21} \Delta t^{4} L^{2} G L^{4}}{5040} - \frac{\sqrt{21} \Delta t^{4} L^{3} G L^{3}}{5040} - \frac{\sqrt{21} \Delta t^{4} L^{4} G L^{2}}{5040} && \\
    & \phantom{M_{49} =} + \frac{\sqrt{21} \Delta t^{4} L^{5} G L}{1008} && \\
\end{flalign*}
\begin{flalign*}
    & M_{50} = - \frac{\sqrt{14} \Delta t^{4} G L^{6}}{3360} - \frac{\sqrt{14} \Delta t^{4} L G L^{5}}{3360} - \frac{\sqrt{14} \Delta t^{4} L^{2} G L^{4}}{3360} - \frac{\sqrt{14} \Delta t^{4} L^{3} G L^{3}}{3360} + \frac{\sqrt{14} \Delta t^{4} L^{4} G L^{2}}{840} && \\
\end{flalign*}
\begin{flalign*}
    & M_{51} = - \frac{\sqrt{210} \Delta t^{4} G L^{6}}{10080} - \frac{\sqrt{210} \Delta t^{4} L G L^{5}}{10080} - \frac{\sqrt{210} \Delta t^{4} L^{2} G L^{4}}{10080} + \frac{\sqrt{210} \Delta t^{4} L^{3} G L^{3}}{3360} && \\
\end{flalign*}
\begin{flalign*}
    & M_{52} = - \frac{\sqrt{105} \Delta t^{4} G L^{6}}{5040} - \frac{\sqrt{105} \Delta t^{4} L G L^{5}}{5040} + \frac{\sqrt{105} \Delta t^{4} L^{2} G L^{4}}{2520} && \\
\end{flalign*}
\begin{flalign*}
    & M_{53} = - \frac{\sqrt{35} \Delta t^{4} G L^{6}}{1680} + \frac{\sqrt{35} \Delta t^{4} L G L^{5}}{1680} && \\
\end{flalign*}
\begin{flalign*}
    & M_{54} = \frac{\sqrt{70} \Delta t^{4} L^{8}}{1680} && \\
\end{flalign*}
}

%% file: SME_6.tex
{\fontsize{9pt}{12pt}\selectfont
\begin{flalign*}
    & P_{0, 0} = \frac{\Delta t^{6} G^{6}}{720} + \frac{\Delta t^{5} G^{5}}{120} + \frac{\Delta t^{4} G^{4}}{24} + \frac{\Delta t^{3} G^{3}}{6} + \frac{\Delta t^{2} G^{2}}{2} && \\
    & \phantom{P_{0, 0} =} + \Delta t G + 1 && \\
\end{flalign*}
\begin{flalign*}
    & P_{1, 0} = \frac{\Delta t^{\frac{11}{2}} G L G^{4}}{720} + \frac{\Delta t^{\frac{11}{2}} G^{2} L G^{3}}{720} + \frac{\Delta t^{\frac{11}{2}} G^{3} L G^{2}}{720} + \frac{\Delta t^{\frac{11}{2}} G^{4} L G}{720} + \frac{\Delta t^{\frac{11}{2}} G^{5} L}{720} && \\
    & \phantom{P_{1, 0} =} + \frac{\Delta t^{\frac{11}{2}} L G^{5}}{720} + \frac{\Delta t^{\frac{9}{2}} G L G^{3}}{120} + \frac{\Delta t^{\frac{9}{2}} G^{2} L G^{2}}{120} + \frac{\Delta t^{\frac{9}{2}} G^{3} L G}{120} + \frac{\Delta t^{\frac{9}{2}} G^{4} L}{120} && \\
    & \phantom{P_{1, 0} =} + \frac{\Delta t^{\frac{9}{2}} L G^{4}}{120} + \frac{\Delta t^{\frac{7}{2}} G L G^{2}}{24} + \frac{\Delta t^{\frac{7}{2}} G^{2} L G}{24} + \frac{\Delta t^{\frac{7}{2}} G^{3} L}{24} + \frac{\Delta t^{\frac{7}{2}} L G^{3}}{24} && \\
    & \phantom{P_{1, 0} =} + \frac{\Delta t^{\frac{5}{2}} G L G}{6} + \frac{\Delta t^{\frac{5}{2}} G^{2} L}{6} + \frac{\Delta t^{\frac{5}{2}} L G^{2}}{6} + \frac{\Delta t^{\frac{3}{2}} G L}{2} + \frac{\Delta t^{\frac{3}{2}} L G}{2} && \\
    & \phantom{P_{1, 0} =} + \sqrt{\Delta t} L && \\
\end{flalign*}
\begin{flalign*}
    & P_{2, 0} = \frac{\sqrt{2} \Delta t^{6} G L G L G^{3}}{5040} + \frac{\sqrt{2} \Delta t^{6} G L G^{2} L G^{2}}{5040} + \frac{\sqrt{2} \Delta t^{6} G L G^{3} L G}{5040} + \frac{\sqrt{2} \Delta t^{6} G L G^{4} L}{5040} + \frac{\sqrt{2} \Delta t^{6} G L^{2} G^{4}}{5040} && \\
    & \phantom{P_{2, 0} =} + \frac{\sqrt{2} \Delta t^{6} G^{2} L G L G^{2}}{5040} + \frac{\sqrt{2} \Delta t^{6} G^{2} L G^{2} L G}{5040} + \frac{\sqrt{2} \Delta t^{6} G^{2} L G^{3} L}{5040} + \frac{\sqrt{2} \Delta t^{6} G^{2} L^{2} G^{3}}{5040} + \frac{\sqrt{2} \Delta t^{6} G^{3} L G L G}{5040} && \\
    & \phantom{P_{2, 0} =} + \frac{\sqrt{2} \Delta t^{6} G^{3} L G^{2} L}{5040} + \frac{\sqrt{2} \Delta t^{6} G^{3} L^{2} G^{2}}{5040} + \frac{\sqrt{2} \Delta t^{6} G^{4} L G L}{5040} + \frac{\sqrt{2} \Delta t^{6} G^{4} L^{2} G}{5040} + \frac{\sqrt{2} \Delta t^{6} G^{5} L^{2}}{5040} && \\
    & \phantom{P_{2, 0} =} + \frac{\sqrt{2} \Delta t^{6} L G L G^{4}}{5040} + \frac{\sqrt{2} \Delta t^{6} L G^{2} L G^{3}}{5040} + \frac{\sqrt{2} \Delta t^{6} L G^{3} L G^{2}}{5040} + \frac{\sqrt{2} \Delta t^{6} L G^{4} L G}{5040} + \frac{\sqrt{2} \Delta t^{6} L G^{5} L}{5040} && \\
    & \phantom{P_{2, 0} =} + \frac{\sqrt{2} \Delta t^{6} L^{2} G^{5}}{5040} + \frac{\sqrt{2} \Delta t^{5} G L G L G^{2}}{720} + \frac{\sqrt{2} \Delta t^{5} G L G^{2} L G}{720} + \frac{\sqrt{2} \Delta t^{5} G L G^{3} L}{720} + \frac{\sqrt{2} \Delta t^{5} G L^{2} G^{3}}{720} && \\
    & \phantom{P_{2, 0} =} + \frac{\sqrt{2} \Delta t^{5} G^{2} L G L G}{720} + \frac{\sqrt{2} \Delta t^{5} G^{2} L G^{2} L}{720} + \frac{\sqrt{2} \Delta t^{5} G^{2} L^{2} G^{2}}{720} + \frac{\sqrt{2} \Delta t^{5} G^{3} L G L}{720} + \frac{\sqrt{2} \Delta t^{5} G^{3} L^{2} G}{720} && \\
    & \phantom{P_{2, 0} =} + \frac{\sqrt{2} \Delta t^{5} G^{4} L^{2}}{720} + \frac{\sqrt{2} \Delta t^{5} L G L G^{3}}{720} + \frac{\sqrt{2} \Delta t^{5} L G^{2} L G^{2}}{720} + \frac{\sqrt{2} \Delta t^{5} L G^{3} L G}{720} + \frac{\sqrt{2} \Delta t^{5} L G^{4} L}{720} && \\
    & \phantom{P_{2, 0} =} + \frac{\sqrt{2} \Delta t^{5} L^{2} G^{4}}{720} + \frac{\sqrt{2} \Delta t^{4} G L G L G}{120} + \frac{\sqrt{2} \Delta t^{4} G L G^{2} L}{120} + \frac{\sqrt{2} \Delta t^{4} G L^{2} G^{2}}{120} + \frac{\sqrt{2} \Delta t^{4} G^{2} L G L}{120} && \\
    & \phantom{P_{2, 0} =} + \frac{\sqrt{2} \Delta t^{4} G^{2} L^{2} G}{120} + \frac{\sqrt{2} \Delta t^{4} G^{3} L^{2}}{120} + \frac{\sqrt{2} \Delta t^{4} L G L G^{2}}{120} + \frac{\sqrt{2} \Delta t^{4} L G^{2} L G}{120} + \frac{\sqrt{2} \Delta t^{4} L G^{3} L}{120} && \\
    & \phantom{P_{2, 0} =} + \frac{\sqrt{2} \Delta t^{4} L^{2} G^{3}}{120} + \frac{\sqrt{2} \Delta t^{3} G L G L}{24} + \frac{\sqrt{2} \Delta t^{3} G L^{2} G}{24} + \frac{\sqrt{2} \Delta t^{3} G^{2} L^{2}}{24} + \frac{\sqrt{2} \Delta t^{3} L G L G}{24} && \\
    & \phantom{P_{2, 0} =} + \frac{\sqrt{2} \Delta t^{3} L G^{2} L}{24} + \frac{\sqrt{2} \Delta t^{3} L^{2} G^{2}}{24} + \frac{\sqrt{2} \Delta t^{2} G L^{2}}{6} + \frac{\sqrt{2} \Delta t^{2} L G L}{6} + \frac{\sqrt{2} \Delta t^{2} L^{2} G}{6} && \\
    & \phantom{P_{2, 0} =} + \frac{\sqrt{2} \Delta t L^{2}}{2} && \\
\end{flalign*}
\begin{flalign*}
    & P_{3, 0} = \frac{\sqrt{6} \Delta t^{\frac{11}{2}} G L G L G L G}{5040} + \frac{\sqrt{6} \Delta t^{\frac{11}{2}} G L G L G^{2} L}{5040} + \frac{\sqrt{6} \Delta t^{\frac{11}{2}} G L G L^{2} G^{2}}{5040} + \frac{\sqrt{6} \Delta t^{\frac{11}{2}} G L G^{2} L G L}{5040} + \frac{\sqrt{6} \Delta t^{\frac{11}{2}} G L G^{2} L^{2} G}{5040} && \\
    & \phantom{P_{3, 0} =} + \frac{\sqrt{6} \Delta t^{\frac{11}{2}} G L G^{3} L^{2}}{5040} + \frac{\sqrt{6} \Delta t^{\frac{11}{2}} G L^{2} G L G^{2}}{5040} + \frac{\sqrt{6} \Delta t^{\frac{11}{2}} G L^{2} G^{2} L G}{5040} + \frac{\sqrt{6} \Delta t^{\frac{11}{2}} G L^{2} G^{3} L}{5040} + \frac{\sqrt{6} \Delta t^{\frac{11}{2}} G L^{3} G^{3}}{5040} && \\
    & \phantom{P_{3, 0} =} + \frac{\sqrt{6} \Delta t^{\frac{11}{2}} G^{2} L G L G L}{5040} + \frac{\sqrt{6} \Delta t^{\frac{11}{2}} G^{2} L G L^{2} G}{5040} + \frac{\sqrt{6} \Delta t^{\frac{11}{2}} G^{2} L G^{2} L^{2}}{5040} + \frac{\sqrt{6} \Delta t^{\frac{11}{2}} G^{2} L^{2} G L G}{5040} + \frac{\sqrt{6} \Delta t^{\frac{11}{2}} G^{2} L^{2} G^{2} L}{5040} && \\
    & \phantom{P_{3, 0} =} + \frac{\sqrt{6} \Delta t^{\frac{11}{2}} G^{2} L^{3} G^{2}}{5040} + \frac{\sqrt{6} \Delta t^{\frac{11}{2}} G^{3} L G L^{2}}{5040} + \frac{\sqrt{6} \Delta t^{\frac{11}{2}} G^{3} L^{2} G L}{5040} + \frac{\sqrt{6} \Delta t^{\frac{11}{2}} G^{3} L^{3} G}{5040} + \frac{\sqrt{6} \Delta t^{\frac{11}{2}} G^{4} L^{3}}{5040} && \\
    & \phantom{P_{3, 0} =} + \frac{\sqrt{6} \Delta t^{\frac{11}{2}} L G L G L G^{2}}{5040} + \frac{\sqrt{6} \Delta t^{\frac{11}{2}} L G L G^{2} L G}{5040} + \frac{\sqrt{6} \Delta t^{\frac{11}{2}} L G L G^{3} L}{5040} + \frac{\sqrt{6} \Delta t^{\frac{11}{2}} L G L^{2} G^{3}}{5040} + \frac{\sqrt{6} \Delta t^{\frac{11}{2}} L G^{2} L G L G}{5040} && \\
    & \phantom{P_{3, 0} =} + \frac{\sqrt{6} \Delta t^{\frac{11}{2}} L G^{2} L G^{2} L}{5040} + \frac{\sqrt{6} \Delta t^{\frac{11}{2}} L G^{2} L^{2} G^{2}}{5040} + \frac{\sqrt{6} \Delta t^{\frac{11}{2}} L G^{3} L G L}{5040} + \frac{\sqrt{6} \Delta t^{\frac{11}{2}} L G^{3} L^{2} G}{5040} + \frac{\sqrt{6} \Delta t^{\frac{11}{2}} L G^{4} L^{2}}{5040} && \\
    & \phantom{P_{3, 0} =} + \frac{\sqrt{6} \Delta t^{\frac{11}{2}} L^{2} G L G^{3}}{5040} + \frac{\sqrt{6} \Delta t^{\frac{11}{2}} L^{2} G^{2} L G^{2}}{5040} + \frac{\sqrt{6} \Delta t^{\frac{11}{2}} L^{2} G^{3} L G}{5040} + \frac{\sqrt{6} \Delta t^{\frac{11}{2}} L^{2} G^{4} L}{5040} + \frac{\sqrt{6} \Delta t^{\frac{11}{2}} L^{3} G^{4}}{5040} && \\
    & \phantom{P_{3, 0} =} + \frac{\sqrt{6} \Delta t^{\frac{9}{2}} G L G L G L}{720} + \frac{\sqrt{6} \Delta t^{\frac{9}{2}} G L G L^{2} G}{720} + \frac{\sqrt{6} \Delta t^{\frac{9}{2}} G L G^{2} L^{2}}{720} + \frac{\sqrt{6} \Delta t^{\frac{9}{2}} G L^{2} G L G}{720} + \frac{\sqrt{6} \Delta t^{\frac{9}{2}} G L^{2} G^{2} L}{720} && \\
    & \phantom{P_{3, 0} =} + \frac{\sqrt{6} \Delta t^{\frac{9}{2}} G L^{3} G^{2}}{720} + \frac{\sqrt{6} \Delta t^{\frac{9}{2}} G^{2} L G L^{2}}{720} + \frac{\sqrt{6} \Delta t^{\frac{9}{2}} G^{2} L^{2} G L}{720} + \frac{\sqrt{6} \Delta t^{\frac{9}{2}} G^{2} L^{3} G}{720} + \frac{\sqrt{6} \Delta t^{\frac{9}{2}} G^{3} L^{3}}{720} && \\
    & \phantom{P_{3, 0} =} + \frac{\sqrt{6} \Delta t^{\frac{9}{2}} L G L G L G}{720} + \frac{\sqrt{6} \Delta t^{\frac{9}{2}} L G L G^{2} L}{720} + \frac{\sqrt{6} \Delta t^{\frac{9}{2}} L G L^{2} G^{2}}{720} + \frac{\sqrt{6} \Delta t^{\frac{9}{2}} L G^{2} L G L}{720} + \frac{\sqrt{6} \Delta t^{\frac{9}{2}} L G^{2} L^{2} G}{720} && \\
    & \phantom{P_{3, 0} =} + \frac{\sqrt{6} \Delta t^{\frac{9}{2}} L G^{3} L^{2}}{720} + \frac{\sqrt{6} \Delta t^{\frac{9}{2}} L^{2} G L G^{2}}{720} + \frac{\sqrt{6} \Delta t^{\frac{9}{2}} L^{2} G^{2} L G}{720} + \frac{\sqrt{6} \Delta t^{\frac{9}{2}} L^{2} G^{3} L}{720} + \frac{\sqrt{6} \Delta t^{\frac{9}{2}} L^{3} G^{3}}{720} && \\
    & \phantom{P_{3, 0} =} + \frac{\sqrt{6} \Delta t^{\frac{7}{2}} G L G L^{2}}{120} + \frac{\sqrt{6} \Delta t^{\frac{7}{2}} G L^{2} G L}{120} + \frac{\sqrt{6} \Delta t^{\frac{7}{2}} G L^{3} G}{120} + \frac{\sqrt{6} \Delta t^{\frac{7}{2}} G^{2} L^{3}}{120} + \frac{\sqrt{6} \Delta t^{\frac{7}{2}} L G L G L}{120} && \\
    & \phantom{P_{3, 0} =} + \frac{\sqrt{6} \Delta t^{\frac{7}{2}} L G L^{2} G}{120} + \frac{\sqrt{6} \Delta t^{\frac{7}{2}} L G^{2} L^{2}}{120} + \frac{\sqrt{6} \Delta t^{\frac{7}{2}} L^{2} G L G}{120} + \frac{\sqrt{6} \Delta t^{\frac{7}{2}} L^{2} G^{2} L}{120} + \frac{\sqrt{6} \Delta t^{\frac{7}{2}} L^{3} G^{2}}{120} && \\
    & \phantom{P_{3, 0} =} + \frac{\sqrt{6} \Delta t^{\frac{5}{2}} G L^{3}}{24} + \frac{\sqrt{6} \Delta t^{\frac{5}{2}} L G L^{2}}{24} + \frac{\sqrt{6} \Delta t^{\frac{5}{2}} L^{2} G L}{24} + \frac{\sqrt{6} \Delta t^{\frac{5}{2}} L^{3} G}{24} + \frac{\sqrt{6} \Delta t^{\frac{3}{2}} L^{3}}{6} && \\
\end{flalign*}
\begin{flalign*}
    & P_{4, 0} = \frac{\sqrt{6} \Delta t^{6} G L G L G L G L}{20160} + \frac{\sqrt{6} \Delta t^{6} G L G L G L^{2} G}{20160} + \frac{\sqrt{6} \Delta t^{6} G L G L G^{2} L^{2}}{20160} + \frac{\sqrt{6} \Delta t^{6} G L G L^{2} G L G}{20160} + \frac{\sqrt{6} \Delta t^{6} G L G L^{2} G^{2} L}{20160} && \\
    & \phantom{P_{4, 0} =} + \frac{\sqrt{6} \Delta t^{6} G L G L^{3} G^{2}}{20160} + \frac{\sqrt{6} \Delta t^{6} G L G^{2} L G L^{2}}{20160} + \frac{\sqrt{6} \Delta t^{6} G L G^{2} L^{2} G L}{20160} + \frac{\sqrt{6} \Delta t^{6} G L G^{2} L^{3} G}{20160} + \frac{\sqrt{6} \Delta t^{6} G L G^{3} L^{3}}{20160} && \\
    & \phantom{P_{4, 0} =} + \frac{\sqrt{6} \Delta t^{6} G L^{2} G L G L G}{20160} + \frac{\sqrt{6} \Delta t^{6} G L^{2} G L G^{2} L}{20160} + \frac{\sqrt{6} \Delta t^{6} G L^{2} G L^{2} G^{2}}{20160} + \frac{\sqrt{6} \Delta t^{6} G L^{2} G^{2} L G L}{20160} + \frac{\sqrt{6} \Delta t^{6} G L^{2} G^{2} L^{2} G}{20160} && \\
    & \phantom{P_{4, 0} =} + \frac{\sqrt{6} \Delta t^{6} G L^{2} G^{3} L^{2}}{20160} + \frac{\sqrt{6} \Delta t^{6} G L^{3} G L G^{2}}{20160} + \frac{\sqrt{6} \Delta t^{6} G L^{3} G^{2} L G}{20160} + \frac{\sqrt{6} \Delta t^{6} G L^{3} G^{3} L}{20160} + \frac{\sqrt{6} \Delta t^{6} G L^{4} G^{3}}{20160} && \\
    & \phantom{P_{4, 0} =} + \frac{\sqrt{6} \Delta t^{6} G^{2} L G L G L^{2}}{20160} + \frac{\sqrt{6} \Delta t^{6} G^{2} L G L^{2} G L}{20160} + \frac{\sqrt{6} \Delta t^{6} G^{2} L G L^{3} G}{20160} + \frac{\sqrt{6} \Delta t^{6} G^{2} L G^{2} L^{3}}{20160} + \frac{\sqrt{6} \Delta t^{6} G^{2} L^{2} G L G L}{20160} && \\
    & \phantom{P_{4, 0} =} + \frac{\sqrt{6} \Delta t^{6} G^{2} L^{2} G L^{2} G}{20160} + \frac{\sqrt{6} \Delta t^{6} G^{2} L^{2} G^{2} L^{2}}{20160} + \frac{\sqrt{6} \Delta t^{6} G^{2} L^{3} G L G}{20160} + \frac{\sqrt{6} \Delta t^{6} G^{2} L^{3} G^{2} L}{20160} + \frac{\sqrt{6} \Delta t^{6} G^{2} L^{4} G^{2}}{20160} && \\
    & \phantom{P_{4, 0} =} + \frac{\sqrt{6} \Delta t^{6} G^{3} L G L^{3}}{20160} + \frac{\sqrt{6} \Delta t^{6} G^{3} L^{2} G L^{2}}{20160} + \frac{\sqrt{6} \Delta t^{6} G^{3} L^{3} G L}{20160} + \frac{\sqrt{6} \Delta t^{6} G^{3} L^{4} G}{20160} + \frac{\sqrt{6} \Delta t^{6} G^{4} L^{4}}{20160} && \\
    & \phantom{P_{4, 0} =} + \frac{\sqrt{6} \Delta t^{6} L G L G L G L G}{20160} + \frac{\sqrt{6} \Delta t^{6} L G L G L G^{2} L}{20160} + \frac{\sqrt{6} \Delta t^{6} L G L G L^{2} G^{2}}{20160} + \frac{\sqrt{6} \Delta t^{6} L G L G^{2} L G L}{20160} + \frac{\sqrt{6} \Delta t^{6} L G L G^{2} L^{2} G}{20160} && \\
    & \phantom{P_{4, 0} =} + \frac{\sqrt{6} \Delta t^{6} L G L G^{3} L^{2}}{20160} + \frac{\sqrt{6} \Delta t^{6} L G L^{2} G L G^{2}}{20160} + \frac{\sqrt{6} \Delta t^{6} L G L^{2} G^{2} L G}{20160} + \frac{\sqrt{6} \Delta t^{6} L G L^{2} G^{3} L}{20160} + \frac{\sqrt{6} \Delta t^{6} L G L^{3} G^{3}}{20160} && \\
    & \phantom{P_{4, 0} =} + \frac{\sqrt{6} \Delta t^{6} L G^{2} L G L G L}{20160} + \frac{\sqrt{6} \Delta t^{6} L G^{2} L G L^{2} G}{20160} + \frac{\sqrt{6} \Delta t^{6} L G^{2} L G^{2} L^{2}}{20160} + \frac{\sqrt{6} \Delta t^{6} L G^{2} L^{2} G L G}{20160} + \frac{\sqrt{6} \Delta t^{6} L G^{2} L^{2} G^{2} L}{20160} && \\
    & \phantom{P_{4, 0} =} + \frac{\sqrt{6} \Delta t^{6} L G^{2} L^{3} G^{2}}{20160} + \frac{\sqrt{6} \Delta t^{6} L G^{3} L G L^{2}}{20160} + \frac{\sqrt{6} \Delta t^{6} L G^{3} L^{2} G L}{20160} + \frac{\sqrt{6} \Delta t^{6} L G^{3} L^{3} G}{20160} + \frac{\sqrt{6} \Delta t^{6} L G^{4} L^{3}}{20160} && \\
    & \phantom{P_{4, 0} =} + \frac{\sqrt{6} \Delta t^{6} L^{2} G L G L G^{2}}{20160} + \frac{\sqrt{6} \Delta t^{6} L^{2} G L G^{2} L G}{20160} + \frac{\sqrt{6} \Delta t^{6} L^{2} G L G^{3} L}{20160} + \frac{\sqrt{6} \Delta t^{6} L^{2} G L^{2} G^{3}}{20160} + \frac{\sqrt{6} \Delta t^{6} L^{2} G^{2} L G L G}{20160} && \\
    & \phantom{P_{4, 0} =} + \frac{\sqrt{6} \Delta t^{6} L^{2} G^{2} L G^{2} L}{20160} + \frac{\sqrt{6} \Delta t^{6} L^{2} G^{2} L^{2} G^{2}}{20160} + \frac{\sqrt{6} \Delta t^{6} L^{2} G^{3} L G L}{20160} + \frac{\sqrt{6} \Delta t^{6} L^{2} G^{3} L^{2} G}{20160} + \frac{\sqrt{6} \Delta t^{6} L^{2} G^{4} L^{2}}{20160} && \\
    & \phantom{P_{4, 0} =} + \frac{\sqrt{6} \Delta t^{6} L^{3} G L G^{3}}{20160} + \frac{\sqrt{6} \Delta t^{6} L^{3} G^{2} L G^{2}}{20160} + \frac{\sqrt{6} \Delta t^{6} L^{3} G^{3} L G}{20160} + \frac{\sqrt{6} \Delta t^{6} L^{3} G^{4} L}{20160} + \frac{\sqrt{6} \Delta t^{6} L^{4} G^{4}}{20160} && \\
    & \phantom{P_{4, 0} =} + \frac{\sqrt{6} \Delta t^{5} G L G L G L^{2}}{2520} + \frac{\sqrt{6} \Delta t^{5} G L G L^{2} G L}{2520} + \frac{\sqrt{6} \Delta t^{5} G L G L^{3} G}{2520} + \frac{\sqrt{6} \Delta t^{5} G L G^{2} L^{3}}{2520} + \frac{\sqrt{6} \Delta t^{5} G L^{2} G L G L}{2520} && \\
    & \phantom{P_{4, 0} =} + \frac{\sqrt{6} \Delta t^{5} G L^{2} G L^{2} G}{2520} + \frac{\sqrt{6} \Delta t^{5} G L^{2} G^{2} L^{2}}{2520} + \frac{\sqrt{6} \Delta t^{5} G L^{3} G L G}{2520} + \frac{\sqrt{6} \Delta t^{5} G L^{3} G^{2} L}{2520} + \frac{\sqrt{6} \Delta t^{5} G L^{4} G^{2}}{2520} && \\
    & \phantom{P_{4, 0} =} + \frac{\sqrt{6} \Delta t^{5} G^{2} L G L^{3}}{2520} + \frac{\sqrt{6} \Delta t^{5} G^{2} L^{2} G L^{2}}{2520} + \frac{\sqrt{6} \Delta t^{5} G^{2} L^{3} G L}{2520} + \frac{\sqrt{6} \Delta t^{5} G^{2} L^{4} G}{2520} + \frac{\sqrt{6} \Delta t^{5} G^{3} L^{4}}{2520} && \\
    & \phantom{P_{4, 0} =} + \frac{\sqrt{6} \Delta t^{5} L G L G L G L}{2520} + \frac{\sqrt{6} \Delta t^{5} L G L G L^{2} G}{2520} + \frac{\sqrt{6} \Delta t^{5} L G L G^{2} L^{2}}{2520} + \frac{\sqrt{6} \Delta t^{5} L G L^{2} G L G}{2520} + \frac{\sqrt{6} \Delta t^{5} L G L^{2} G^{2} L}{2520} && \\
    & \phantom{P_{4, 0} =} + \frac{\sqrt{6} \Delta t^{5} L G L^{3} G^{2}}{2520} + \frac{\sqrt{6} \Delta t^{5} L G^{2} L G L^{2}}{2520} + \frac{\sqrt{6} \Delta t^{5} L G^{2} L^{2} G L}{2520} + \frac{\sqrt{6} \Delta t^{5} L G^{2} L^{3} G}{2520} + \frac{\sqrt{6} \Delta t^{5} L G^{3} L^{3}}{2520} && \\
    & \phantom{P_{4, 0} =} + \frac{\sqrt{6} \Delta t^{5} L^{2} G L G L G}{2520} + \frac{\sqrt{6} \Delta t^{5} L^{2} G L G^{2} L}{2520} + \frac{\sqrt{6} \Delta t^{5} L^{2} G L^{2} G^{2}}{2520} + \frac{\sqrt{6} \Delta t^{5} L^{2} G^{2} L G L}{2520} + \frac{\sqrt{6} \Delta t^{5} L^{2} G^{2} L^{2} G}{2520} && \\
    & \phantom{P_{4, 0} =} + \frac{\sqrt{6} \Delta t^{5} L^{2} G^{3} L^{2}}{2520} + \frac{\sqrt{6} \Delta t^{5} L^{3} G L G^{2}}{2520} + \frac{\sqrt{6} \Delta t^{5} L^{3} G^{2} L G}{2520} + \frac{\sqrt{6} \Delta t^{5} L^{3} G^{3} L}{2520} + \frac{\sqrt{6} \Delta t^{5} L^{4} G^{3}}{2520} && \\
    & \phantom{P_{4, 0} =} + \frac{\sqrt{6} \Delta t^{4} G L G L^{3}}{360} + \frac{\sqrt{6} \Delta t^{4} G L^{2} G L^{2}}{360} + \frac{\sqrt{6} \Delta t^{4} G L^{3} G L}{360} + \frac{\sqrt{6} \Delta t^{4} G L^{4} G}{360} + \frac{\sqrt{6} \Delta t^{4} G^{2} L^{4}}{360} && \\
    & \phantom{P_{4, 0} =} + \frac{\sqrt{6} \Delta t^{4} L G L G L^{2}}{360} + \frac{\sqrt{6} \Delta t^{4} L G L^{2} G L}{360} + \frac{\sqrt{6} \Delta t^{4} L G L^{3} G}{360} + \frac{\sqrt{6} \Delta t^{4} L G^{2} L^{3}}{360} + \frac{\sqrt{6} \Delta t^{4} L^{2} G L G L}{360} && \\
    & \phantom{P_{4, 0} =} + \frac{\sqrt{6} \Delta t^{4} L^{2} G L^{2} G}{360} + \frac{\sqrt{6} \Delta t^{4} L^{2} G^{2} L^{2}}{360} + \frac{\sqrt{6} \Delta t^{4} L^{3} G L G}{360} + \frac{\sqrt{6} \Delta t^{4} L^{3} G^{2} L}{360} + \frac{\sqrt{6} \Delta t^{4} L^{4} G^{2}}{360} && \\
    & \phantom{P_{4, 0} =} + \frac{\sqrt{6} \Delta t^{3} G L^{4}}{60} + \frac{\sqrt{6} \Delta t^{3} L G L^{3}}{60} + \frac{\sqrt{6} \Delta t^{3} L^{2} G L^{2}}{60} + \frac{\sqrt{6} \Delta t^{3} L^{3} G L}{60} + \frac{\sqrt{6} \Delta t^{3} L^{4} G}{60} && \\
    & \phantom{P_{4, 0} =} + \frac{\sqrt{6} \Delta t^{2} L^{4}}{12} && \\
\end{flalign*}
\begin{flalign*}
    & P_{5, 0} = \frac{\sqrt{30} \Delta t^{\frac{11}{2}} G L G L G L^{3}}{20160} + \frac{\sqrt{30} \Delta t^{\frac{11}{2}} G L G L^{2} G L^{2}}{20160} + \frac{\sqrt{30} \Delta t^{\frac{11}{2}} G L G L^{3} G L}{20160} + \frac{\sqrt{30} \Delta t^{\frac{11}{2}} G L G L^{4} G}{20160} + \frac{\sqrt{30} \Delta t^{\frac{11}{2}} G L G^{2} L^{4}}{20160} && \\
    & \phantom{P_{5, 0} =} + \frac{\sqrt{30} \Delta t^{\frac{11}{2}} G L^{2} G L G L^{2}}{20160} + \frac{\sqrt{30} \Delta t^{\frac{11}{2}} G L^{2} G L^{2} G L}{20160} + \frac{\sqrt{30} \Delta t^{\frac{11}{2}} G L^{2} G L^{3} G}{20160} + \frac{\sqrt{30} \Delta t^{\frac{11}{2}} G L^{2} G^{2} L^{3}}{20160} + \frac{\sqrt{30} \Delta t^{\frac{11}{2}} G L^{3} G L G L}{20160} && \\
    & \phantom{P_{5, 0} =} + \frac{\sqrt{30} \Delta t^{\frac{11}{2}} G L^{3} G L^{2} G}{20160} + \frac{\sqrt{30} \Delta t^{\frac{11}{2}} G L^{3} G^{2} L^{2}}{20160} + \frac{\sqrt{30} \Delta t^{\frac{11}{2}} G L^{4} G L G}{20160} + \frac{\sqrt{30} \Delta t^{\frac{11}{2}} G L^{4} G^{2} L}{20160} + \frac{\sqrt{30} \Delta t^{\frac{11}{2}} G L^{5} G^{2}}{20160} && \\
    & \phantom{P_{5, 0} =} + \frac{\sqrt{30} \Delta t^{\frac{11}{2}} G^{2} L G L^{4}}{20160} + \frac{\sqrt{30} \Delta t^{\frac{11}{2}} G^{2} L^{2} G L^{3}}{20160} + \frac{\sqrt{30} \Delta t^{\frac{11}{2}} G^{2} L^{3} G L^{2}}{20160} + \frac{\sqrt{30} \Delta t^{\frac{11}{2}} G^{2} L^{4} G L}{20160} + \frac{\sqrt{30} \Delta t^{\frac{11}{2}} G^{2} L^{5} G}{20160} && \\
    & \phantom{P_{5, 0} =} + \frac{\sqrt{30} \Delta t^{\frac{11}{2}} G^{3} L^{5}}{20160} + \frac{\sqrt{30} \Delta t^{\frac{11}{2}} L G L G L G L^{2}}{20160} + \frac{\sqrt{30} \Delta t^{\frac{11}{2}} L G L G L^{2} G L}{20160} + \frac{\sqrt{30} \Delta t^{\frac{11}{2}} L G L G L^{3} G}{20160} + \frac{\sqrt{30} \Delta t^{\frac{11}{2}} L G L G^{2} L^{3}}{20160} && \\
    & \phantom{P_{5, 0} =} + \frac{\sqrt{30} \Delta t^{\frac{11}{2}} L G L^{2} G L G L}{20160} + \frac{\sqrt{30} \Delta t^{\frac{11}{2}} L G L^{2} G L^{2} G}{20160} + \frac{\sqrt{30} \Delta t^{\frac{11}{2}} L G L^{2} G^{2} L^{2}}{20160} + \frac{\sqrt{30} \Delta t^{\frac{11}{2}} L G L^{3} G L G}{20160} + \frac{\sqrt{30} \Delta t^{\frac{11}{2}} L G L^{3} G^{2} L}{20160} && \\
    & \phantom{P_{5, 0} =} + \frac{\sqrt{30} \Delta t^{\frac{11}{2}} L G L^{4} G^{2}}{20160} + \frac{\sqrt{30} \Delta t^{\frac{11}{2}} L G^{2} L G L^{3}}{20160} + \frac{\sqrt{30} \Delta t^{\frac{11}{2}} L G^{2} L^{2} G L^{2}}{20160} + \frac{\sqrt{30} \Delta t^{\frac{11}{2}} L G^{2} L^{3} G L}{20160} + \frac{\sqrt{30} \Delta t^{\frac{11}{2}} L G^{2} L^{4} G}{20160} && \\
    & \phantom{P_{5, 0} =} + \frac{\sqrt{30} \Delta t^{\frac{11}{2}} L G^{3} L^{4}}{20160} + \frac{\sqrt{30} \Delta t^{\frac{11}{2}} L^{2} G L G L G L}{20160} + \frac{\sqrt{30} \Delta t^{\frac{11}{2}} L^{2} G L G L^{2} G}{20160} + \frac{\sqrt{30} \Delta t^{\frac{11}{2}} L^{2} G L G^{2} L^{2}}{20160} + \frac{\sqrt{30} \Delta t^{\frac{11}{2}} L^{2} G L^{2} G L G}{20160} && \\
    & \phantom{P_{5, 0} =} + \frac{\sqrt{30} \Delta t^{\frac{11}{2}} L^{2} G L^{2} G^{2} L}{20160} + \frac{\sqrt{30} \Delta t^{\frac{11}{2}} L^{2} G L^{3} G^{2}}{20160} + \frac{\sqrt{30} \Delta t^{\frac{11}{2}} L^{2} G^{2} L G L^{2}}{20160} + \frac{\sqrt{30} \Delta t^{\frac{11}{2}} L^{2} G^{2} L^{2} G L}{20160} + \frac{\sqrt{30} \Delta t^{\frac{11}{2}} L^{2} G^{2} L^{3} G}{20160} && \\
    & \phantom{P_{5, 0} =} + \frac{\sqrt{30} \Delta t^{\frac{11}{2}} L^{2} G^{3} L^{3}}{20160} + \frac{\sqrt{30} \Delta t^{\frac{11}{2}} L^{3} G L G L G}{20160} + \frac{\sqrt{30} \Delta t^{\frac{11}{2}} L^{3} G L G^{2} L}{20160} + \frac{\sqrt{30} \Delta t^{\frac{11}{2}} L^{3} G L^{2} G^{2}}{20160} + \frac{\sqrt{30} \Delta t^{\frac{11}{2}} L^{3} G^{2} L G L}{20160} && \\
    & \phantom{P_{5, 0} =} + \frac{\sqrt{30} \Delta t^{\frac{11}{2}} L^{3} G^{2} L^{2} G}{20160} + \frac{\sqrt{30} \Delta t^{\frac{11}{2}} L^{3} G^{3} L^{2}}{20160} + \frac{\sqrt{30} \Delta t^{\frac{11}{2}} L^{4} G L G^{2}}{20160} + \frac{\sqrt{30} \Delta t^{\frac{11}{2}} L^{4} G^{2} L G}{20160} + \frac{\sqrt{30} \Delta t^{\frac{11}{2}} L^{4} G^{3} L}{20160} && \\
    & \phantom{P_{5, 0} =} + \frac{\sqrt{30} \Delta t^{\frac{11}{2}} L^{5} G^{3}}{20160} + \frac{\sqrt{30} \Delta t^{\frac{9}{2}} G L G L^{4}}{2520} + \frac{\sqrt{30} \Delta t^{\frac{9}{2}} G L^{2} G L^{3}}{2520} + \frac{\sqrt{30} \Delta t^{\frac{9}{2}} G L^{3} G L^{2}}{2520} + \frac{\sqrt{30} \Delta t^{\frac{9}{2}} G L^{4} G L}{2520} && \\
    & \phantom{P_{5, 0} =} + \frac{\sqrt{30} \Delta t^{\frac{9}{2}} G L^{5} G}{2520} + \frac{\sqrt{30} \Delta t^{\frac{9}{2}} G^{2} L^{5}}{2520} + \frac{\sqrt{30} \Delta t^{\frac{9}{2}} L G L G L^{3}}{2520} + \frac{\sqrt{30} \Delta t^{\frac{9}{2}} L G L^{2} G L^{2}}{2520} + \frac{\sqrt{30} \Delta t^{\frac{9}{2}} L G L^{3} G L}{2520} && \\
    & \phantom{P_{5, 0} =} + \frac{\sqrt{30} \Delta t^{\frac{9}{2}} L G L^{4} G}{2520} + \frac{\sqrt{30} \Delta t^{\frac{9}{2}} L G^{2} L^{4}}{2520} + \frac{\sqrt{30} \Delta t^{\frac{9}{2}} L^{2} G L G L^{2}}{2520} + \frac{\sqrt{30} \Delta t^{\frac{9}{2}} L^{2} G L^{2} G L}{2520} + \frac{\sqrt{30} \Delta t^{\frac{9}{2}} L^{2} G L^{3} G}{2520} && \\
    & \phantom{P_{5, 0} =} + \frac{\sqrt{30} \Delta t^{\frac{9}{2}} L^{2} G^{2} L^{3}}{2520} + \frac{\sqrt{30} \Delta t^{\frac{9}{2}} L^{3} G L G L}{2520} + \frac{\sqrt{30} \Delta t^{\frac{9}{2}} L^{3} G L^{2} G}{2520} + \frac{\sqrt{30} \Delta t^{\frac{9}{2}} L^{3} G^{2} L^{2}}{2520} + \frac{\sqrt{30} \Delta t^{\frac{9}{2}} L^{4} G L G}{2520} && \\
    & \phantom{P_{5, 0} =} + \frac{\sqrt{30} \Delta t^{\frac{9}{2}} L^{4} G^{2} L}{2520} + \frac{\sqrt{30} \Delta t^{\frac{9}{2}} L^{5} G^{2}}{2520} + \frac{\sqrt{30} \Delta t^{\frac{7}{2}} G L^{5}}{360} + \frac{\sqrt{30} \Delta t^{\frac{7}{2}} L G L^{4}}{360} + \frac{\sqrt{30} \Delta t^{\frac{7}{2}} L^{2} G L^{3}}{360} && \\
    & \phantom{P_{5, 0} =} + \frac{\sqrt{30} \Delta t^{\frac{7}{2}} L^{3} G L^{2}}{360} + \frac{\sqrt{30} \Delta t^{\frac{7}{2}} L^{4} G L}{360} + \frac{\sqrt{30} \Delta t^{\frac{7}{2}} L^{5} G}{360} + \frac{\sqrt{30} \Delta t^{\frac{5}{2}} L^{5}}{60} && \\
\end{flalign*}
\begin{flalign*}
    & P_{6, 0} = \frac{\sqrt{5} \Delta t^{6} G L G L G L^{4}}{30240} + \frac{\sqrt{5} \Delta t^{6} G L G L^{2} G L^{3}}{30240} + \frac{\sqrt{5} \Delta t^{6} G L G L^{3} G L^{2}}{30240} + \frac{\sqrt{5} \Delta t^{6} G L G L^{4} G L}{30240} + \frac{\sqrt{5} \Delta t^{6} G L G L^{5} G}{30240} && \\
    & \phantom{P_{6, 0} =} + \frac{\sqrt{5} \Delta t^{6} G L G^{2} L^{5}}{30240} + \frac{\sqrt{5} \Delta t^{6} G L^{2} G L G L^{3}}{30240} + \frac{\sqrt{5} \Delta t^{6} G L^{2} G L^{2} G L^{2}}{30240} + \frac{\sqrt{5} \Delta t^{6} G L^{2} G L^{3} G L}{30240} + \frac{\sqrt{5} \Delta t^{6} G L^{2} G L^{4} G}{30240} && \\
    & \phantom{P_{6, 0} =} + \frac{\sqrt{5} \Delta t^{6} G L^{2} G^{2} L^{4}}{30240} + \frac{\sqrt{5} \Delta t^{6} G L^{3} G L G L^{2}}{30240} + \frac{\sqrt{5} \Delta t^{6} G L^{3} G L^{2} G L}{30240} + \frac{\sqrt{5} \Delta t^{6} G L^{3} G L^{3} G}{30240} + \frac{\sqrt{5} \Delta t^{6} G L^{3} G^{2} L^{3}}{30240} && \\
    & \phantom{P_{6, 0} =} + \frac{\sqrt{5} \Delta t^{6} G L^{4} G L G L}{30240} + \frac{\sqrt{5} \Delta t^{6} G L^{4} G L^{2} G}{30240} + \frac{\sqrt{5} \Delta t^{6} G L^{4} G^{2} L^{2}}{30240} + \frac{\sqrt{5} \Delta t^{6} G L^{5} G L G}{30240} + \frac{\sqrt{5} \Delta t^{6} G L^{5} G^{2} L}{30240} && \\
    & \phantom{P_{6, 0} =} + \frac{\sqrt{5} \Delta t^{6} G L^{6} G^{2}}{30240} + \frac{\sqrt{5} \Delta t^{6} G^{2} L G L^{5}}{30240} + \frac{\sqrt{5} \Delta t^{6} G^{2} L^{2} G L^{4}}{30240} + \frac{\sqrt{5} \Delta t^{6} G^{2} L^{3} G L^{3}}{30240} + \frac{\sqrt{5} \Delta t^{6} G^{2} L^{4} G L^{2}}{30240} && \\
    & \phantom{P_{6, 0} =} + \frac{\sqrt{5} \Delta t^{6} G^{2} L^{5} G L}{30240} + \frac{\sqrt{5} \Delta t^{6} G^{2} L^{6} G}{30240} + \frac{\sqrt{5} \Delta t^{6} G^{3} L^{6}}{30240} + \frac{\sqrt{5} \Delta t^{6} L G L G L G L^{3}}{30240} + \frac{\sqrt{5} \Delta t^{6} L G L G L^{2} G L^{2}}{30240} && \\
    & \phantom{P_{6, 0} =} + \frac{\sqrt{5} \Delta t^{6} L G L G L^{3} G L}{30240} + \frac{\sqrt{5} \Delta t^{6} L G L G L^{4} G}{30240} + \frac{\sqrt{5} \Delta t^{6} L G L G^{2} L^{4}}{30240} + \frac{\sqrt{5} \Delta t^{6} L G L^{2} G L G L^{2}}{30240} + \frac{\sqrt{5} \Delta t^{6} L G L^{2} G L^{2} G L}{30240} && \\
    & \phantom{P_{6, 0} =} + \frac{\sqrt{5} \Delta t^{6} L G L^{2} G L^{3} G}{30240} + \frac{\sqrt{5} \Delta t^{6} L G L^{2} G^{2} L^{3}}{30240} + \frac{\sqrt{5} \Delta t^{6} L G L^{3} G L G L}{30240} + \frac{\sqrt{5} \Delta t^{6} L G L^{3} G L^{2} G}{30240} + \frac{\sqrt{5} \Delta t^{6} L G L^{3} G^{2} L^{2}}{30240} && \\
    & \phantom{P_{6, 0} =} + \frac{\sqrt{5} \Delta t^{6} L G L^{4} G L G}{30240} + \frac{\sqrt{5} \Delta t^{6} L G L^{4} G^{2} L}{30240} + \frac{\sqrt{5} \Delta t^{6} L G L^{5} G^{2}}{30240} + \frac{\sqrt{5} \Delta t^{6} L G^{2} L G L^{4}}{30240} + \frac{\sqrt{5} \Delta t^{6} L G^{2} L^{2} G L^{3}}{30240} && \\
    & \phantom{P_{6, 0} =} + \frac{\sqrt{5} \Delta t^{6} L G^{2} L^{3} G L^{2}}{30240} + \frac{\sqrt{5} \Delta t^{6} L G^{2} L^{4} G L}{30240} + \frac{\sqrt{5} \Delta t^{6} L G^{2} L^{5} G}{30240} + \frac{\sqrt{5} \Delta t^{6} L G^{3} L^{5}}{30240} + \frac{\sqrt{5} \Delta t^{6} L^{2} G L G L G L^{2}}{30240} && \\
    & \phantom{P_{6, 0} =} + \frac{\sqrt{5} \Delta t^{6} L^{2} G L G L^{2} G L}{30240} + \frac{\sqrt{5} \Delta t^{6} L^{2} G L G L^{3} G}{30240} + \frac{\sqrt{5} \Delta t^{6} L^{2} G L G^{2} L^{3}}{30240} + \frac{\sqrt{5} \Delta t^{6} L^{2} G L^{2} G L G L}{30240} + \frac{\sqrt{5} \Delta t^{6} L^{2} G L^{2} G L^{2} G}{30240} && \\
    & \phantom{P_{6, 0} =} + \frac{\sqrt{5} \Delta t^{6} L^{2} G L^{2} G^{2} L^{2}}{30240} + \frac{\sqrt{5} \Delta t^{6} L^{2} G L^{3} G L G}{30240} + \frac{\sqrt{5} \Delta t^{6} L^{2} G L^{3} G^{2} L}{30240} + \frac{\sqrt{5} \Delta t^{6} L^{2} G L^{4} G^{2}}{30240} + \frac{\sqrt{5} \Delta t^{6} L^{2} G^{2} L G L^{3}}{30240} && \\
    & \phantom{P_{6, 0} =} + \frac{\sqrt{5} \Delta t^{6} L^{2} G^{2} L^{2} G L^{2}}{30240} + \frac{\sqrt{5} \Delta t^{6} L^{2} G^{2} L^{3} G L}{30240} + \frac{\sqrt{5} \Delta t^{6} L^{2} G^{2} L^{4} G}{30240} + \frac{\sqrt{5} \Delta t^{6} L^{2} G^{3} L^{4}}{30240} + \frac{\sqrt{5} \Delta t^{6} L^{3} G L G L G L}{30240} && \\
    & \phantom{P_{6, 0} =} + \frac{\sqrt{5} \Delta t^{6} L^{3} G L G L^{2} G}{30240} + \frac{\sqrt{5} \Delta t^{6} L^{3} G L G^{2} L^{2}}{30240} + \frac{\sqrt{5} \Delta t^{6} L^{3} G L^{2} G L G}{30240} + \frac{\sqrt{5} \Delta t^{6} L^{3} G L^{2} G^{2} L}{30240} + \frac{\sqrt{5} \Delta t^{6} L^{3} G L^{3} G^{2}}{30240} && \\
    & \phantom{P_{6, 0} =} + \frac{\sqrt{5} \Delta t^{6} L^{3} G^{2} L G L^{2}}{30240} + \frac{\sqrt{5} \Delta t^{6} L^{3} G^{2} L^{2} G L}{30240} + \frac{\sqrt{5} \Delta t^{6} L^{3} G^{2} L^{3} G}{30240} + \frac{\sqrt{5} \Delta t^{6} L^{3} G^{3} L^{3}}{30240} + \frac{\sqrt{5} \Delta t^{6} L^{4} G L G L G}{30240} && \\
    & \phantom{P_{6, 0} =} + \frac{\sqrt{5} \Delta t^{6} L^{4} G L G^{2} L}{30240} + \frac{\sqrt{5} \Delta t^{6} L^{4} G L^{2} G^{2}}{30240} + \frac{\sqrt{5} \Delta t^{6} L^{4} G^{2} L G L}{30240} + \frac{\sqrt{5} \Delta t^{6} L^{4} G^{2} L^{2} G}{30240} + \frac{\sqrt{5} \Delta t^{6} L^{4} G^{3} L^{2}}{30240} && \\
    & \phantom{P_{6, 0} =} + \frac{\sqrt{5} \Delta t^{6} L^{5} G L G^{2}}{30240} + \frac{\sqrt{5} \Delta t^{6} L^{5} G^{2} L G}{30240} + \frac{\sqrt{5} \Delta t^{6} L^{5} G^{3} L}{30240} + \frac{\sqrt{5} \Delta t^{6} L^{6} G^{3}}{30240} + \frac{\sqrt{5} \Delta t^{5} G L G L^{5}}{3360} && \\
    & \phantom{P_{6, 0} =} + \frac{\sqrt{5} \Delta t^{5} G L^{2} G L^{4}}{3360} + \frac{\sqrt{5} \Delta t^{5} G L^{3} G L^{3}}{3360} + \frac{\sqrt{5} \Delta t^{5} G L^{4} G L^{2}}{3360} + \frac{\sqrt{5} \Delta t^{5} G L^{5} G L}{3360} + \frac{\sqrt{5} \Delta t^{5} G L^{6} G}{3360} && \\
    & \phantom{P_{6, 0} =} + \frac{\sqrt{5} \Delta t^{5} G^{2} L^{6}}{3360} + \frac{\sqrt{5} \Delta t^{5} L G L G L^{4}}{3360} + \frac{\sqrt{5} \Delta t^{5} L G L^{2} G L^{3}}{3360} + \frac{\sqrt{5} \Delta t^{5} L G L^{3} G L^{2}}{3360} + \frac{\sqrt{5} \Delta t^{5} L G L^{4} G L}{3360} && \\
    & \phantom{P_{6, 0} =} + \frac{\sqrt{5} \Delta t^{5} L G L^{5} G}{3360} + \frac{\sqrt{5} \Delta t^{5} L G^{2} L^{5}}{3360} + \frac{\sqrt{5} \Delta t^{5} L^{2} G L G L^{3}}{3360} + \frac{\sqrt{5} \Delta t^{5} L^{2} G L^{2} G L^{2}}{3360} + \frac{\sqrt{5} \Delta t^{5} L^{2} G L^{3} G L}{3360} && \\
    & \phantom{P_{6, 0} =} + \frac{\sqrt{5} \Delta t^{5} L^{2} G L^{4} G}{3360} + \frac{\sqrt{5} \Delta t^{5} L^{2} G^{2} L^{4}}{3360} + \frac{\sqrt{5} \Delta t^{5} L^{3} G L G L^{2}}{3360} + \frac{\sqrt{5} \Delta t^{5} L^{3} G L^{2} G L}{3360} + \frac{\sqrt{5} \Delta t^{5} L^{3} G L^{3} G}{3360} && \\
    & \phantom{P_{6, 0} =} + \frac{\sqrt{5} \Delta t^{5} L^{3} G^{2} L^{3}}{3360} + \frac{\sqrt{5} \Delta t^{5} L^{4} G L G L}{3360} + \frac{\sqrt{5} \Delta t^{5} L^{4} G L^{2} G}{3360} + \frac{\sqrt{5} \Delta t^{5} L^{4} G^{2} L^{2}}{3360} + \frac{\sqrt{5} \Delta t^{5} L^{5} G L G}{3360} && \\
    & \phantom{P_{6, 0} =} + \frac{\sqrt{5} \Delta t^{5} L^{5} G^{2} L}{3360} + \frac{\sqrt{5} \Delta t^{5} L^{6} G^{2}}{3360} + \frac{\sqrt{5} \Delta t^{4} G L^{6}}{420} + \frac{\sqrt{5} \Delta t^{4} L G L^{5}}{420} + \frac{\sqrt{5} \Delta t^{4} L^{2} G L^{4}}{420} && \\
    & \phantom{P_{6, 0} =} + \frac{\sqrt{5} \Delta t^{4} L^{3} G L^{3}}{420} + \frac{\sqrt{5} \Delta t^{4} L^{4} G L^{2}}{420} + \frac{\sqrt{5} \Delta t^{4} L^{5} G L}{420} + \frac{\sqrt{5} \Delta t^{4} L^{6} G}{420} + \frac{\sqrt{5} \Delta t^{3} L^{6}}{60} && \\
\end{flalign*}
\begin{flalign*}
    & P_{7, 0} = \frac{\sqrt{35} \Delta t^{\frac{11}{2}} G L G L^{6}}{30240} + \frac{\sqrt{35} \Delta t^{\frac{11}{2}} G L^{2} G L^{5}}{30240} + \frac{\sqrt{35} \Delta t^{\frac{11}{2}} G L^{3} G L^{4}}{30240} + \frac{\sqrt{35} \Delta t^{\frac{11}{2}} G L^{4} G L^{3}}{30240} + \frac{\sqrt{35} \Delta t^{\frac{11}{2}} G L^{5} G L^{2}}{30240} && \\
    & \phantom{P_{7, 0} =} + \frac{\sqrt{35} \Delta t^{\frac{11}{2}} G L^{6} G L}{30240} + \frac{\sqrt{35} \Delta t^{\frac{11}{2}} G L^{7} G}{30240} + \frac{\sqrt{35} \Delta t^{\frac{11}{2}} G^{2} L^{7}}{30240} + \frac{\sqrt{35} \Delta t^{\frac{11}{2}} L G L G L^{5}}{30240} + \frac{\sqrt{35} \Delta t^{\frac{11}{2}} L G L^{2} G L^{4}}{30240} && \\
    & \phantom{P_{7, 0} =} + \frac{\sqrt{35} \Delta t^{\frac{11}{2}} L G L^{3} G L^{3}}{30240} + \frac{\sqrt{35} \Delta t^{\frac{11}{2}} L G L^{4} G L^{2}}{30240} + \frac{\sqrt{35} \Delta t^{\frac{11}{2}} L G L^{5} G L}{30240} + \frac{\sqrt{35} \Delta t^{\frac{11}{2}} L G L^{6} G}{30240} + \frac{\sqrt{35} \Delta t^{\frac{11}{2}} L G^{2} L^{6}}{30240} && \\
    & \phantom{P_{7, 0} =} + \frac{\sqrt{35} \Delta t^{\frac{11}{2}} L^{2} G L G L^{4}}{30240} + \frac{\sqrt{35} \Delta t^{\frac{11}{2}} L^{2} G L^{2} G L^{3}}{30240} + \frac{\sqrt{35} \Delta t^{\frac{11}{2}} L^{2} G L^{3} G L^{2}}{30240} + \frac{\sqrt{35} \Delta t^{\frac{11}{2}} L^{2} G L^{4} G L}{30240} + \frac{\sqrt{35} \Delta t^{\frac{11}{2}} L^{2} G L^{5} G}{30240} && \\
    & \phantom{P_{7, 0} =} + \frac{\sqrt{35} \Delta t^{\frac{11}{2}} L^{2} G^{2} L^{5}}{30240} + \frac{\sqrt{35} \Delta t^{\frac{11}{2}} L^{3} G L G L^{3}}{30240} + \frac{\sqrt{35} \Delta t^{\frac{11}{2}} L^{3} G L^{2} G L^{2}}{30240} + \frac{\sqrt{35} \Delta t^{\frac{11}{2}} L^{3} G L^{3} G L}{30240} + \frac{\sqrt{35} \Delta t^{\frac{11}{2}} L^{3} G L^{4} G}{30240} && \\
    & \phantom{P_{7, 0} =} + \frac{\sqrt{35} \Delta t^{\frac{11}{2}} L^{3} G^{2} L^{4}}{30240} + \frac{\sqrt{35} \Delta t^{\frac{11}{2}} L^{4} G L G L^{2}}{30240} + \frac{\sqrt{35} \Delta t^{\frac{11}{2}} L^{4} G L^{2} G L}{30240} + \frac{\sqrt{35} \Delta t^{\frac{11}{2}} L^{4} G L^{3} G}{30240} + \frac{\sqrt{35} \Delta t^{\frac{11}{2}} L^{4} G^{2} L^{3}}{30240} && \\
    & \phantom{P_{7, 0} =} + \frac{\sqrt{35} \Delta t^{\frac{11}{2}} L^{5} G L G L}{30240} + \frac{\sqrt{35} \Delta t^{\frac{11}{2}} L^{5} G L^{2} G}{30240} + \frac{\sqrt{35} \Delta t^{\frac{11}{2}} L^{5} G^{2} L^{2}}{30240} + \frac{\sqrt{35} \Delta t^{\frac{11}{2}} L^{6} G L G}{30240} + \frac{\sqrt{35} \Delta t^{\frac{11}{2}} L^{6} G^{2} L}{30240} && \\
    & \phantom{P_{7, 0} =} + \frac{\sqrt{35} \Delta t^{\frac{11}{2}} L^{7} G^{2}}{30240} + \frac{\sqrt{35} \Delta t^{\frac{9}{2}} G L^{7}}{3360} + \frac{\sqrt{35} \Delta t^{\frac{9}{2}} L G L^{6}}{3360} + \frac{\sqrt{35} \Delta t^{\frac{9}{2}} L^{2} G L^{5}}{3360} + \frac{\sqrt{35} \Delta t^{\frac{9}{2}} L^{3} G L^{4}}{3360} && \\
    & \phantom{P_{7, 0} =} + \frac{\sqrt{35} \Delta t^{\frac{9}{2}} L^{4} G L^{3}}{3360} + \frac{\sqrt{35} \Delta t^{\frac{9}{2}} L^{5} G L^{2}}{3360} + \frac{\sqrt{35} \Delta t^{\frac{9}{2}} L^{6} G L}{3360} + \frac{\sqrt{35} \Delta t^{\frac{9}{2}} L^{7} G}{3360} + \frac{\sqrt{35} \Delta t^{\frac{7}{2}} L^{7}}{420} && \\
\end{flalign*}
\begin{flalign*}
    & P_{8, 0} = \frac{\sqrt{70} \Delta t^{6} G L G L^{7}}{151200} + \frac{\sqrt{70} \Delta t^{6} G L^{2} G L^{6}}{151200} + \frac{\sqrt{70} \Delta t^{6} G L^{3} G L^{5}}{151200} + \frac{\sqrt{70} \Delta t^{6} G L^{4} G L^{4}}{151200} + \frac{\sqrt{70} \Delta t^{6} G L^{5} G L^{3}}{151200} && \\
    & \phantom{P_{8, 0} =} + \frac{\sqrt{70} \Delta t^{6} G L^{6} G L^{2}}{151200} + \frac{\sqrt{70} \Delta t^{6} G L^{7} G L}{151200} + \frac{\sqrt{70} \Delta t^{6} G L^{8} G}{151200} + \frac{\sqrt{70} \Delta t^{6} G^{2} L^{8}}{151200} + \frac{\sqrt{70} \Delta t^{6} L G L G L^{6}}{151200} && \\
    & \phantom{P_{8, 0} =} + \frac{\sqrt{70} \Delta t^{6} L G L^{2} G L^{5}}{151200} + \frac{\sqrt{70} \Delta t^{6} L G L^{3} G L^{4}}{151200} + \frac{\sqrt{70} \Delta t^{6} L G L^{4} G L^{3}}{151200} + \frac{\sqrt{70} \Delta t^{6} L G L^{5} G L^{2}}{151200} + \frac{\sqrt{70} \Delta t^{6} L G L^{6} G L}{151200} && \\
    & \phantom{P_{8, 0} =} + \frac{\sqrt{70} \Delta t^{6} L G L^{7} G}{151200} + \frac{\sqrt{70} \Delta t^{6} L G^{2} L^{7}}{151200} + \frac{\sqrt{70} \Delta t^{6} L^{2} G L G L^{5}}{151200} + \frac{\sqrt{70} \Delta t^{6} L^{2} G L^{2} G L^{4}}{151200} + \frac{\sqrt{70} \Delta t^{6} L^{2} G L^{3} G L^{3}}{151200} && \\
    & \phantom{P_{8, 0} =} + \frac{\sqrt{70} \Delta t^{6} L^{2} G L^{4} G L^{2}}{151200} + \frac{\sqrt{70} \Delta t^{6} L^{2} G L^{5} G L}{151200} + \frac{\sqrt{70} \Delta t^{6} L^{2} G L^{6} G}{151200} + \frac{\sqrt{70} \Delta t^{6} L^{2} G^{2} L^{6}}{151200} + \frac{\sqrt{70} \Delta t^{6} L^{3} G L G L^{4}}{151200} && \\
    & \phantom{P_{8, 0} =} + \frac{\sqrt{70} \Delta t^{6} L^{3} G L^{2} G L^{3}}{151200} + \frac{\sqrt{70} \Delta t^{6} L^{3} G L^{3} G L^{2}}{151200} + \frac{\sqrt{70} \Delta t^{6} L^{3} G L^{4} G L}{151200} + \frac{\sqrt{70} \Delta t^{6} L^{3} G L^{5} G}{151200} + \frac{\sqrt{70} \Delta t^{6} L^{3} G^{2} L^{5}}{151200} && \\
    & \phantom{P_{8, 0} =} + \frac{\sqrt{70} \Delta t^{6} L^{4} G L G L^{3}}{151200} + \frac{\sqrt{70} \Delta t^{6} L^{4} G L^{2} G L^{2}}{151200} + \frac{\sqrt{70} \Delta t^{6} L^{4} G L^{3} G L}{151200} + \frac{\sqrt{70} \Delta t^{6} L^{4} G L^{4} G}{151200} + \frac{\sqrt{70} \Delta t^{6} L^{4} G^{2} L^{4}}{151200} && \\
    & \phantom{P_{8, 0} =} + \frac{\sqrt{70} \Delta t^{6} L^{5} G L G L^{2}}{151200} + \frac{\sqrt{70} \Delta t^{6} L^{5} G L^{2} G L}{151200} + \frac{\sqrt{70} \Delta t^{6} L^{5} G L^{3} G}{151200} + \frac{\sqrt{70} \Delta t^{6} L^{5} G^{2} L^{3}}{151200} + \frac{\sqrt{70} \Delta t^{6} L^{6} G L G L}{151200} && \\
    & \phantom{P_{8, 0} =} + \frac{\sqrt{70} \Delta t^{6} L^{6} G L^{2} G}{151200} + \frac{\sqrt{70} \Delta t^{6} L^{6} G^{2} L^{2}}{151200} + \frac{\sqrt{70} \Delta t^{6} L^{7} G L G}{151200} + \frac{\sqrt{70} \Delta t^{6} L^{7} G^{2} L}{151200} + \frac{\sqrt{70} \Delta t^{6} L^{8} G^{2}}{151200} && \\
    & \phantom{P_{8, 0} =} + \frac{\sqrt{70} \Delta t^{5} G L^{8}}{15120} + \frac{\sqrt{70} \Delta t^{5} L G L^{7}}{15120} + \frac{\sqrt{70} \Delta t^{5} L^{2} G L^{6}}{15120} + \frac{\sqrt{70} \Delta t^{5} L^{3} G L^{5}}{15120} + \frac{\sqrt{70} \Delta t^{5} L^{4} G L^{4}}{15120} && \\
    & \phantom{P_{8, 0} =} + \frac{\sqrt{70} \Delta t^{5} L^{5} G L^{3}}{15120} + \frac{\sqrt{70} \Delta t^{5} L^{6} G L^{2}}{15120} + \frac{\sqrt{70} \Delta t^{5} L^{7} G L}{15120} + \frac{\sqrt{70} \Delta t^{5} L^{8} G}{15120} + \frac{\sqrt{70} \Delta t^{4} L^{8}}{1680} && \\
\end{flalign*}
\begin{flalign*}
    & P_{9, 0} = \frac{\sqrt{70} \Delta t^{\frac{11}{2}} G L^{9}}{50400} + \frac{\sqrt{70} \Delta t^{\frac{11}{2}} L G L^{8}}{50400} + \frac{\sqrt{70} \Delta t^{\frac{11}{2}} L^{2} G L^{7}}{50400} + \frac{\sqrt{70} \Delta t^{\frac{11}{2}} L^{3} G L^{6}}{50400} + \frac{\sqrt{70} \Delta t^{\frac{11}{2}} L^{4} G L^{5}}{50400} && \\
    & \phantom{P_{9, 0} =} + \frac{\sqrt{70} \Delta t^{\frac{11}{2}} L^{5} G L^{4}}{50400} + \frac{\sqrt{70} \Delta t^{\frac{11}{2}} L^{6} G L^{3}}{50400} + \frac{\sqrt{70} \Delta t^{\frac{11}{2}} L^{7} G L^{2}}{50400} + \frac{\sqrt{70} \Delta t^{\frac{11}{2}} L^{8} G L}{50400} + \frac{\sqrt{70} \Delta t^{\frac{11}{2}} L^{9} G}{50400} && \\
    & \phantom{P_{9, 0} =} + \frac{\sqrt{70} \Delta t^{\frac{9}{2}} L^{9}}{5040} && \\
\end{flalign*}
\begin{flalign*}
    & P_{10, 0} = \frac{\sqrt{7} \Delta t^{6} G L^{10}}{55440} + \frac{\sqrt{7} \Delta t^{6} L G L^{9}}{55440} + \frac{\sqrt{7} \Delta t^{6} L^{2} G L^{8}}{55440} + \frac{\sqrt{7} \Delta t^{6} L^{3} G L^{7}}{55440} + \frac{\sqrt{7} \Delta t^{6} L^{4} G L^{6}}{55440} && \\
    & \phantom{P_{10, 0} =} + \frac{\sqrt{7} \Delta t^{6} L^{5} G L^{5}}{55440} + \frac{\sqrt{7} \Delta t^{6} L^{6} G L^{4}}{55440} + \frac{\sqrt{7} \Delta t^{6} L^{7} G L^{3}}{55440} + \frac{\sqrt{7} \Delta t^{6} L^{8} G L^{2}}{55440} + \frac{\sqrt{7} \Delta t^{6} L^{9} G L}{55440} && \\
    & \phantom{P_{10, 0} =} + \frac{\sqrt{7} \Delta t^{6} L^{10} G}{55440} + \frac{\sqrt{7} \Delta t^{5} L^{10}}{5040} && \\
\end{flalign*}
\begin{flalign*}
    & P_{11, 0} = \frac{\sqrt{77} \Delta t^{\frac{11}{2}} L^{11}}{55440} && \\
\end{flalign*}
\begin{flalign*}
    & P_{12, 0} = \frac{\sqrt{231} \Delta t^{6} L^{12}}{332640} && \\
\end{flalign*}
\begin{flalign*}
    & P_{0, 1} = \frac{\sqrt{3} \Delta t^{\frac{9}{2}} G L G^{3}}{360} - \frac{\sqrt{3} \Delta t^{\frac{9}{2}} G^{3} L G}{360} - \frac{\sqrt{3} \Delta t^{\frac{9}{2}} G^{4} L}{180} + \frac{\sqrt{3} \Delta t^{\frac{9}{2}} L G^{4}}{180} + \frac{\sqrt{3} \Delta t^{\frac{7}{2}} G L G^{2}}{120} && \\
    & \phantom{P_{0, 1} =} - \frac{\sqrt{3} \Delta t^{\frac{7}{2}} G^{2} L G}{120} - \frac{\sqrt{3} \Delta t^{\frac{7}{2}} G^{3} L}{40} + \frac{\sqrt{3} \Delta t^{\frac{7}{2}} L G^{3}}{40} - \frac{\sqrt{3} \Delta t^{\frac{5}{2}} G^{2} L}{12} + \frac{\sqrt{3} \Delta t^{\frac{5}{2}} L G^{2}}{12} && \\
    & \phantom{P_{0, 1} =} - \frac{\sqrt{3} \Delta t^{\frac{3}{2}} G L}{6} + \frac{\sqrt{3} \Delta t^{\frac{3}{2}} L G}{6} && \\
\end{flalign*}
\begin{flalign*}
    & P_{1, 1} = - \frac{\sqrt{3} \Delta t^{4} G L G^{2} L}{360} + \frac{\sqrt{3} \Delta t^{4} G L^{2} G^{2}}{360} - \frac{\sqrt{3} \Delta t^{4} G^{2} L G L}{180} - \frac{\sqrt{3} \Delta t^{4} G^{2} L^{2} G}{360} - \frac{\sqrt{3} \Delta t^{4} G^{3} L^{2}}{120} && \\
    & \phantom{P_{1, 1} =} + \frac{\sqrt{3} \Delta t^{4} L G L G^{2}}{180} + \frac{\sqrt{3} \Delta t^{4} L G^{2} L G}{360} + \frac{\sqrt{3} \Delta t^{4} L^{2} G^{3}}{120} - \frac{\sqrt{3} \Delta t^{3} G L G L}{60} - \frac{\sqrt{3} \Delta t^{3} G^{2} L^{2}}{30} && \\
    & \phantom{P_{1, 1} =} + \frac{\sqrt{3} \Delta t^{3} L G L G}{60} + \frac{\sqrt{3} \Delta t^{3} L^{2} G^{2}}{30} - \frac{\sqrt{3} \Delta t^{2} G L^{2}}{12} + \frac{\sqrt{3} \Delta t^{2} L^{2} G}{12} && \\
\end{flalign*}
\begin{flalign*}
    & P_{2, 1} = - \frac{\sqrt{6} \Delta t^{\frac{9}{2}} G L G L G L}{1680} - \frac{\sqrt{6} \Delta t^{\frac{9}{2}} G L G L^{2} G}{5040} - \frac{\sqrt{6} \Delta t^{\frac{9}{2}} G L G^{2} L^{2}}{1008} + \frac{\sqrt{6} \Delta t^{\frac{9}{2}} G L^{2} G L G}{5040} - \frac{\sqrt{6} \Delta t^{\frac{9}{2}} G L^{2} G^{2} L}{5040} && \\
    & \phantom{P_{2, 1} =} + \frac{\sqrt{6} \Delta t^{\frac{9}{2}} G L^{3} G^{2}}{1680} - \frac{\sqrt{6} \Delta t^{\frac{9}{2}} G^{2} L G L^{2}}{720} - \frac{\sqrt{6} \Delta t^{\frac{9}{2}} G^{2} L^{2} G L}{1008} - \frac{\sqrt{6} \Delta t^{\frac{9}{2}} G^{2} L^{3} G}{1680} - \frac{\sqrt{6} \Delta t^{\frac{9}{2}} G^{3} L^{3}}{560} && \\
    & \phantom{P_{2, 1} =} + \frac{\sqrt{6} \Delta t^{\frac{9}{2}} L G L G L G}{1680} + \frac{\sqrt{6} \Delta t^{\frac{9}{2}} L G L G^{2} L}{5040} + \frac{\sqrt{6} \Delta t^{\frac{9}{2}} L G L^{2} G^{2}}{1008} - \frac{\sqrt{6} \Delta t^{\frac{9}{2}} L G^{2} L G L}{5040} + \frac{\sqrt{6} \Delta t^{\frac{9}{2}} L G^{2} L^{2} G}{5040} && \\
    & \phantom{P_{2, 1} =} - \frac{\sqrt{6} \Delta t^{\frac{9}{2}} L G^{3} L^{2}}{1680} + \frac{\sqrt{6} \Delta t^{\frac{9}{2}} L^{2} G L G^{2}}{720} + \frac{\sqrt{6} \Delta t^{\frac{9}{2}} L^{2} G^{2} L G}{1008} + \frac{\sqrt{6} \Delta t^{\frac{9}{2}} L^{2} G^{3} L}{1680} + \frac{\sqrt{6} \Delta t^{\frac{9}{2}} L^{3} G^{3}}{560} && \\
    & \phantom{P_{2, 1} =} - \frac{\sqrt{6} \Delta t^{\frac{7}{2}} G L G L^{2}}{180} - \frac{\sqrt{6} \Delta t^{\frac{7}{2}} G L^{2} G L}{360} - \frac{\sqrt{6} \Delta t^{\frac{7}{2}} G^{2} L^{3}}{120} + \frac{\sqrt{6} \Delta t^{\frac{7}{2}} L G L^{2} G}{360} - \frac{\sqrt{6} \Delta t^{\frac{7}{2}} L G^{2} L^{2}}{360} && \\
    & \phantom{P_{2, 1} =} + \frac{\sqrt{6} \Delta t^{\frac{7}{2}} L^{2} G L G}{180} + \frac{\sqrt{6} \Delta t^{\frac{7}{2}} L^{2} G^{2} L}{360} + \frac{\sqrt{6} \Delta t^{\frac{7}{2}} L^{3} G^{2}}{120} - \frac{\sqrt{6} \Delta t^{\frac{5}{2}} G L^{3}}{40} - \frac{\sqrt{6} \Delta t^{\frac{5}{2}} L G L^{2}}{120} && \\
    & \phantom{P_{2, 1} =} + \frac{\sqrt{6} \Delta t^{\frac{5}{2}} L^{2} G L}{120} + \frac{\sqrt{6} \Delta t^{\frac{5}{2}} L^{3} G}{40} && \\
\end{flalign*}
\begin{flalign*}
    & P_{3, 1} = - \frac{\sqrt{2} \Delta t^{4} G L G L^{3}}{280} - \frac{\sqrt{2} \Delta t^{4} G L^{2} G L^{2}}{420} - \frac{\sqrt{2} \Delta t^{4} G L^{3} G L}{840} - \frac{\sqrt{2} \Delta t^{4} G^{2} L^{4}}{210} - \frac{\sqrt{2} \Delta t^{4} L G L G L^{2}}{840} && \\
    & \phantom{P_{3, 1} =} + \frac{\sqrt{2} \Delta t^{4} L G L^{3} G}{840} - \frac{\sqrt{2} \Delta t^{4} L G^{2} L^{3}}{420} + \frac{\sqrt{2} \Delta t^{4} L^{2} G L G L}{840} + \frac{\sqrt{2} \Delta t^{4} L^{2} G L^{2} G}{420} + \frac{\sqrt{2} \Delta t^{4} L^{3} G L G}{280} && \\
    & \phantom{P_{3, 1} =} + \frac{\sqrt{2} \Delta t^{4} L^{3} G^{2} L}{420} + \frac{\sqrt{2} \Delta t^{4} L^{4} G^{2}}{210} - \frac{\sqrt{2} \Delta t^{3} G L^{4}}{60} - \frac{\sqrt{2} \Delta t^{3} L G L^{3}}{120} + \frac{\sqrt{2} \Delta t^{3} L^{3} G L}{120} && \\
    & \phantom{P_{3, 1} =} + \frac{\sqrt{2} \Delta t^{3} L^{4} G}{60} && \\
\end{flalign*}
\begin{flalign*}
    & P_{4, 1} = - \frac{\sqrt{2} \Delta t^{\frac{9}{2}} G L G L^{4}}{840} - \frac{\sqrt{2} \Delta t^{\frac{9}{2}} G L^{2} G L^{3}}{1120} - \frac{\sqrt{2} \Delta t^{\frac{9}{2}} G L^{3} G L^{2}}{1680} - \frac{\sqrt{2} \Delta t^{\frac{9}{2}} G L^{4} G L}{3360} - \frac{\sqrt{2} \Delta t^{\frac{9}{2}} G^{2} L^{5}}{672} && \\
    & \phantom{P_{4, 1} =} - \frac{\sqrt{2} \Delta t^{\frac{9}{2}} L G L G L^{3}}{1680} - \frac{\sqrt{2} \Delta t^{\frac{9}{2}} L G L^{2} G L^{2}}{3360} + \frac{\sqrt{2} \Delta t^{\frac{9}{2}} L G L^{4} G}{3360} - \frac{\sqrt{2} \Delta t^{\frac{9}{2}} L G^{2} L^{4}}{1120} + \frac{\sqrt{2} \Delta t^{\frac{9}{2}} L^{2} G L^{2} G L}{3360} && \\
    & \phantom{P_{4, 1} =} + \frac{\sqrt{2} \Delta t^{\frac{9}{2}} L^{2} G L^{3} G}{1680} - \frac{\sqrt{2} \Delta t^{\frac{9}{2}} L^{2} G^{2} L^{3}}{3360} + \frac{\sqrt{2} \Delta t^{\frac{9}{2}} L^{3} G L G L}{1680} + \frac{\sqrt{2} \Delta t^{\frac{9}{2}} L^{3} G L^{2} G}{1120} + \frac{\sqrt{2} \Delta t^{\frac{9}{2}} L^{3} G^{2} L^{2}}{3360} && \\
    & \phantom{P_{4, 1} =} + \frac{\sqrt{2} \Delta t^{\frac{9}{2}} L^{4} G L G}{840} + \frac{\sqrt{2} \Delta t^{\frac{9}{2}} L^{4} G^{2} L}{1120} + \frac{\sqrt{2} \Delta t^{\frac{9}{2}} L^{5} G^{2}}{672} - \frac{\sqrt{2} \Delta t^{\frac{7}{2}} G L^{5}}{168} - \frac{\sqrt{2} \Delta t^{\frac{7}{2}} L G L^{4}}{280} && \\
    & \phantom{P_{4, 1} =} - \frac{\sqrt{2} \Delta t^{\frac{7}{2}} L^{2} G L^{3}}{840} + \frac{\sqrt{2} \Delta t^{\frac{7}{2}} L^{3} G L^{2}}{840} + \frac{\sqrt{2} \Delta t^{\frac{7}{2}} L^{4} G L}{280} + \frac{\sqrt{2} \Delta t^{\frac{7}{2}} L^{5} G}{168} && \\
\end{flalign*}
\begin{flalign*}
    & P_{5, 1} = - \frac{\sqrt{10} \Delta t^{4} G L^{6}}{1120} - \frac{\sqrt{10} \Delta t^{4} L G L^{5}}{1680} - \frac{\sqrt{10} \Delta t^{4} L^{2} G L^{4}}{3360} + \frac{\sqrt{10} \Delta t^{4} L^{4} G L^{2}}{3360} + \frac{\sqrt{10} \Delta t^{4} L^{5} G L}{1680} && \\
    & \phantom{P_{5, 1} =} + \frac{\sqrt{10} \Delta t^{4} L^{6} G}{1120} && \\
\end{flalign*}
\begin{flalign*}
    & P_{6, 1} = - \frac{\sqrt{15} \Delta t^{\frac{9}{2}} G L^{7}}{4320} - \frac{\sqrt{15} \Delta t^{\frac{9}{2}} L G L^{6}}{6048} - \frac{\sqrt{15} \Delta t^{\frac{9}{2}} L^{2} G L^{5}}{10080} - \frac{\sqrt{15} \Delta t^{\frac{9}{2}} L^{3} G L^{4}}{30240} + \frac{\sqrt{15} \Delta t^{\frac{9}{2}} L^{4} G L^{3}}{30240} && \\
    & \phantom{P_{6, 1} =} + \frac{\sqrt{15} \Delta t^{\frac{9}{2}} L^{5} G L^{2}}{10080} + \frac{\sqrt{15} \Delta t^{\frac{9}{2}} L^{6} G L}{6048} + \frac{\sqrt{15} \Delta t^{\frac{9}{2}} L^{7} G}{4320} && \\
\end{flalign*}
\begin{flalign*}
    & P_{0, 2} = \frac{\sqrt{5} \Delta t^{\frac{7}{2}} G L G^{2}}{120} + \frac{\sqrt{5} \Delta t^{\frac{7}{2}} G^{2} L G}{120} - \frac{\sqrt{5} \Delta t^{\frac{7}{2}} G^{3} L}{120} - \frac{\sqrt{5} \Delta t^{\frac{7}{2}} L G^{3}}{120} + \frac{\sqrt{5} \Delta t^{\frac{5}{2}} G L G}{30} && \\
    & \phantom{P_{0, 2} =} - \frac{\sqrt{5} \Delta t^{\frac{5}{2}} G^{2} L}{60} - \frac{\sqrt{5} \Delta t^{\frac{5}{2}} L G^{2}}{60} && \\
\end{flalign*}
\begin{flalign*}
    & P_{1, 2} = \frac{\sqrt{5} \Delta t^{4} G L G L G}{420} - \frac{\sqrt{5} \Delta t^{4} G L G^{2} L}{840} + \frac{\sqrt{5} \Delta t^{4} G L^{2} G^{2}}{280} + \frac{\sqrt{5} \Delta t^{4} G^{2} L^{2} G}{280} - \frac{\sqrt{5} \Delta t^{4} G^{3} L^{2}}{840} && \\
    & \phantom{P_{1, 2} =} - \frac{\sqrt{5} \Delta t^{4} L G^{2} L G}{840} - \frac{\sqrt{5} \Delta t^{4} L G^{3} L}{210} - \frac{\sqrt{5} \Delta t^{4} L^{2} G^{3}}{840} + \frac{\sqrt{5} \Delta t^{3} G L^{2} G}{60} - \frac{\sqrt{5} \Delta t^{3} L G^{2} L}{60} && \\
    & \phantom{P_{1, 2} =} + \frac{\sqrt{5} \Delta t^{2} G L^{2}}{60} - \frac{\sqrt{5} \Delta t^{2} L G L}{30} + \frac{\sqrt{5} \Delta t^{2} L^{2} G}{60} && \\
\end{flalign*}
\begin{flalign*}
    & P_{2, 2} = \frac{\sqrt{10} \Delta t^{\frac{7}{2}} G L^{2} G L}{840} + \frac{\sqrt{10} \Delta t^{\frac{7}{2}} G L^{3} G}{210} + \frac{\sqrt{10} \Delta t^{\frac{7}{2}} G^{2} L^{3}}{840} - \frac{\sqrt{10} \Delta t^{\frac{7}{2}} L G L G L}{420} + \frac{\sqrt{10} \Delta t^{\frac{7}{2}} L G L^{2} G}{840} && \\
    & \phantom{P_{2, 2} =} - \frac{\sqrt{10} \Delta t^{\frac{7}{2}} L G^{2} L^{2}}{280} - \frac{\sqrt{10} \Delta t^{\frac{7}{2}} L^{2} G^{2} L}{280} + \frac{\sqrt{10} \Delta t^{\frac{7}{2}} L^{3} G^{2}}{840} + \frac{\sqrt{10} \Delta t^{\frac{5}{2}} G L^{3}}{120} - \frac{\sqrt{10} \Delta t^{\frac{5}{2}} L G L^{2}}{120} && \\
    & \phantom{P_{2, 2} =} - \frac{\sqrt{10} \Delta t^{\frac{5}{2}} L^{2} G L}{120} + \frac{\sqrt{10} \Delta t^{\frac{5}{2}} L^{3} G}{120} && \\
\end{flalign*}
\begin{flalign*}
    & P_{3, 2} = \frac{\sqrt{30} \Delta t^{4} G L G L^{3}}{10080} + \frac{\sqrt{30} \Delta t^{4} G L^{2} G L^{2}}{10080} + \frac{\sqrt{30} \Delta t^{4} G L^{3} G L}{2520} + \frac{\sqrt{30} \Delta t^{4} G L^{4} G}{1008} + \frac{\sqrt{30} \Delta t^{4} G^{2} L^{4}}{2520} && \\
    & \phantom{P_{3, 2} =} - \frac{\sqrt{30} \Delta t^{4} L G L G L^{2}}{2016} - \frac{\sqrt{30} \Delta t^{4} L G L^{2} G L}{5040} + \frac{\sqrt{30} \Delta t^{4} L G L^{3} G}{2520} - \frac{\sqrt{30} \Delta t^{4} L G^{2} L^{3}}{2016} - \frac{\sqrt{30} \Delta t^{4} L^{2} G L G L}{2016} && \\
    & \phantom{P_{3, 2} =} + \frac{\sqrt{30} \Delta t^{4} L^{2} G L^{2} G}{10080} - \frac{\sqrt{30} \Delta t^{4} L^{2} G^{2} L^{2}}{1260} + \frac{\sqrt{30} \Delta t^{4} L^{3} G L G}{10080} - \frac{\sqrt{30} \Delta t^{4} L^{3} G^{2} L}{2016} + \frac{\sqrt{30} \Delta t^{4} L^{4} G^{2}}{2520} && \\
    & \phantom{P_{3, 2} =} + \frac{\sqrt{30} \Delta t^{3} G L^{4}}{420} - \frac{\sqrt{30} \Delta t^{3} L G L^{3}}{840} - \frac{\sqrt{30} \Delta t^{3} L^{2} G L^{2}}{420} - \frac{\sqrt{30} \Delta t^{3} L^{3} G L}{840} + \frac{\sqrt{30} \Delta t^{3} L^{4} G}{420} && \\
\end{flalign*}
\begin{flalign*}
    & P_{4, 2} = \frac{\sqrt{30} \Delta t^{\frac{7}{2}} G L^{5}}{1008} - \frac{\sqrt{30} \Delta t^{\frac{7}{2}} L G L^{4}}{5040} - \frac{\sqrt{30} \Delta t^{\frac{7}{2}} L^{2} G L^{3}}{1260} - \frac{\sqrt{30} \Delta t^{\frac{7}{2}} L^{3} G L^{2}}{1260} - \frac{\sqrt{30} \Delta t^{\frac{7}{2}} L^{4} G L}{5040} && \\
    & \phantom{P_{4, 2} =} + \frac{\sqrt{30} \Delta t^{\frac{7}{2}} L^{5} G}{1008} && \\
\end{flalign*}
\begin{flalign*}
    & P_{5, 2} = \frac{5 \sqrt{6} \Delta t^{4} G L^{6}}{6048} - \frac{\sqrt{6} \Delta t^{4} L^{2} G L^{4}}{2016} - \frac{\sqrt{6} \Delta t^{4} L^{3} G L^{3}}{1512} - \frac{\sqrt{6} \Delta t^{4} L^{4} G L^{2}}{2016} + \frac{5 \sqrt{6} \Delta t^{4} L^{6} G}{6048} && \\
\end{flalign*}
\begin{flalign*}
    & P_{0, 3} = - \frac{\sqrt{5} \Delta t^{4} G L G L G}{840} - \frac{\sqrt{5} \Delta t^{4} G L G^{2} L}{280} + \frac{\sqrt{5} \Delta t^{4} G L^{2} G^{2}}{420} + \frac{\sqrt{5} \Delta t^{4} G^{2} L^{2} G}{420} + \frac{\sqrt{5} \Delta t^{4} G^{3} L^{2}}{210} && \\
    & \phantom{P_{0, 3} =} - \frac{\sqrt{5} \Delta t^{4} L G^{2} L G}{280} - \frac{\sqrt{5} \Delta t^{4} L G^{3} L}{168} + \frac{\sqrt{5} \Delta t^{4} L^{2} G^{3}}{210} - \frac{\sqrt{5} \Delta t^{3} G L G L}{120} + \frac{\sqrt{5} \Delta t^{3} G L^{2} G}{120} && \\
    & \phantom{P_{0, 3} =} + \frac{\sqrt{5} \Delta t^{3} G^{2} L^{2}}{60} - \frac{\sqrt{5} \Delta t^{3} L G L G}{120} - \frac{\sqrt{5} \Delta t^{3} L G^{2} L}{40} + \frac{\sqrt{5} \Delta t^{3} L^{2} G^{2}}{60} + \frac{\sqrt{5} \Delta t^{2} G L^{2}}{30} && \\
    & \phantom{P_{0, 3} =} - \frac{\sqrt{5} \Delta t^{2} L G L}{15} + \frac{\sqrt{5} \Delta t^{2} L^{2} G}{30} && \\
\end{flalign*}
\begin{flalign*}
    & P_{1, 3} = - \frac{\sqrt{5} \Delta t^{\frac{7}{2}} G L^{2} G L}{840} + \frac{\sqrt{5} \Delta t^{\frac{7}{2}} G L^{3} G}{280} + \frac{\sqrt{5} \Delta t^{\frac{7}{2}} G^{2} L^{3}}{140} - \frac{\sqrt{5} \Delta t^{\frac{7}{2}} L G L G L}{168} - \frac{\sqrt{5} \Delta t^{\frac{7}{2}} L G L^{2} G}{840} && \\
    & \phantom{P_{1, 3} =} - \frac{\sqrt{5} \Delta t^{\frac{7}{2}} L G^{2} L^{2}}{210} - \frac{\sqrt{5} \Delta t^{\frac{7}{2}} L^{2} G^{2} L}{210} + \frac{\sqrt{5} \Delta t^{\frac{7}{2}} L^{3} G^{2}}{140} + \frac{\sqrt{5} \Delta t^{\frac{5}{2}} G L^{3}}{60} - \frac{\sqrt{5} \Delta t^{\frac{5}{2}} L G L^{2}}{60} && \\
    & \phantom{P_{1, 3} =} - \frac{\sqrt{5} \Delta t^{\frac{5}{2}} L^{2} G L}{60} + \frac{\sqrt{5} \Delta t^{\frac{5}{2}} L^{3} G}{60} && \\
\end{flalign*}
\begin{flalign*}
    & P_{2, 3} = \frac{\sqrt{10} \Delta t^{4} G L G L^{3}}{2240} - \frac{\sqrt{10} \Delta t^{4} G L^{2} G L^{2}}{6720} + \frac{\sqrt{10} \Delta t^{4} G L^{4} G}{1120} + \frac{\sqrt{10} \Delta t^{4} G^{2} L^{4}}{560} - \frac{\sqrt{10} \Delta t^{4} L G L G L^{2}}{960} && \\
    & \phantom{P_{2, 3} =} - \frac{\sqrt{10} \Delta t^{4} L G L^{2} G L}{1120} - \frac{\sqrt{10} \Delta t^{4} L G^{2} L^{3}}{2240} - \frac{\sqrt{10} \Delta t^{4} L^{2} G L G L}{960} - \frac{\sqrt{10} \Delta t^{4} L^{2} G L^{2} G}{6720} - \frac{\sqrt{10} \Delta t^{4} L^{2} G^{2} L^{2}}{840} && \\
    & \phantom{P_{2, 3} =} + \frac{\sqrt{10} \Delta t^{4} L^{3} G L G}{2240} - \frac{\sqrt{10} \Delta t^{4} L^{3} G^{2} L}{2240} + \frac{\sqrt{10} \Delta t^{4} L^{4} G^{2}}{560} + \frac{\sqrt{10} \Delta t^{3} G L^{4}}{210} - \frac{\sqrt{10} \Delta t^{3} L G L^{3}}{420} && \\
    & \phantom{P_{2, 3} =} - \frac{\sqrt{10} \Delta t^{3} L^{2} G L^{2}}{210} - \frac{\sqrt{10} \Delta t^{3} L^{3} G L}{420} + \frac{\sqrt{10} \Delta t^{3} L^{4} G}{210} && \\
\end{flalign*}
\begin{flalign*}
    & P_{3, 3} = \frac{\sqrt{30} \Delta t^{\frac{7}{2}} G L^{5}}{1008} - \frac{\sqrt{30} \Delta t^{\frac{7}{2}} L G L^{4}}{5040} - \frac{\sqrt{30} \Delta t^{\frac{7}{2}} L^{2} G L^{3}}{1260} - \frac{\sqrt{30} \Delta t^{\frac{7}{2}} L^{3} G L^{2}}{1260} - \frac{\sqrt{30} \Delta t^{\frac{7}{2}} L^{4} G L}{5040} && \\
    & \phantom{P_{3, 3} =} + \frac{\sqrt{30} \Delta t^{\frac{7}{2}} L^{5} G}{1008} && \\
\end{flalign*}
\begin{flalign*}
    & P_{4, 3} = \frac{\sqrt{30} \Delta t^{4} G L^{6}}{3024} - \frac{\sqrt{30} \Delta t^{4} L^{2} G L^{4}}{5040} - \frac{\sqrt{30} \Delta t^{4} L^{3} G L^{3}}{3780} - \frac{\sqrt{30} \Delta t^{4} L^{4} G L^{2}}{5040} + \frac{\sqrt{30} \Delta t^{4} L^{6} G}{3024} && \\
\end{flalign*}
\begin{flalign*}
    & P_{0, 4} = - \frac{\sqrt{7} \Delta t^{\frac{7}{2}} G L G^{2}}{280} + \frac{\sqrt{7} \Delta t^{\frac{7}{2}} G^{2} L G}{280} - \frac{\sqrt{7} \Delta t^{\frac{7}{2}} G^{3} L}{840} + \frac{\sqrt{7} \Delta t^{\frac{7}{2}} L G^{3}}{840} && \\
\end{flalign*}
\begin{flalign*}
    & P_{1, 4} = - \frac{\sqrt{7} \Delta t^{3} G L G L}{210} + \frac{\sqrt{7} \Delta t^{3} G^{2} L^{2}}{420} + \frac{\sqrt{7} \Delta t^{3} L G L G}{210} - \frac{\sqrt{7} \Delta t^{3} L^{2} G^{2}}{420} && \\
\end{flalign*}
\begin{flalign*}
    & P_{2, 4} = - \frac{\sqrt{14} \Delta t^{\frac{7}{2}} G L G L^{2}}{1680} - \frac{\sqrt{14} \Delta t^{\frac{7}{2}} G L^{2} G L}{560} + \frac{\sqrt{14} \Delta t^{\frac{7}{2}} G^{2} L^{3}}{1680} + \frac{\sqrt{14} \Delta t^{\frac{7}{2}} L G L^{2} G}{560} + \frac{\sqrt{14} \Delta t^{\frac{7}{2}} L G^{2} L^{2}}{840} && \\
    & \phantom{P_{2, 4} =} + \frac{\sqrt{14} \Delta t^{\frac{7}{2}} L^{2} G L G}{1680} - \frac{\sqrt{14} \Delta t^{\frac{7}{2}} L^{2} G^{2} L}{840} - \frac{\sqrt{14} \Delta t^{\frac{7}{2}} L^{3} G^{2}}{1680} - \frac{\sqrt{14} \Delta t^{\frac{5}{2}} G L^{3}}{840} + \frac{\sqrt{14} \Delta t^{\frac{5}{2}} L G L^{2}}{280} && \\
    & \phantom{P_{2, 4} =} - \frac{\sqrt{14} \Delta t^{\frac{5}{2}} L^{2} G L}{280} + \frac{\sqrt{14} \Delta t^{\frac{5}{2}} L^{3} G}{840} && \\
\end{flalign*}
\begin{flalign*}
    & P_{3, 4} = - \frac{\sqrt{42} \Delta t^{3} G L^{4}}{1680} + \frac{\sqrt{42} \Delta t^{3} L G L^{3}}{840} - \frac{\sqrt{42} \Delta t^{3} L^{3} G L}{840} + \frac{\sqrt{42} \Delta t^{3} L^{4} G}{1680} && \\
\end{flalign*}
\begin{flalign*}
    & P_{4, 4} = - \frac{\sqrt{42} \Delta t^{\frac{7}{2}} G L^{5}}{3024} + \frac{\sqrt{42} \Delta t^{\frac{7}{2}} L G L^{4}}{2160} + \frac{\sqrt{42} \Delta t^{\frac{7}{2}} L^{2} G L^{3}}{3780} - \frac{\sqrt{42} \Delta t^{\frac{7}{2}} L^{3} G L^{2}}{3780} - \frac{\sqrt{42} \Delta t^{\frac{7}{2}} L^{4} G L}{2160} && \\
    & \phantom{P_{4, 4} =} + \frac{\sqrt{42} \Delta t^{\frac{7}{2}} L^{5} G}{3024} && \\
\end{flalign*}
\begin{flalign*}
    & P_{0, 5} = - \frac{\sqrt{70} \Delta t^{3} G L G L}{420} + \frac{\sqrt{70} \Delta t^{3} G^{2} L^{2}}{840} + \frac{\sqrt{70} \Delta t^{3} L G L G}{420} - \frac{\sqrt{70} \Delta t^{3} L^{2} G^{2}}{840} && \\
\end{flalign*}
\begin{flalign*}
    & P_{1, 5} = - \frac{\sqrt{70} \Delta t^{\frac{7}{2}} G L^{2} G L}{840} + \frac{\sqrt{70} \Delta t^{\frac{7}{2}} L G L^{2} G}{840} + \frac{\sqrt{70} \Delta t^{\frac{7}{2}} L G^{2} L^{2}}{840} - \frac{\sqrt{70} \Delta t^{\frac{7}{2}} L^{2} G^{2} L}{840} - \frac{\sqrt{70} \Delta t^{\frac{5}{2}} G L^{3}}{840} && \\
    & \phantom{P_{1, 5} =} + \frac{\sqrt{70} \Delta t^{\frac{5}{2}} L G L^{2}}{280} - \frac{\sqrt{70} \Delta t^{\frac{5}{2}} L^{2} G L}{280} + \frac{\sqrt{70} \Delta t^{\frac{5}{2}} L^{3} G}{840} && \\
\end{flalign*}
\begin{flalign*}
    & P_{2, 5} = - \frac{\sqrt{35} \Delta t^{3} G L^{4}}{840} + \frac{\sqrt{35} \Delta t^{3} L G L^{3}}{420} - \frac{\sqrt{35} \Delta t^{3} L^{3} G L}{420} + \frac{\sqrt{35} \Delta t^{3} L^{4} G}{840} && \\
\end{flalign*}
\begin{flalign*}
    & P_{3, 5} = - \frac{\sqrt{105} \Delta t^{\frac{7}{2}} G L^{5}}{3024} + \frac{\sqrt{105} \Delta t^{\frac{7}{2}} L G L^{4}}{2160} + \frac{\sqrt{105} \Delta t^{\frac{7}{2}} L^{2} G L^{3}}{3780} - \frac{\sqrt{105} \Delta t^{\frac{7}{2}} L^{3} G L^{2}}{3780} - \frac{\sqrt{105} \Delta t^{\frac{7}{2}} L^{4} G L}{2160} && \\
    & \phantom{P_{3, 5} =} + \frac{\sqrt{105} \Delta t^{\frac{7}{2}} L^{5} G}{3024} && \\
\end{flalign*}
\begin{flalign*}
    & P_{0, 6} = \frac{\sqrt{210} \Delta t^{\frac{7}{2}} G L G L^{2}}{1680} - \frac{\sqrt{210} \Delta t^{\frac{7}{2}} G L^{2} G L}{1680} - \frac{\sqrt{210} \Delta t^{\frac{7}{2}} G^{2} L^{3}}{1680} + \frac{\sqrt{210} \Delta t^{\frac{7}{2}} L G L^{2} G}{1680} + \frac{\sqrt{210} \Delta t^{\frac{7}{2}} L G^{2} L^{2}}{840} && \\
    & \phantom{P_{0, 6} =} - \frac{\sqrt{210} \Delta t^{\frac{7}{2}} L^{2} G L G}{1680} - \frac{\sqrt{210} \Delta t^{\frac{7}{2}} L^{2} G^{2} L}{840} + \frac{\sqrt{210} \Delta t^{\frac{7}{2}} L^{3} G^{2}}{1680} - \frac{\sqrt{210} \Delta t^{\frac{5}{2}} G L^{3}}{840} + \frac{\sqrt{210} \Delta t^{\frac{5}{2}} L G L^{2}}{280} && \\
    & \phantom{P_{0, 6} =} - \frac{\sqrt{210} \Delta t^{\frac{5}{2}} L^{2} G L}{280} + \frac{\sqrt{210} \Delta t^{\frac{5}{2}} L^{3} G}{840} && \\
\end{flalign*}
\begin{flalign*}
    & P_{1, 6} = - \frac{\sqrt{210} \Delta t^{3} G L^{4}}{1680} + \frac{\sqrt{210} \Delta t^{3} L G L^{3}}{840} - \frac{\sqrt{210} \Delta t^{3} L^{3} G L}{840} + \frac{\sqrt{210} \Delta t^{3} L^{4} G}{1680} && \\
\end{flalign*}
\begin{flalign*}
    & P_{2, 6} = - \frac{\sqrt{105} \Delta t^{\frac{7}{2}} G L^{5}}{3024} + \frac{\sqrt{105} \Delta t^{\frac{7}{2}} L G L^{4}}{2160} + \frac{\sqrt{105} \Delta t^{\frac{7}{2}} L^{2} G L^{3}}{3780} - \frac{\sqrt{105} \Delta t^{\frac{7}{2}} L^{3} G L^{2}}{3780} - \frac{\sqrt{105} \Delta t^{\frac{7}{2}} L^{4} G L}{2160} && \\
    & \phantom{P_{2, 6} =} + \frac{\sqrt{105} \Delta t^{\frac{7}{2}} L^{5} G}{3024} && \\
\end{flalign*}
\begin{flalign*}
    & P_{0, 7} = \frac{\sqrt{5} \Delta t^{3} G L G L}{120} - \frac{\sqrt{5} \Delta t^{3} G L^{2} G}{120} + \frac{\sqrt{5} \Delta t^{3} L G L G}{120} - \frac{\sqrt{5} \Delta t^{3} L G^{2} L}{120} && \\
\end{flalign*}
\begin{flalign*}
    & P_{1, 7} = 0 && \\
\end{flalign*}
\begin{flalign*}
    & P_{2, 7} = 0 && \\
\end{flalign*}
\begin{flalign*}
    & P_{0, 8} = 0 && \\
\end{flalign*}
\begin{flalign*}
    & P_{1, 8} = 0 && \\
\end{flalign*}
\begin{flalign*}
    & P_{2, 8} = 0 && \\
\end{flalign*}
\begin{flalign*}
    & P_{3, 8} = \frac{\sqrt{6} \Delta t^{3} G L^{4}}{5040} - \frac{\sqrt{6} \Delta t^{3} L G L^{3}}{1260} + \frac{\sqrt{6} \Delta t^{3} L^{2} G L^{2}}{840} - \frac{\sqrt{6} \Delta t^{3} L^{3} G L}{1260} + \frac{\sqrt{6} \Delta t^{3} L^{4} G}{5040} && \\
\end{flalign*}
\begin{flalign*}
    & P_{0, 9} = 0 && \\
\end{flalign*}
\begin{flalign*}
    & P_{1, 9} = 0 && \\
\end{flalign*}
\begin{flalign*}
    & P_{2, 9} = \frac{\Delta t^{3} G L^{4}}{840} - \frac{\Delta t^{3} L G L^{3}}{210} + \frac{\Delta t^{3} L^{2} G L^{2}}{140} - \frac{\Delta t^{3} L^{3} G L}{210} + \frac{\Delta t^{3} L^{4} G}{840} && \\
\end{flalign*}
\begin{flalign*}
    & P_{0, 10} = 0 && \\
\end{flalign*}
\begin{flalign*}
    & P_{1, 10} = \frac{\sqrt{14} \Delta t^{3} G L^{4}}{1680} - \frac{\sqrt{14} \Delta t^{3} L G L^{3}}{420} + \frac{\sqrt{14} \Delta t^{3} L^{2} G L^{2}}{280} - \frac{\sqrt{14} \Delta t^{3} L^{3} G L}{420} + \frac{\sqrt{14} \Delta t^{3} L^{4} G}{1680} && \\
\end{flalign*}
\begin{flalign*}
    & P_{0, 11} = \frac{\sqrt{21} \Delta t^{3} G L^{4}}{1260} - \frac{\sqrt{21} \Delta t^{3} L G L^{3}}{315} + \frac{\sqrt{21} \Delta t^{3} L^{2} G L^{2}}{210} - \frac{\sqrt{21} \Delta t^{3} L^{3} G L}{315} + \frac{\sqrt{21} \Delta t^{3} L^{4} G}{1260} && \\
\end{flalign*}
}